\documentclass[]{aastex631}

\newcommand{\E}[1]{\times 10^{#1}}
\newcommand{\tw}{$^{12}$CO }
\newcommand{\tht}{$^{13}$CO } 
\newcommand{\ei}{C$^{18}$O }
\newcommand{\msun}{$M_\odot$}
\newcommand{\hii}{H~\textsc{ii} }

\shorttitle{Sample article}
\shortauthors{Guo et al.}
\graphicspath{{./}{figures/}}

\begin{document}

\title{CO(J = 1-0) Observations toward the Filamentary Cloud in the Galactic Region of  153\fdg60 $\leqslant l \leqslant$ 156\fdg50 and 1\fdg85 $\leqslant b \leqslant$ 3\fdg50}

\correspondingauthor{Weihua Guo, Xuepeng Chen}
\email{whguo@pmo.ac.cn, xpchen@pmo.ac.cn}

\author{Weihua Guo}
\affiliation{Purple Mountain Observatory and Key Laboratory of Radio Astronomy, Chinese Academy of Sciences,
10 Yuanhua Road, Nanjing 210034, China}
\affiliation{School of Astronomy and Space Science, University of Science and Technology of China, Hefei, Anhui 230026, China}

\author{Xuepeng Chen}
\affiliation{Purple Mountain Observatory and Key Laboratory of Radio Astronomy, Chinese Academy of Sciences,
10 Yuanhua Road, Nanjing 210034, China}
\affiliation{School of Astronomy and Space Science, University of Science and Technology of China, Hefei, Anhui 230026, China}

\author{Jiancheng Feng}
\affiliation{Purple Mountain Observatory and Key Laboratory of Radio Astronomy, Chinese Academy of Sciences,
10 Yuanhua Road, Nanjing 210034, China}
\affiliation{School of Astronomy and Space Science, University of Science and Technology of China, Hefei, Anhui 230026, China}

\author{Li Sun}
\affiliation{Purple Mountain Observatory and Key Laboratory of Radio Astronomy, Chinese Academy of Sciences,
10 Yuanhua Road, Nanjing 210034, China}
\affiliation{School of Astronomy and Space Science, University of Science and Technology of China, Hefei, Anhui 230026, China}

\author{Shiyu Zhang}
\affiliation{Purple Mountain Observatory and Key Laboratory of Radio Astronomy, Chinese Academy of Sciences,
10 Yuanhua Road, Nanjing 210034, China}
\affiliation{School of Astronomy and Space Science, University of Science and Technology of China, Hefei, Anhui 230026, China}

\author{Chen Wang}
\affiliation{Purple Mountain Observatory and Key Laboratory of Radio Astronomy, Chinese Academy of Sciences,
10 Yuanhua Road, Nanjing 210034, China}

\author{Yang Su}
\affiliation{Purple Mountain Observatory and Key Laboratory of Radio Astronomy, Chinese Academy of Sciences,
10 Yuanhua Road, Nanjing 210034, China}
\affiliation{School of Astronomy and Space Science, University of Science and Technology of China, Hefei, Anhui 230026, China}

\author{Yan Sun}
\affiliation{Purple Mountain Observatory and Key Laboratory of Radio Astronomy, Chinese Academy of Sciences,
10 Yuanhua Road, Nanjing 210034, China}
\affiliation{School of Astronomy and Space Science, University of Science and Technology of China, Hefei, Anhui 230026, China}

\author{Qingzeng Yan}
\affiliation{Purple Mountain Observatory and Key Laboratory of Radio Astronomy, Chinese Academy of Sciences,
10 Yuanhua Road, Nanjing 210034, China}

\author{Shaobo Zhang}
\affiliation{Purple Mountain Observatory and Key Laboratory of Radio Astronomy, Chinese Academy of Sciences,
10 Yuanhua Road, Nanjing 210034, China}

\author{Xin Zhou}
\affiliation{Purple Mountain Observatory and Key Laboratory of Radio Astronomy, Chinese Academy of Sciences,
10 Yuanhua Road, Nanjing 210034, China}

\author{MiaoMiao Zhang}
\affiliation{Purple Mountain Observatory and Key Laboratory of Radio Astronomy, Chinese Academy of Sciences,
10 Yuanhua Road, Nanjing 210034, China}

\author{Min Fang}
\affiliation{Purple Mountain Observatory and Key Laboratory of Radio Astronomy, Chinese Academy of Sciences,
10 Yuanhua Road, Nanjing 210034, China}

\author{Ji Yang}
\affiliation{Purple Mountain Observatory and Key Laboratory of Radio Astronomy, Chinese Academy of Sciences,
10 Yuanhua Road, Nanjing 210034, China}
\affiliation{School of Astronomy and Space Science, University of Science and Technology of China, Hefei, Anhui 230026, China}

\begin{abstract}

	   We present observations of $J$=1-0 transition lines of ${ }^{12} \mathrm{CO}$, 
	   ${ }^{13} \mathrm{CO}$, and $\mathrm{C}^{18} \mathrm{O}$ towards the Galactic region of 
	   153\fdg60$ \leqslant l \leqslant$ 156\fdg50 and 1\fdg85$ \leqslant b \leqslant$ 3\fdg50, 
	   using the Purple Mountain Observatory (PMO) 13.7 m millimeter telescope. 
	   Based on the \tht data,  one main filament and five sub-filaments are found together as a network structure
	   in the velocity interval of  $[-42.5, -30.0] \,\mathrm{km} \mathrm{\,s}^{-1}$.
	   The kinematic distance of this molecular cloud (MC) is estimated to be
	   $\sim4.5  \mathrm{\,kpc}$. 
	   The median length, width, excitation temperature, line mass of these filaments are $\sim49 \mathrm{\,pc}$, $\sim2.9 \mathrm{\,pc}$,
	   $\sim8.9 \mathrm{\,K}$, and $\sim39 \,M_{\odot} \mathrm{pc}^{-1}$, respectively.
	   The velocity structures along these filaments exhibit oscillatory patterns,  which are likely caused by 
	   the fragmentation or accretion process along these filaments. 
	   The maximum accretion rate is estimated to be as high as $\sim700 \,M_{\odot} \mathrm{pc}^{-1}$.
	   A total of $\sim162$ \tht clumps and  $\sim 103$ young stellar objects (YSOs) are identified in this region. Most of the clumps are in gravitationally bound states. 
	   Three \hii regions (G154.359+2.606, SH2-211, SH2-212) are found to be located in the apexes of the filaments. 
	   Intense star forming activities are found along the entire filamentary cloud. 
	   The observed results may help us to better understand the link between filaments and massive star formation. 
\end{abstract}


\section{Introduction} \label{sec:intro}	
	
	Surveys at multi-wavelengths, e.g., infrared surveys \citep{2010A&A...518L.102A,2014prpl.conf...27A}, high-resolution HI surveys \citep{2020A&A...642A.163S}, and CO surveys \citep[][]{2021A&A...655L...2C,2021ApJS..257...51Y}, have revealed that filamentary structures are ubiquitous in cold Galactic interstellar medium (ISM). The filaments are found to have lengths ranging from $0.1\,\mathrm{pc}$ to $1\,\mathrm{kpc}$
	, containing gas mass of $\sim 10^{3} - 10^{5} M_{\odot}$ \citep{2014A&A...568A..73R,2018A&A...610A..77H,2021arXiv211101057S}. As part of the ISM cycle, dense molecular filament plays a key role as it connects several important processes, such as the compression of diffuse gas, the fragmentation of cloud structures, and the formation of dense cores. For example, about $75\%$ pre- and proto-stellar cores are found to be located in filaments in the $Herschel$ Gould Belt Survey  \citep[e.g.,][]{2014prpl.conf...27A,2015A&A...584A..91K}.
	 Large-scale filaments containing dense molecular gas appear to form massive stars and clusters, e.g., Nessie \citep{2010ApJ...719L.185J,2019AAS...23312706J,2018A&A...619A.166M}, Orion \citep{1981PhDT.........3B,1999ApJ...510L..49J,2008PASJ...60..407T}, California \citep{2009ApJ...703...52L,2017A&A...606A.100L,2021ApJ...921...23G}, and NGC 6334 \citep{2013A&A...554A..42R,2018PASJ...70S..41F}, though the relevant mechanisms are still in study. 
	 \cite{2018ARA&A..56...41M} suggested an evolutionary picture for the formation of high-mass stars, which involves scenarios of a global hierarchical collapse and clump-fed accretion. 
	 A typical example is the DR21 ridge, the most massive cloud structure in the Cygnus-X region \citep[e.g.,][]{2010A&A...520A..49S}.
	 Compared to the Taurus filament B211/B213 \citep{2013A&A...550A..38P}, it has a more spherical/elliptical geometry with a few sub-filaments connected to the hub ridge.  	 The mass-flows transported along sub-filaments toward the hub ridge are considered to account for the mass growth of the massive clumps and cores therein \citep{2012A&A...543L...3H}. 
	 The observations toward other filamentary clouds, e.g., Monoceros R2 \citep{2019A&A...629A..81T}, G22 \citep{2018ApJ...852...12Y}, and W33 \citep{2021A&A...646A.137L}, support this scenario.
	 In the hub ridges, with the material convergence, the mass accumulates and density increases, mini-starburst activities, with  $\Sigma_{\mathrm{SFR}}> 1 \,M_{\odot} \mathrm{yr}^{-1} \mathrm{kpc}^{-2}$ and $\Sigma_{M_{\mathrm{gas}}}> 100 \,M_{\odot} \mathrm{pc}^{-2}$ \citep{2016ApJ...833...23N}, can even take place.  
	 The filaments IRDC G035.39–00.33 \citep{2011A&A...535A..76N} and W43-MM1 ridge \citep{2014A&A...570A..15L} were found to have a SFE of $\sim 15 \%$ within an area of $\sim 8 \mathrm{\,pc}^{2}$ and a SFE of $\sim 6 \%$ within an area of $\sim 8 \mathrm{\,pc}^{3}$, respectively. Thus the two regions were suggested to form starburst clusters. 
	 
	 Investigating the multiscale kinematics will provide crucial clues to understand the relation between filamentary structures and high-mass star formation.
	 It is necessary to search for more filamentary clouds to reveal their gas dynamics in terms of filaments fragmentation, cores accretion, high-mass star formation.	
	 The Milky Way Imaging Scroll Painting (MWISP)  project\footnote{\url{http://www.radioast.nsdc.cn/mwisp.php}}
	 is an unbiased, high sensitivity CO multi-line survey towards the northern Galactic plane \citep{2019ApJS..240....9S,2021ApJS..256...32S}, which provides us a good opportunity to study large-scale filamentary clouds \citep[e.g.,][]{2021ApJ...921...23G,2021ApJS..257...51Y}.
	 In this work, we present the observations of the three CO isotope lines towards the Galactic region of 153\fdg6 $\leqslant l \leqslant$ 156\fdg5 and 1\fdg85 $\leqslant b \leqslant$ 3\fdg5.  
	 The observations are described in Section 2. Section 3 presents the main results including the basic physical parameters of the MC, the process of the discrimination of the filaments, the physical properties and kinematic structures of the filaments, and the star formation activities therein. Section 4 discuss these results and a summary is made in Section 5.
	 
\section{Observations and Data Reduction} \label{sec-data}

	As part of the MWISP, the observations were performed with  
	the PMO 13.7 m millimeter telescope from April 2012 to February 2017.
	The three isotope emission lines of \tw at $115.271 \mathrm{\,GHz}$, \tht at $110.201 \mathrm{\,GHz}$, and \ei at $109.782 \mathrm{\,GHz}$ were simultaneously observed with the 3$\times$3 beams Superconducting Spectroscopic Array Receiver (SSAR) system in the sideband separation mode\citep{2012ITTST...2..593S}. 
	The ${ }^{12} \mathrm{CO}$ was at the upper sideband $(\mathrm{USB}),$ and ${ }^{13} \mathrm{CO}$ 
	and $\mathrm{C}^{18} \mathrm{O}$ were at the lower sideband $(\mathrm{LSB})$. 
	Typical system temperatures were around $\sim 270\mathrm{\,K}$ for the USB and around $175\mathrm{\,K}$ for the LSB, respectively. 
		
	In the position-switch On-The-Fly (OTF) mode, the telescope
	scanned the sky along both the longitude and latitude directions at a constant rate of $50^{\prime \prime} \mathrm{\,s}^{-1}$.
	The half-power beamwidth (HPBW) is approximately 52$\arcsec$ at $115.2\mathrm{\,GHz}$ and 55$\arcsec$ at $110.2\mathrm{\,GHz}$. 
	The beam efficiencies ($\eta_{\rm MB}$) are 44$\%$ for $115.2 \mathrm{\,GHz}$ and 48$\%$ for $110.2 \mathrm{\,GHz}$, respectively. 
	The pointing uncertainty is approximately within $5^{\prime \prime}$. 
	The total 1GHz bandwidth is separated into 16384 channels, 
	resulting in a resolution of 61 kHz in frequency, which is  corresponding to $\sim 0.16 \mathrm{\,km}\mathrm{\,s}^{-1}$ for $^{12}$CO, $\sim 0.17 \mathrm{\,km}\mathrm{\,s}^{-1}$ for 
	\tht and \ei in velocity, respectively. 
	Velocities were all given with respect to the local standard of rest (LSR). 
	The rms noises levels are $\sim 0.5 \mathrm{\,K}$ for \tw, and $\sim 0.3 \mathrm{\,K}$ for \tht and $\mathrm{C}^{18} \mathrm{O}$, respectively.
	More information about the telescope can be found in the telescope status 
	report\footnote{\url{http://www.radioast.csdb.cn/zhuangtaibaogao.php}}. 
	All data were reduced by the dedicated pipelines based on C,  IDL,  and Python by MWISP working group\footnote{\url{http://www.iram.fr/IRAMFR/GILDAS}} and the GILDAS/CLASS package by IRAM \citep{2005sf2a.conf..721P}.
	After removing bad channels and correcting the first order (linear) base line fitting, 
	the data were regridded into the standard FITS files  with a pixel size of $30^{\prime \prime} \times 30^{\prime \prime}$ 
	(approximately half of the beam size). 
	
\section{Results}\label{sec-Result}

\subsection{Filamentary Network Cloud}\label{sec-cloud}

\begin{figure}[htpb] 
             \centering
             \includegraphics[width=1.0\textwidth]{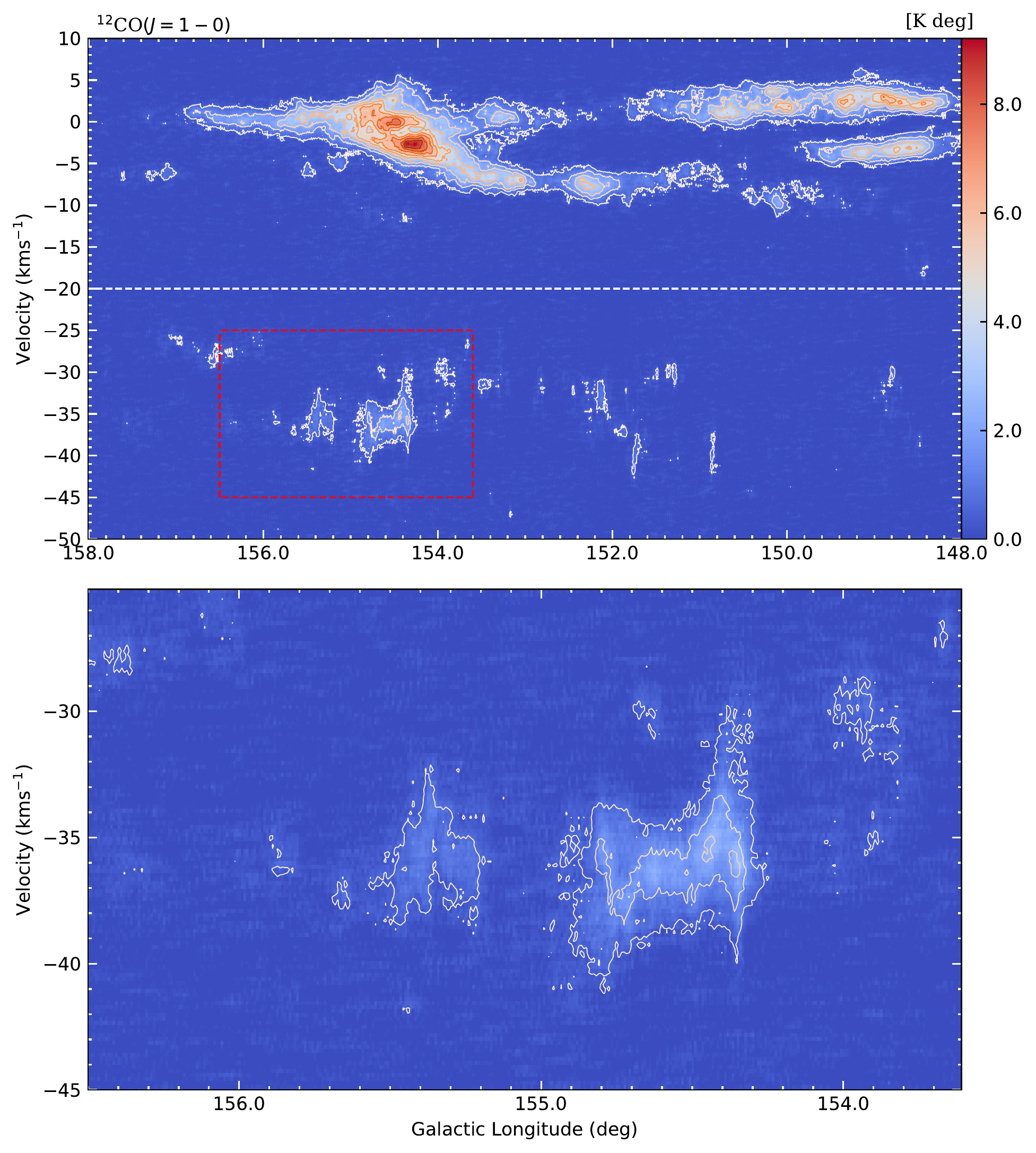}
             \caption{Upper panel: The LV map of \tw
             towards the region of $148\degr \leqslant l \leqslant 158\degr$ 
             and $-50 \leqslant v \leqslant 10 \mathrm{\,km}\mathrm{\,s}^{-1}$.
             The latitude is integrated from 1\fdg85 to 3\fdg5.   
             The integrated noise level($\sigma$) is about $0.051 \mathrm{K\,deg}$.
             The contours are set from $10 \sigma$ to peak value ($9.2\mathrm{K\,deg}$) by a step of $10\%$.
             The horizontal white dashed line separates the velocity intervals into Local Arm and Perseus/Outer Arm \citep[see][]{2001ApJ...547..792D,2019ApJ...885..131R}
             .
             The red rectangle indicates the region focused in this work.   
             Below panel: The zoom-in map of the red rectangle  
             with a velocity range of $[-45,-25] \mathrm{\,km}\mathrm{\,s}^{-1}$ and a longitude range of [153\fdg6, 156\fdg5].
             }\label{fig-lv}
        \end{figure}

      As shown in Figure \ref{fig-lv}, we construct the Longitude-Velocity (LV) diagram by summing the emission along the latitude interval of $[1\fdg85, 3\fdg5]$.
	  The CO emission mainly appears in the velocity interval of $[-45, 5] \mathrm{\,km}\mathrm{\,s}^{-1}$. 
	  According to the analysis in \cite{2001ApJ...547..792D}, the emission in $[-15,5] \mathrm{\,km}\mathrm{\,s}^{-1}$ is generally attributed to the Local arm. 
	  \cite{2017ApJ...838...49X} partially presented the MWISP CO data analysis in this interval (towards the region of $147\fdg75 \leqslant l \leqslant 152\fdg0 , 1\fdg5 \leqslant b \leqslant 5\fdg25$). 
	  In this work, we focus on the CO emission in the velocity interval of [-42.5, -30.0]$\mathrm{\,km}\mathrm{\,s}^{-1}$.
	  The zoom-in LV diagram shows that the emission  
	  could be further divided into 
	  eastern\footnote{All the directions mentioned in this work are in the Galactic coordinate system.} 
	  ($l \sim$[155\fdg0, 156\fdg5]) and western ($l \sim$[154\fdg0, 155\fdg0]) regions.

\begin{figure}[ht]
		\centering
		\includegraphics[width=1.0\textwidth]{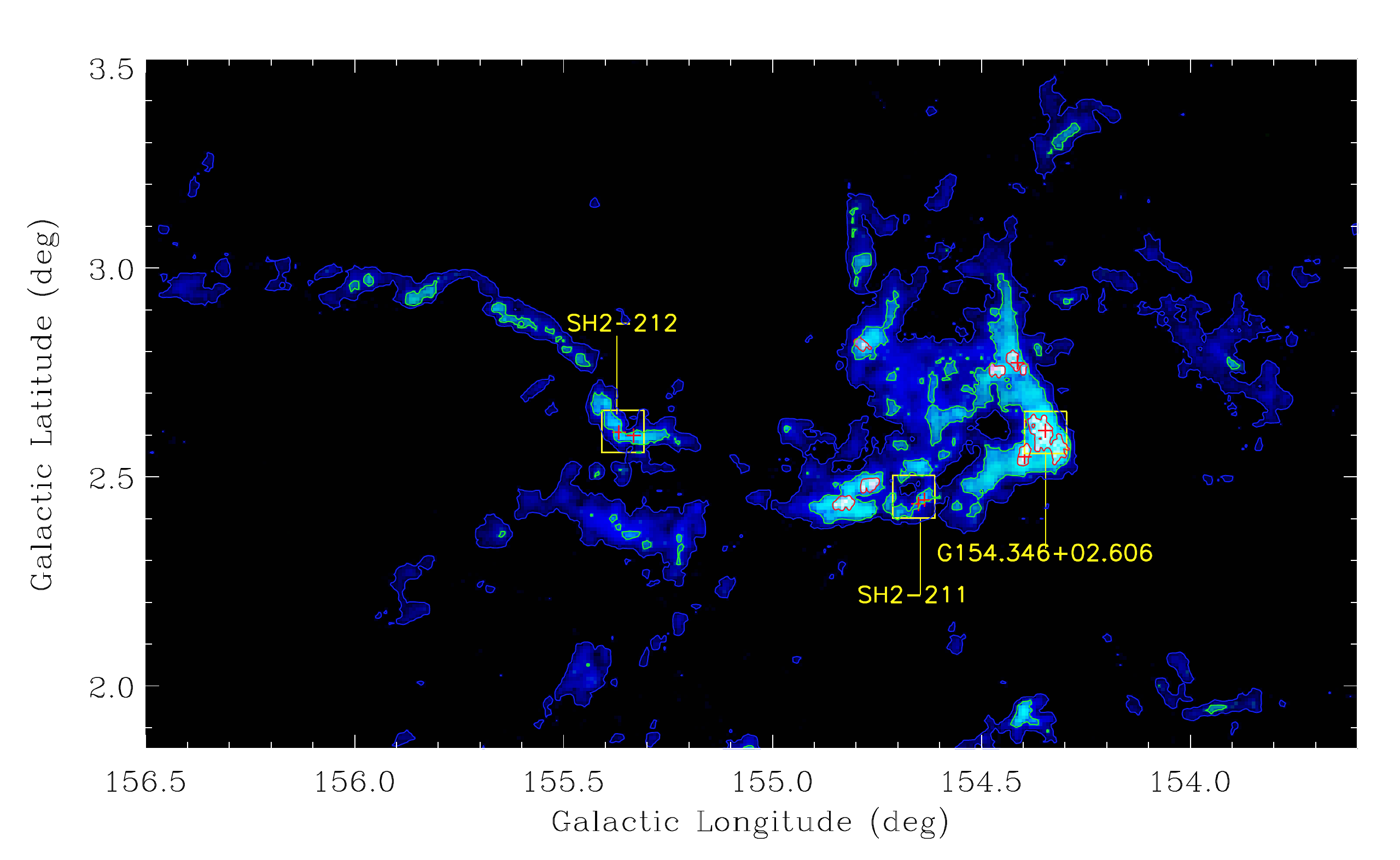}
		\caption{The three-color image of \tw (blue), \tht (green), and \ei(red) in the velocity interval of $[-42.5,-30.0] \mathrm{\,km} \mathrm{\,s}^{-1}$. For \tw and $^{13}$CO,  the pixels with emission above 3 $\sigma$  with at least 3 continuous channels are used for integration. 
	  	The pixels of C$^{18}$O are obtained by DBSCAN algorithm (see Section \ref{sec-cloud}). 
	  	The blue, green and red contours are the 3$\sigma$, 3$\sigma$, and 1$\sigma$ boundaries for \tw, $^{13}$CO and C$^{18}$O, respectively. 
		The yellow squares and the red plusses represent the locations of \hii regions and IRAS sources, respectively.    
		}
		\label{fig-thrcol}
	 \end{figure}

	  Figure \ref{fig-thrcol} shows the velocity-integrated intensity image for the three CO lines.
	  For $^{12}$CO and $^{13}$CO, to obtain clear maps, we use  pixels with emission above $3 \,\sigma$  and at least $3$ continuous channels for integration.
	  For C$^{18}$O, in order to retain as many valid signals as possible, we apply for the DBSCAN algorithm \citep[connectivity 1 and MinPts 4, see][for details]{2020ApJ...898...80Y} to make a mask.
	  The parameters of cutoff threshold and the peak brightness temperature are selected as $1.5 \,\sigma$ and $>3 \,\sigma$ to filter noise in this process. Then pixels within the mask greater than $2 \,\sigma$ are used for integration. 
	  As shown in Figure \ref{fig-thrcol}, the three-color intensity image clearly shows the relative spatial distribution traced by the three isotope CO lines, respectively.
	  
	  In the western region, 
	  the emission of \tw and \tht outlines a 
	  network structure. Compared with the relatively diffuse emission traced by \tw, the \tht emission shows the distinct skeletons of the network structure, while the \ei emission is mainly concentrated at the main node structure. The peak intensity of the \tw emission appears at the position of (l, b) $\sim$ (154\fdg358, 2\fdg600).
	  In the eastern region, two isolated filamentary structures are separately distributed up and down. 
	  The upper one extends about 1$\degr$ along the direction from southwest to northeast with respect to the 
	  Galactic plane by an inclination angle of about 45$\degr$. 
	  The emission in the southwestern part seems brighter than the northeastern part. 
	  The lower one is roughly parallel to the Galactic plane.
	  
	  Table\ref{tb-IRAS} lists \hii regions and IRAS sources associated with the MC. The association between the \hii regions and the MC is based on the consistency of the velocities of the ionized gas and the CO gas.
	  For IRAS sources, four of them are selected from the catalog of \cite{2002ApJS..141..157Y}. 
	  The rest three are selected from the analysis of previous studies \citep{2001ApJ...555..613I,2005ApJS..161..361K}, e.g., according to correlation between the luminosities of IR and $\mathrm{H} \alpha$, the IRAS04368+5021
	  and IRAS04329+5045 are identified to be associated with the \hii regions of SH2-211 and SH2-212 respectively.

	\begin{deluxetable*}{lcccc}
		\tabletypesize{\small} 
		\setlength{\tabcolsep}{0.25in}
		\tablecaption{associated\hii regions and IRAS sources \label{tb-IRAS}}
		\tablewidth{0pt}
		\tablehead{ \colhead{Name} & \colhead{$l$} & \colhead{$b$}   & \colhead{Velocity} & \colhead{references for association} \\
			& \colhead{($\degr$)} & \colhead{($\degr$)}  & \colhead{($\mathrm{\,km}\mathrm{\,s}^{-1}$)} & 
			}
		\startdata
		G154.346+2.60   &  154.346   &2.606 &	$-43.60 \pm 24.00$ \tablenotemark{a} &	\cite{2015ApJS..221...26A}   \\
		 SH2-211               &  154.661    &	2.452    &	  $-35.22 \pm 28.50$ \tablenotemark{a} &	\cite{2011ApJ...738...27B}   \\
		                       &            &	           &	 $-33.00$ \tablenotemark{b} &	\cite{1991PASP..103..843P}   \\		 
		 SH2-212                &  155.357    &	2.609   &	  $-43.95 \pm 25.57 $ \tablenotemark{a}&	\cite{2011ApJ...738...27B}    \\
		                       &             &	        &	  $-38.4$ \tablenotemark{b}&	\cite{1991PASP..103..843P}   \\	
		                       &             &	        &	  $-40.1 \pm 1.0$ \tablenotemark{c}&	\cite{1989ApJS...71..469L}   \\			                       
		IRAS04324+5102          &  154.397    &	2.548    &	  	$-37.55$ &	\cite{2002ApJS..141..157Y}   \\
		IRAS04324+5106          &  154.347    &	2.606    &	  	$-36.02$ &	\cite{2002ApJS..141..157Y}   \\
		          &      &	    &	  	$-37.3$ \tablenotemark{d}& \cite{Pirogov1999}	   \\		
		IRAS04335+5110          &  154.414    &	2.774    &	  	$-34.89$ &	\cite{2002ApJS..141..157Y}   \\
		IRAS04366+5022          &  155.331    &	2.596    &	  	$-34.00$ &	\cite{2002ApJS..141..157Y}   \\	
		IRAS04368+5021          &  155.367    &	2.607    &	  	 $\sim$ &	\cite{2001ApJ...555..613I}   \\	
		IRAS04329+5045          &  154.650   &   2.435   &        $\sim$ &    \cite{2001ApJ...555..613I}   \\ 
		IRAS04329+5047          &  154.637   &   2.445   &        $-37.1$ \tablenotemark{e}&    \cite{1990ApJ...352..139S}   \\
		          &     &     &        $-38.3$ \tablenotemark{d}&    \cite{Pirogov1999}   \\
		\enddata 
		\tablenotetext{}{a: velocity of hydrogen radio recombination lines ($v_{\rm LSR,\mathrm{HRRLs}}$) $\pm$ Full width at half height (FWHM).  b: $v_{\rm LSR,\mathrm{ H} \alpha}$.  c: $v_{\rm LSR,\mathrm{ H109} \alpha}$. d: $v_{\rm {LSR,\mathrm{ HCN}}(J=1-0)}$. e: $v_{\rm LSR,\mathrm{ CO}}$}
	\end{deluxetable*}

\subsection{Distance of the cloud}\label{Distance}

	  In the second quadrant of the Milky Way, the systemic velocity ($v_{\rm LSR}$) of molecular gas increases negatively from  $\sim 0 \mathrm{\,km}\mathrm{\,s}^{-1}$ to 
	  trace subsequently the Local arm, the Perseus Arm, the Outer Arm, and the Scutum-Centaurus Arm 
	  \citep{2008AJ....135.1301V, 2015ApJ...798L..27S,2019ApJ...885..131R}.  
	  \cite{2016ApJS..224....7D, 2017ApJS..229...24D} analyzed the MC structures in the region of
	  l $\sim$ [100\fdg0, 150\fdg0], b $\sim$ [-3\fdg0, 5\fdg0],  using the CO data from the MWISP survey. 
	  According to the tendency of LV map \cite[see, e.g., Figure 5 in][]{2016ApJS..224....7D}, 
	  the filamentary MC in this work is located in the Outer arm.
	  According to previous optical observations of stars associated with the \hii regions, the distance of IRAS$04324+5106$ was suggested to be $\sim 6 \mathrm{\,kpc}$ \citep{1993ApJ...407..657C}. The distance of SH2-212 was suggested to range from $\sim 4.81 \mathrm{\,kpc}$ to $\sim 6.8 \mathrm{\,kpc}$ \citep{1979A&AS...38..197M,2011MNRAS.411.2530J,2015AJ....149..127L,2020A&A...633A..99C}.
	  For SH2-211, the spectroscopic distance was relatively uncertain, ranging from $\sim 3 \mathrm{\,kpc}$ to $\sim 7.8 \mathrm{\,kpc}$  \citep{1984A&A...139L...5C,1990AJ.....99..846H,2015AJ....150..147F}.

\begin{figure}[ht]
        \centering
        \includegraphics[width=0.85\textwidth]{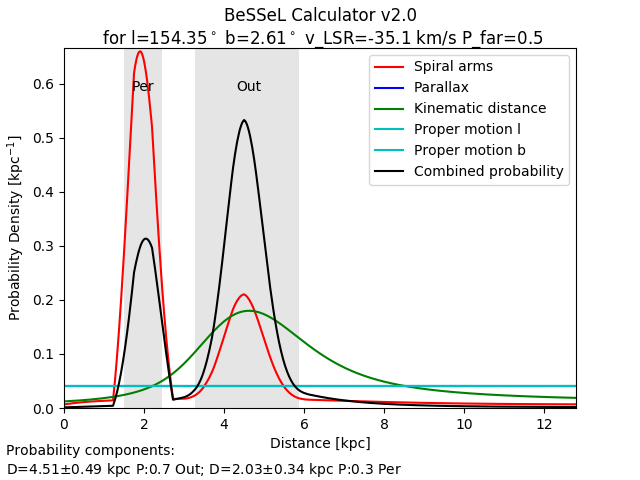}
        \caption{Distance probability density function of the position NO.1 (see Table\ref{tb-distance}),  using the Bayesian distance calculator \citep{2016ApJ...823...77R}.
        }
        \label{fig-3d-dis}
        \end{figure}

	  We try to estimate the distances using the Bayesian distance calculator provided by \cite{2019ApJ...885..131R}.
	  We select nine pixels with strong radiation and apply single Gaussian fits to the spectral lines to derive the peak velocities. The rotation model from \cite{2019ApJ...885..131R} is then applied to each coordinate and peak velocity to get the distance. The results are listed in Table \ref{tb-distance}.
	  As shown in Figure \ref{fig-3d-dis}, the position with 
	  strongest emission (named as ``NO.1'') is found to have a higher combined probability ($70\%$) to be associated with portion of the Outer arm at 
	  a distance of $4.51 \pm 0.49 \mathrm{\,kpc}$ than to be associated with the Perseus arm at $2.03 \pm 0.34 \mathrm{\,kpc}$. 
	  The other eight positions go through the similar situation. The average distance is $\sim 4.50 \pm 0.51 \mathrm{\,kpc}$. 
	  We therefor adopt $4.5 \mathrm{\,kpc}$ as the distance of the filamentary cloud, which is compariable with the distance estimated in previous studies (see above).

	\begin{deluxetable}{lllcccc}
		\tabletypesize{\small} 
		\setlength{\tabcolsep}{0.30in}
		\tablecaption{Measured distances of the six selected positions. \label{tb-distance}}
		\tablewidth{0pt}
		\tablehead{ \colhead{NO.} & \colhead{$l$} & \colhead{$b$} & \colhead{$v_{\rm LSR}$} & \colhead{distance}  & \colhead{probability} & \colhead{Arm}\\
			\colhead{} & \colhead{($\degr$)} & \colhead{($\degr$)} & \colhead{(km s$^{-1}$)} & \colhead{(kpc)}  & \colhead{}  & \colhead{}
			}
		\startdata
		1  &   154.35 &   2.58  & -35.1 &  4.50$\pm$0.51 & 0.63 &  Out    \\
		2  &   154.48 &   2.49  & -35.9 &  4.52$\pm$0.51 & 0.68 &  Out    \\  
		3  &   154.43 &   2.76  & -33.8 &  4.48$\pm$0.51 & 0.55 &  Out    \\  
		4  &   154.44 &   2.87  & -35.1 &  4.49$\pm$0.51 & 0.60 &  Out    \\  
		5  &   154.59 &   2.64  & -35.2 &  4.50$\pm$0.51 & 0.66 &  Out    \\  
		6  &   154.84 &   2.43  & -37.2 &  4.53$\pm$0.51 & 0.74 &  Out    \\  
		7  &   154.78 &   2.48  & -35.5 &  4.51$\pm$0.51 & 0.69 &  Out    \\  
		8  &   154.77 &   2.81  & -36.2 &  4.51$\pm$0.51 & 0.68 &  Out    \\  
		9  &   155.35 &   2.59  & -36.6 &  4.50$\pm$0.53 & 0.70 &  Out    \\  
		\enddata 
	\end{deluxetable}

\subsection{Properties of the cloud}\label{Properties}
	 
\begin{figure}[htb]
        \centering
        \includegraphics[width=1.0\textwidth]{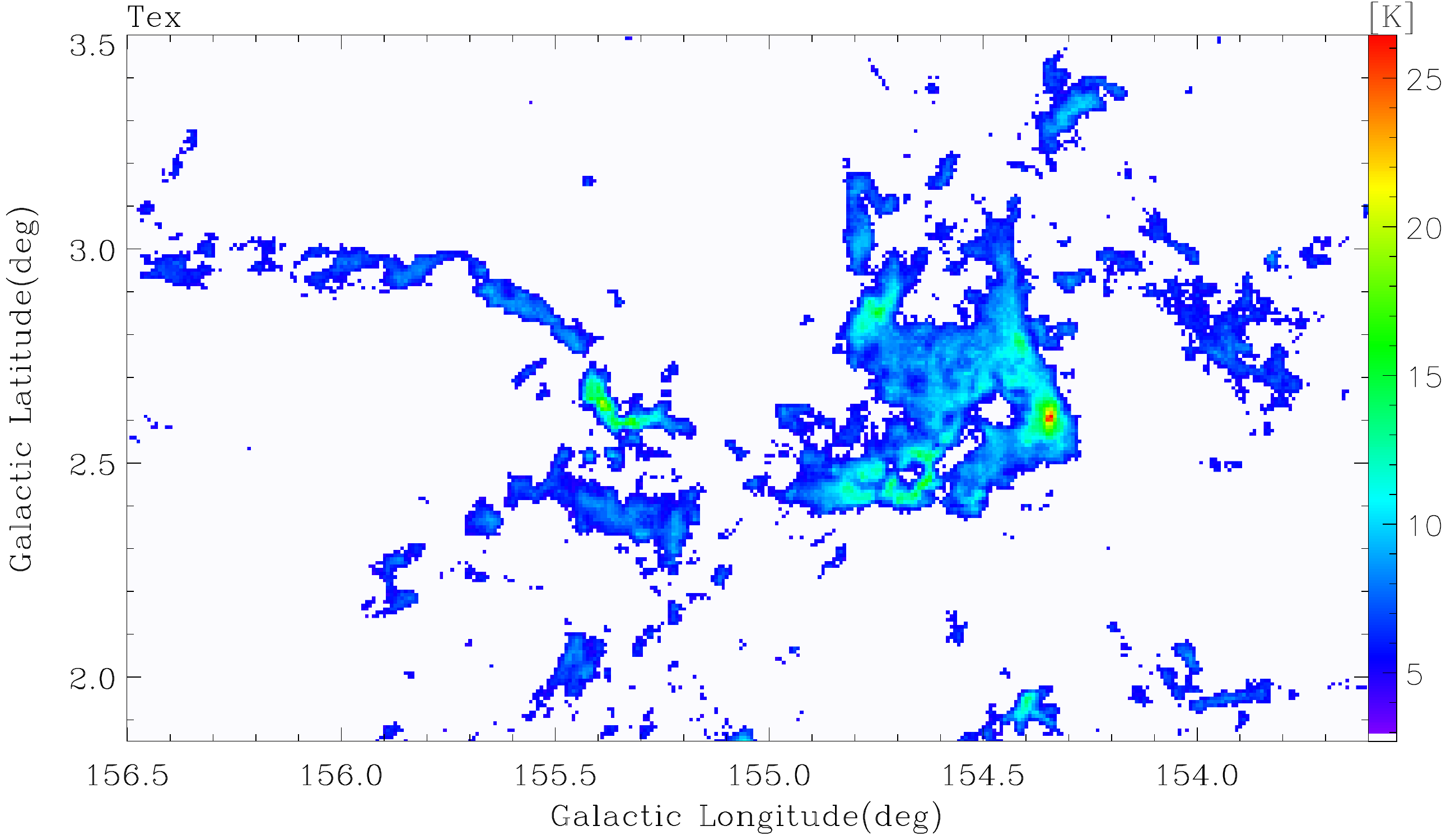}
        \caption{The distribution of the excitation temperature calculated from the peak value of the main beam temperature of \tw in 
        the velocity interval of $[-42.5,-30.0] \mathrm{\,km}\mathrm{\,s}^{-1}$. 
        }
        \label{fig-Tex}
        \end{figure}

	  Assuming \tw is optically thick and the background temperature (${T}_{\rm bg})$ is $2.73 \mathrm{\,K}$, 
	  the excitation temperature ($T_{\mathrm{ex}}$)  could be calculated from the peak value of the main beam temperature of 
	  \tw ($T_{\rm mb, peak,^{12}CO}$) following the method of \citet{1991ApJ...374..540G}. 
	  The formula is simplified as 
	  \begin{equation}
	  T_{\rm ex}=5.532 \left[{\rm log}(1+\frac{5.53}{T_{\rm mb,peak,^{12}CO}+0.819})\right]^{-1}\,({\rm K}).
	  \label{eq:tex}
	  \end{equation} 	
	  The distribution of excitation temperature is shown in Figure \ref{fig-Tex}. 
	  The value ranges from $3.1$ to $26.4 \mathrm{\,K}$. 
	  The position of the peak value is spatially coincident with the \hii region G154.346+02.606.

	  In the assumption of local thermodynamic equilibrium (LET) 
	  and equal $T_{\mathrm{ex}}$ of the isotopologue pair, the $^{13}$CO column density ($N_{\rm ^{13}CO}$) 
	  can be estimated by 
	  \begin{equation}
	  N_{\rm ^{13}CO} = 2.42 \times 10^{14} \cdot \frac{\int T_{\rm mb, ^{13}CO} dv}{1-{\rm exp}(-5.29/T_{\rm ex})}\cdot \frac{\tau_{13}}{1-e^{-\tau_{13}}},\mathrm{where}
	  \label{N13}
	  \end{equation}
	  \begin{equation}
	  \tau_{13}=-\ln \left[1-\frac{T_{\mathrm{mb}, \text { peak },{ }^{13} \mathrm{CO}}}{5.29}\left(\left[e^{5.29 / T_{\mathrm{ex}}}-1\right]^{-1}-0.164\right)^{-1}\right]
	  \end{equation}
	  As emission of the molecular cloud primarily occurs in regions with excitation temperatures greater than 5K,  we set the excitation temperature that below 5K to 5K when estimating optical depth of \tht($\tau_{13}$). We get that the range of $\tau_{13}$ is 0.08 - 1.98.

	  The H$_2$ column density could be derived by multiplying the $^{13}$CO column density with the ratio of 
	  $N_{\rm H_2}/N_{\rm ^{13}CO}$ $\sim$ 7$\times$10$^{5}$ \citep{1982ApJ...262..590F}. 
	  For \ei is always thin, we use a similar but simpler method under the same assumption to derive the \ei column density: 
	  \begin{equation}
	  N_{\rm C^{18}O} = 2.24 \times 10^{14} \cdot \frac{(1+0.88/T_{\rm ex})\int T_{\rm mb, C^{18}O} dv}{1-{\rm exp}(-5.27/T_{\rm ex})}.
	  \label{N18}
	  \end{equation}  
	  
	 The H$_2$ column density could be derived by a conversion ratio of 
	  $N_{\rm H_2}/N_{\rm C^{18}O}$ $\sim$ 7$\times$10$^{6}$ \citep{1995A&A...294..835C}.
	  The H$_2$ column density can also be estimated by multiplying the \tw integrated intensity with 
	  the CO-to-H$_2$ conversion factor $X \sim$ 1.8$\times$10$^{20}$\,cm$^{-2}$\,K$^{-1}$\,km$^{-1}$\,s  \citep{2001ApJ...547..792D}: 
	  \begin{equation}
	  N_{\rm H_2} = X \int T_{\rm mb, ^{12}CO} dv.
	  \label{N12}
	  \end{equation}

\begin{figure}[htb]
        \centering
        \includegraphics[width=1.0\textwidth]{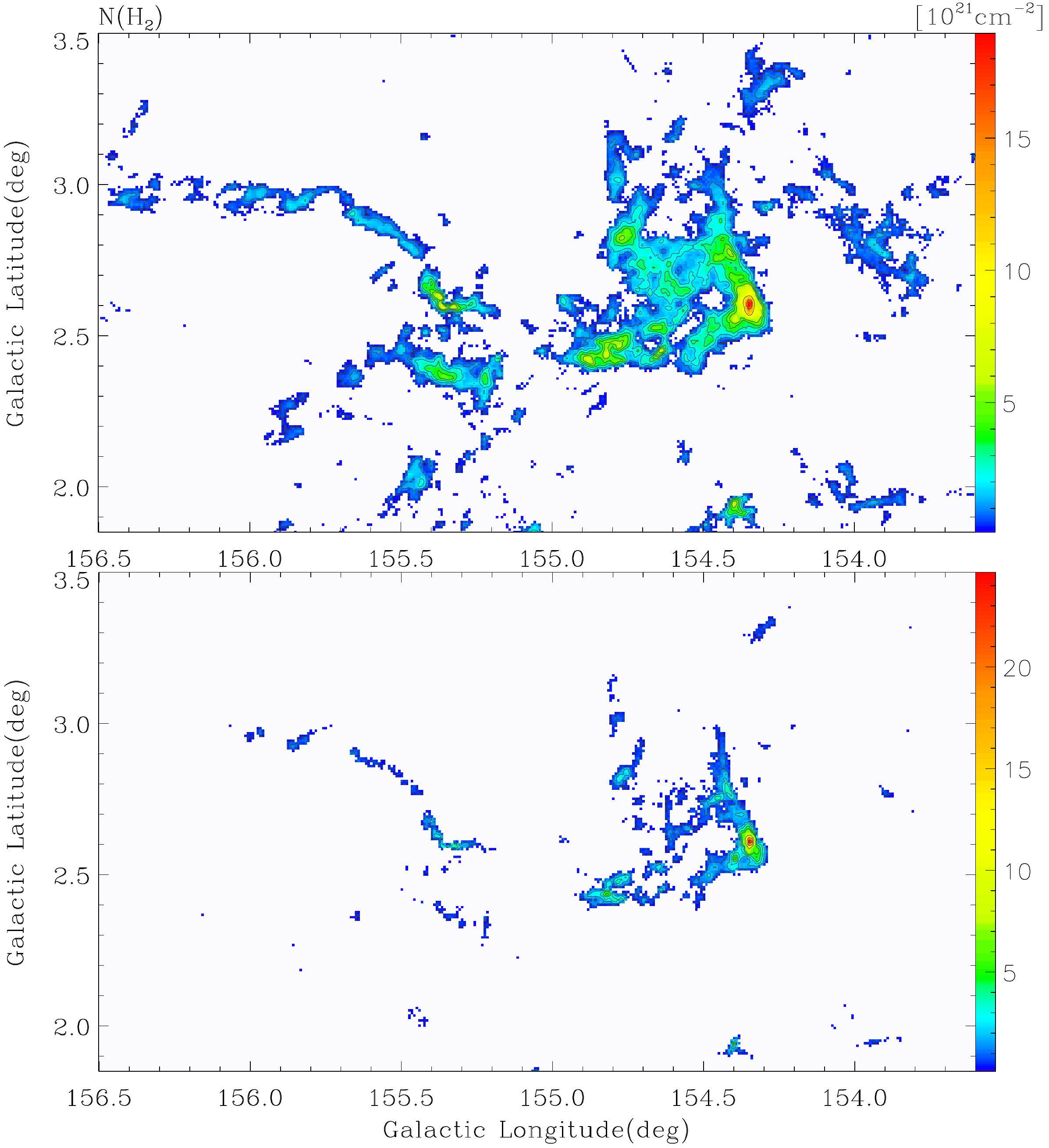}
        \caption{The H$_{2}$ column density maps derived from ${ }^{12} \mathrm{CO}$ (upper panel) and ${ }^{13} \mathrm{CO}$ (bottom panel) in the velocity interval of $[-42.5,-30] \mathrm{\,km}\mathrm{\,s}^{-1}$, respectively. 
        The contours are 0.4 (3 $\sigma$ noise level), 1, 2, 3, 4, 5, 10, 15 $\times$ 10$^{21}$cm$^{-2}$, respectively. The colorbar is in unit of 10$^{21}$cm$^{-2}$.}
        \label{fig-columndensity}
        \end{figure}

	  Figure \ref{fig-columndensity} shows the distributions of the derived H$_2$ column density.   
	  The volume density is obtained by dividing the column density by an equivalent radius, where the equivalent radius can be obtained if the emission area is assumed to be circle in shape.
	  In the estimation process of the mass and volume density,  
	  a mean molecular weight with the value of 2.83 is adopted 
	  \citep{2008A&A...487..993K}. 
	  The projected emission area, column density, volume density and mass are listed in Table\ref{tb-cloud}. 
	  The emission area of \tw ($2347$ arcmin$^{2}$) is much larger than that of \tht ($506$ arcmin$^{2}$) and \ei ($49$ arcmin$^{2}$). 
	  The low-density molecular gas traced by \tw occupies most of the total mass.
	  In the following section \ref{filaments}, we focus on the \tht emission as it better traces the filamentary skeletons.
	  
	  \begin{deluxetable}{ccccccc}
		\tabletypesize{\small}
		\setlength{\tabcolsep}{0.15in}
		\tablecaption{Properties of the cloud \label{tb-cloud}}
		\tablewidth{0pt} 
		\tablehead{ \colhead{Velocity interval}&\colhead{Distance}&\colhead{Molecule Tracer} & \colhead{Area} & \colhead{$N(\rm H_{2})$\tablenotemark{a}} & \colhead{$n(\rm H_{2})$\tablenotemark{a}} & \colhead{$M(\rm H_{2})$\tablenotemark{b}}\\
		\colhead{(km s$^{-1}$)}&\colhead{(kpc)}&\colhead{}& \colhead{(arcmin$^{2}$)} & \colhead{($10^{21}$cm$^{-2}$)} & \colhead{(cm$^{-3}$)} & \colhead{($10^{4}M_{\sun}$)} 
		} 
		\colnumbers
		\startdata 
		{}          &{    }&\tw     &  2347     & 2.85    & 25     & 11.7    \\
		{[-42.5, -30.0]} &4.5  &\tht    &  506     & 3.53   & 68    & 2.96     \\
		{       }&{       }&\ei     &  49      & 1.87    & 110   &  0.42    \\
		\enddata
		\tablecomments{a,  mean value with weight of integrated intensity. b,  the estimated values at the distance of 4.5 kpc.}
	   \end{deluxetable}

\subsection{Filaments}\label{filaments}   
                         
\begin{figure}[htb]
	       \centering
	       \includegraphics[width=1.0\textwidth]{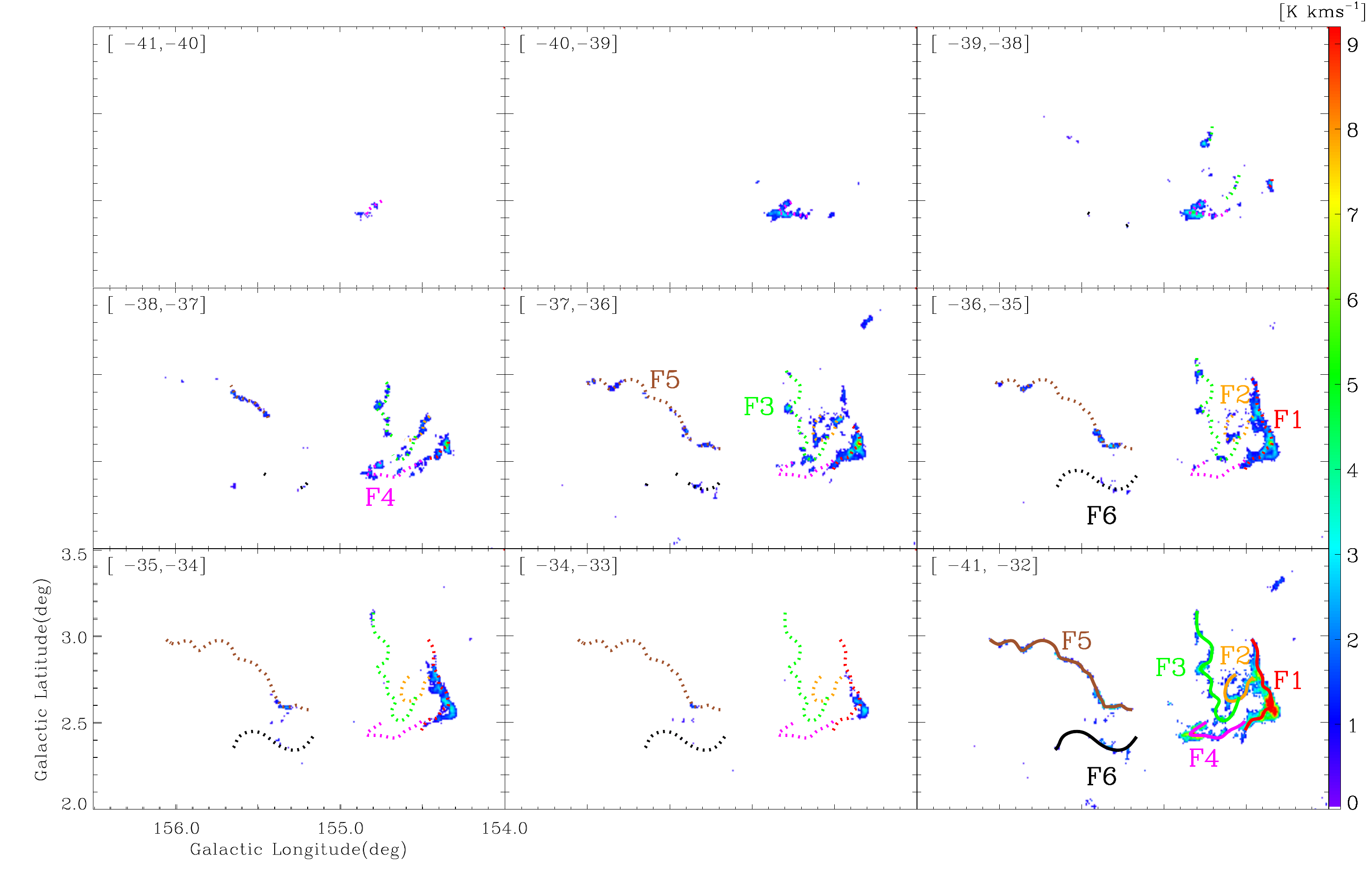}
	       \caption{The \tht channel maps  from $-41   \mathrm{\,km}\mathrm{\,s}^{-1}$ to $-32 \mathrm{\,km}\mathrm{\,s}^{-1}$ by a step of $1 \mathrm{\,km} \mathrm{s}^{-1}$. 
	       Only pixels with at least 3 channels above 3 sigma are kept for integration. The colored dotted lines indicate the filament skeletons identified by visual inspection step by step. The names are marked out beside each filaments. The last panel shows the intensity image integrated within $[-41, -32] \mathrm{\,km}\mathrm{\,s}^{-1}$ that overlapped with the visual inspection filaments (colored solid lines).}
	       \label{fig-cubemoment}
	\end{figure}

\begin{figure}[htb]
	       \centering
	       \includegraphics[width=1.0\textwidth]{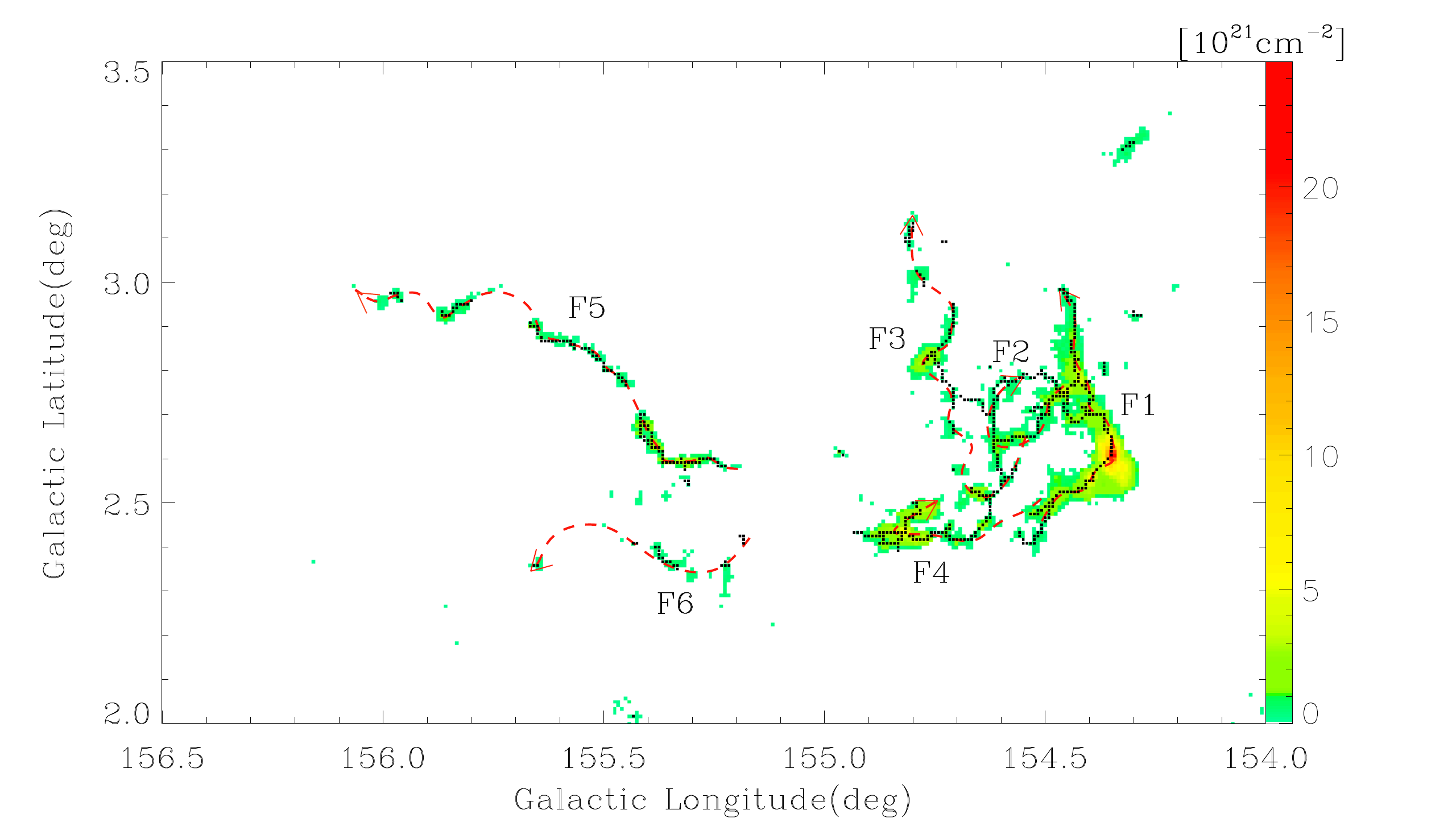} 
	       \caption{The $\mathrm{H}_{2}$ column density map estimated from the ${ }^{13} \mathrm{CO}$ emission.
	       The dotted black lines and the red dashed lines represent the filament skeletons that are extracted by DisPerSE algorithm and 
	       visual inspection, respectively.
	       The colorbar is in unit of 10$^{21}$cm$^{-2}$.}
	       \label{fig-disperse}
	\end{figure}

	    Generally, a filament is characterized by an elongated structure with an aspect ratio greater than $\sim 3-5$  \citep{2014prpl.conf...27A}, and is significantly overdense with respect to its surroundings. The identification of filaments in this work is based on two aspects: the characteristic of morphology and the velocity coherence.
	    As shown in Figure \ref{fig-cubemoment}, through visual inspecting the channel maps, 
	    six skeletons (F1 $\sim$ F6) are carefully depicted and emphasized by different colored dashed lines.
	    To avoid arbitrary judgment, we apply for the Discrete Persistent Structures Extractor (DisPerSE) \citep{2011MNRAS.414..350S}
	    to the H$_{2}$ column density map derived from \tht to further identify the persistent filamentary structures.  
	    In order to not only guarantee a clear skeleton construction, but also retain the information of persistence 
	    in morphology as far as possible,  
	    the persistence threshold is set to be  $3.8\times 10^{20}$\,cm$^{-2}$ ($\sim 4 \,\sigma$), 
	    and the trimBelow threshold is set to be  $9.9\times 10^{20}$\,cm$^{-2}$ ($\sim 10 \,\sigma$).  
	    The results  of a set of 1-pixel wide curves are marked out by black solid lines in Figure \ref{fig-disperse}.
	    
\begin{figure}[ht]
	       \centering
	       \includegraphics[width=1\textwidth]{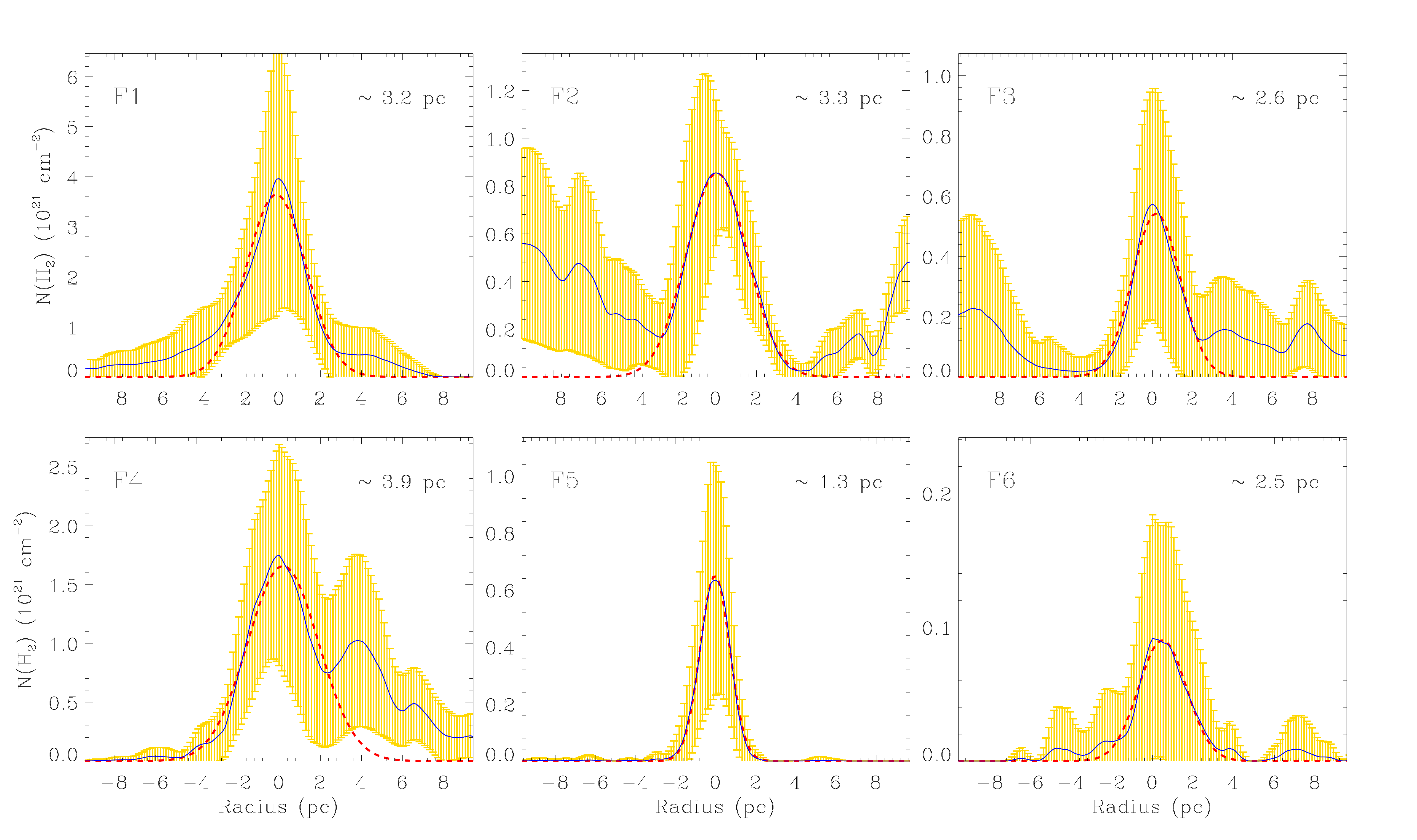} 
	       \caption{The blue curves represent the mean radial column density profiles of the \tht emission.
	       The yellow bars indicate the $\pm$ 1 $\sigma$ dispersion. 
	       The red dashed curves represent the Gaussian fitting results.
	       The peak position in each profile is regarded as the center of the profile.  
	       The widths (deconvolved FWHMs) are marked in the upper right corner of each panel.}
	       \label{fig-profile}
	\end{figure}

\begin{figure}[htpb]
	      \centering
	      \includegraphics[width=0.98\textwidth]{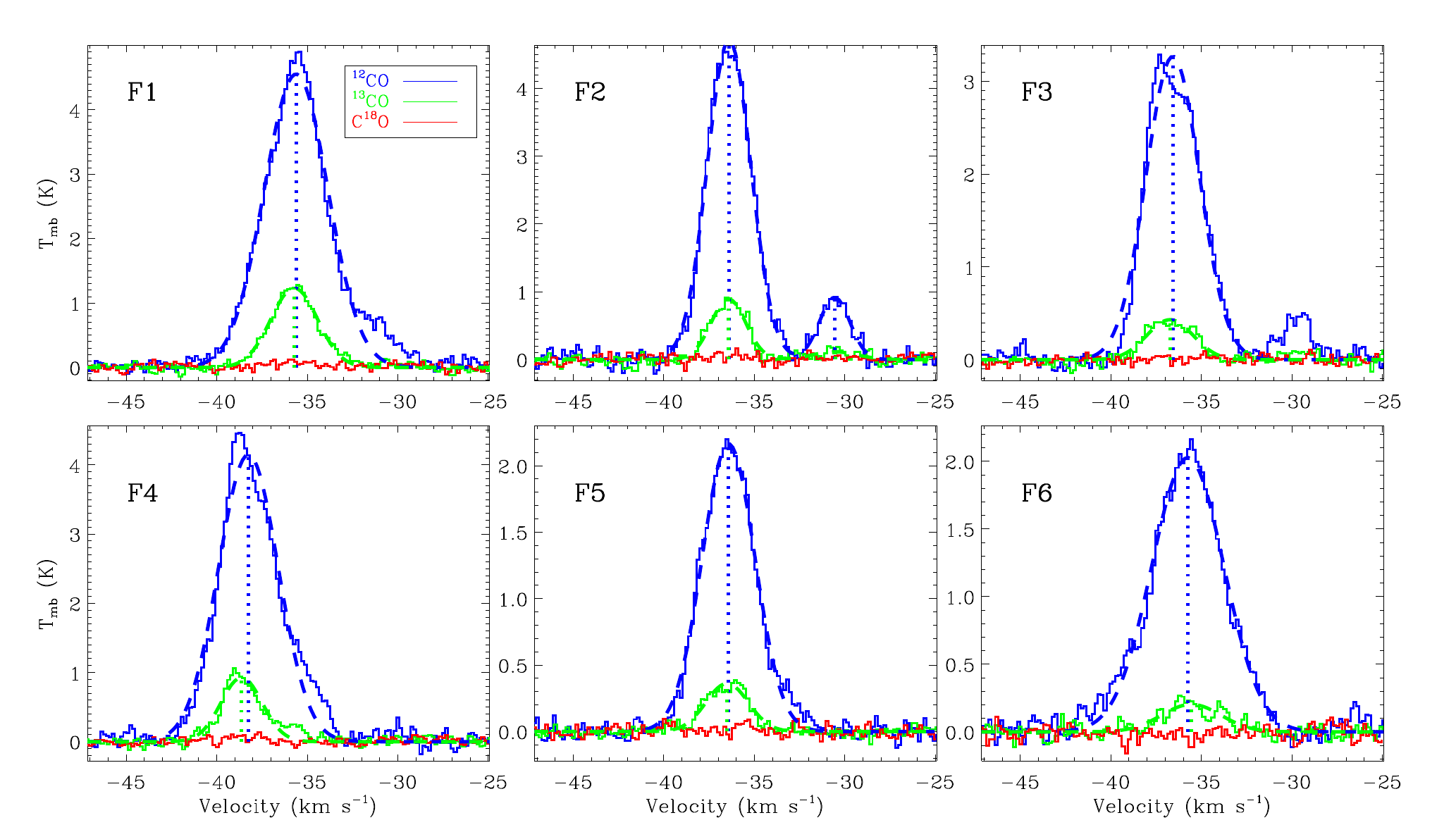}
	      \caption{Averaged spectral lines of \tw (blue), \tht (green), and \ei (red) extracted from the six identified filaments. 
	      The dashed lines with the corresponding colors are the Gaussian fitting results. The vertical dotted lines indicate the fitted central velocity.
	      The average $v_{\mathrm{LSR}}$ of the filaments are obtained from \tht.
	      }
	      \label{fig-fil-spectrum}
	 \end{figure}

	    As shown in Figure \ref{fig-disperse}, 
	    the skeletons obtained by the two methods are in good agreement with each other, especially for F1, F5, and F6.
	    The DisPerSE identification only relies on column density and is not sensitive enough to the velocity coherence. 
	    On the other hand, visual inspection has considered the information of velocity gradients instead of connecting the filaments
	    randomly in spatial distribution. 
	    The mean column density profile of each filament is constructed from radial cuts using a similar procedure 
	    as described in \cite{2011A&A...529L...6A} and \cite{2013A&A...550A..38P}. 
	    The width (deconvolved FWHM) of each filament is derived by applying for Gaussian fitting to the \tht column density profile (see Figure \ref{fig-profile}).
	    The average spectra of the filaments are extracted to get average $v_{\mathrm{LSR}}$ by applying Gaussian fits to \tht (see Figure \ref{fig-fil-spectrum}).
	    Adopting  a distance of $4.5 \mathrm{\,kpc}$, the lengths of these filaments range from $32$ to $90 \mathrm{\,pc}$. 
	    The widths range from $1.3$ to $3.9 \mathrm{\,pc}$. 	
	    We extract each filament along the visual inspection skeleton within the  corresponding width and 
	    estimate the mean $T_{\mathrm{ex}}$, the mean column density,  and the total mass using 
	    the same method as in the calculations of the cloud physical parameters. 
	    The average line mass ($M_{\text {line}}$) is measured by dividing the mass by length.
	    All the results are listed in Table \ref{tb-filaments}. 
	   \begin{deluxetable}{cccccccc}
		\tabletypesize{\small} 
		\setlength{\tabcolsep}{0.15in}
		\tablecaption{Properties of filaments \label{tb-filaments}}
		\tablewidth{0pt}
		\tablehead{ \colhead{Name} & \colhead{Length} & \colhead{Width} & \colhead{$v_{\mathrm{LSR}}$} & \colhead{Mass} & 
			\colhead{$ T_\mathrm{ex} $} &
			\colhead{$ N_\mathrm{H_2} $} & \colhead{$M_\mathrm{line}$} \\
			& \colhead{(pc)} & \colhead{(pc)} & \colhead{kms$^{-1}$}& \colhead{(\msun)} & \colhead{(K)} 
			  & \colhead{($ 10^{21} $ cm$^{-2}$)}  & \colhead{(\msun\,pc$^{-1}$)}}
		\colnumbers
		\startdata
		F1 &  51     & 3.2  & -35.7 & 13650          &  10.0    & 5.74   & 248 \\
		F2 &  32     & 3.3  & -36.4  & 1725          &  9.1     & 1.47   & 49  \\
		F3 &  75     & 2.6  & -36.7  & 2245          &  8.6     & 1.32   & 28 \\
		F4 &  38     & 3.9  & -38.6  & 4045          &  9.6     & 3.17   & 98  \\
		F5 &  90     & 1.3  & -36.5  & 2343          &  6.7     & 1.18   & 24  \\
		F6 &  47     & 2.5  & -35.6  & 227           &  6.6     & 0.41   & 4  \\
		\enddata 
	   \end{deluxetable}

\begin{figure}[htb]
		\centering
    		\includegraphics[width=1.0\textwidth]{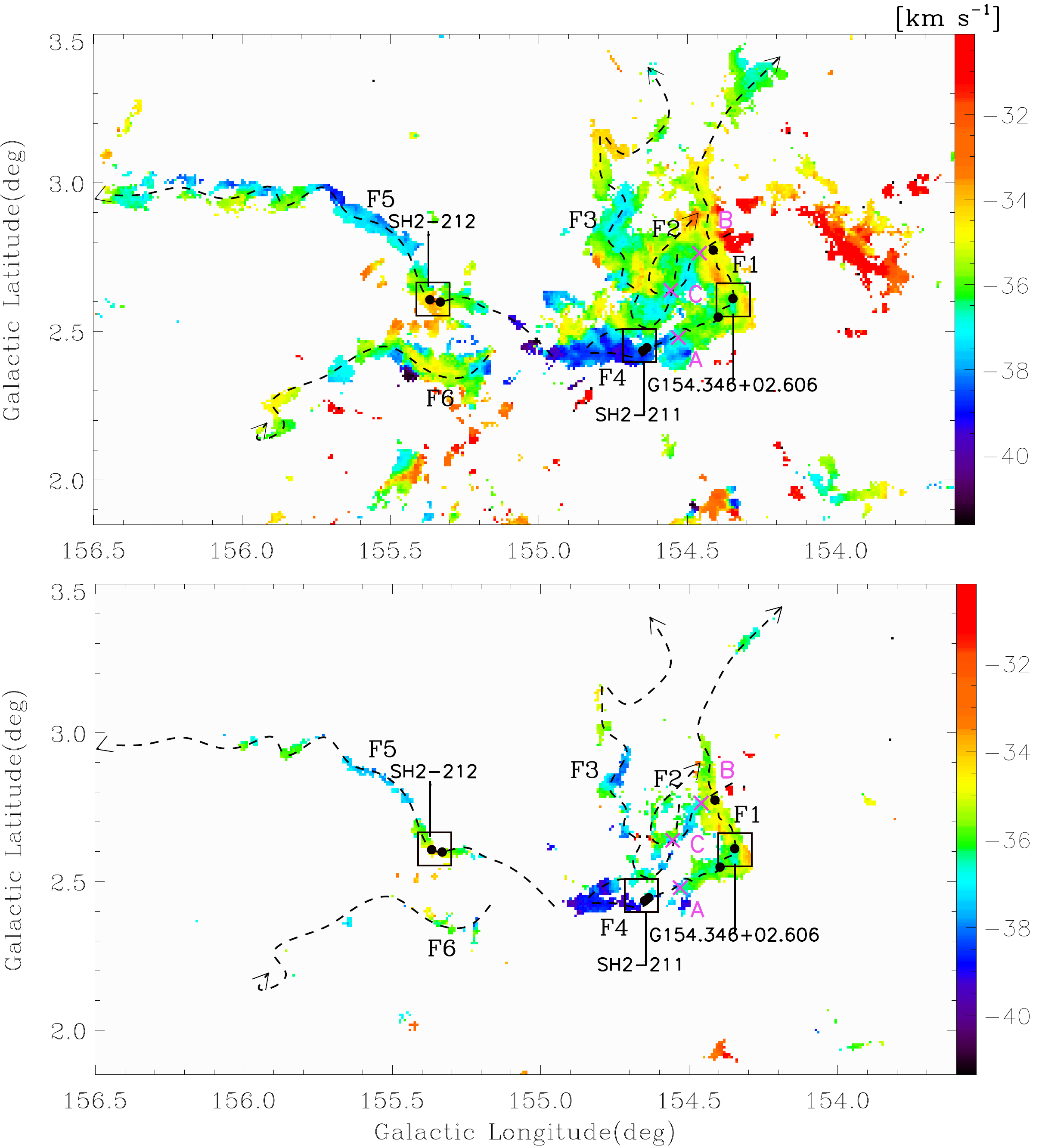} 
		\caption{The velocity-field maps (moment 1) of \tw (upper panel) and \tht (bottom panel). 
		The black arrow lines indicate the extended paths according to the \tw emission. 
		The name of each filament is marked beside.
		The symbols of squares, solid circles, magenta crosses represent
		the associated \hii regions, IRAS sources, and the intersection positions between filaments, respectively. The coordinates $(l, b)$ of the intersection are cross-A $\sim\left(154\fdg531, 2\fdg479\right)$, cross-B $\sim\left(154\fdg452, 2\fdg754\right)$, and cross-C $\sim\left(154\fdg556, 2\fdg638\right)$, respectively.}
	  \label{fig-m11}
	  \end{figure}

	  Figure \ref{fig-m11} shows the velocity field of the observed region.  In morphology, 
	  F1 $\sim$ F4 mesh together in the \tw image, while they are isolated relatively in the \tht image.
	  F1 displays a V-shape opening to east. 
	  F2, F3 and F4 are located to the east of F1 and all the four filaments arrange in the order from Northwest to Southeast. 
	  F1 connects with F4 at the cross-A and connects with F2 at the cross-B.
	  F3 connects with F2 at the cross-C.
	  F5 and F6 are separated from the complex network and continue to extend eastward. 
	  Compared to F1, F2 $\sim$ F6 have lower integrated intensities and similar aspect ratios. 
	  The three \hii regions are 
	  located at the apexes of the filaments, e.g.,  G154.346+2.606 located within F1, SH2-212 located within F5.
	  The seven IRAS sources are located within the filaments. Five of them are associated with \hii regions.
	  For F1$-$F4, the average $v_{\rm LSR}$  changes from redshifted to blueshifted  in a direction from northwest to southeast. 
	  The velocity difference is up to $\sim 10 \mathrm{\,km}\mathrm{\,s}^{-1}$.
	  Correspondingly, as shown in Figure \ref{fig-fil-spectrum} and Table\ref{tb-filaments}, the line center velocities of the four filament change form from $-35.7 \mathrm{\,km}\mathrm{\,s}^{-1}$ to $-38.6 \mathrm{\,km}\mathrm{\,s}^{-1}$ 
	  by a step of $\sim 0.5 \mathrm{\,km}\mathrm{\,s}^{-1}$.
	  In order to better analysis kinematic properties of the six filaments, we extend their paths appropriately based on velocity consistency of \tw, 
	  e.g.,  connecting F4 with F1 and extending the ends of F2, F3, F5 and F6 northward.

	   As shown in Figure \ref{fig-pvfit}, 
	   we extract five PV slices (widths $\sim 2$ pixels) along these extended paths from south end to north end.
	   The filaments  show large-scale kinematic oscillation pattern, e.g., F4+F1,F3, F5.
	   We perform a linear fitting to velocity difference to obtain velocity gradient ($\bigtriangledown v_{\shortparallel}$) when the velocity difference is 
	   greater than $0.5 \mathrm{\,km}\mathrm{\,s}^{-1}$ (3 times velocity resolution) over a length of about 4 beam widths ($\sim 4.5 \mathrm{\,pc}$).
	   For example, around the G154.346+2.606, the velocity difference is $\sim 1.34 \mathrm{\,km}\mathrm{\,s}^{-1}$ along a length of $\sim$ 0\fdg15 in the south side and $\sim 1.27 \mathrm{\,km}\mathrm{\,s}^{-1}$ along a length of $\sim$ 0\fdg13 length in the north side. The fitted velocity gradients are $\sim 1.27 \mathrm{\,km}\mathrm{\,s}^{-1} \mathrm{\,pc}^{-1}$ and $\sim 0.39 \mathrm{\,km}\mathrm{\,s}^{-1} \mathrm{\,pc}^{-1}$, respectively.
	   The feature that velocity gradients with opposite directions existing between the IRAS sources, or \hii regions or intersection points is common along these filaments, which implies ongoing mass-flow process around the star forming zones (see discussion below).
	   The results including the center coordinates and half-length of the segments, the length of the segments, the velocity differences, and the $\bigtriangledown v_{\shortparallel}$ are listed in Table \ref{tb-segment}.

	   In order to investigate the periodicity of the six filaments, 
	   we construct the 1-D second-order structure functions (SF) of velocity and column density. More specifically, we extract slices \footnote{The width of slice is 8 pixels for F1 and 5 pixels for F2 - F6.} along the splines and derive the average values as well as the standard deviations from the intensity-weighted mean velocities of ${ }^{12} \mathrm{CO}$ and the column densities estimated from ${ }^{13} \mathrm{CO}$. The SF is defined as $S_{p}(\ell)=\left\langle|x(r)-x(r+\ell)|^{p}\right\rangle$, where $x(r)$ is the velocity or density at position, $r$, $\ell$ is the spatial displacement from $r$, $p$ is the order\citep{2015ARA&A..53..583H,2020NatAs...4.1064H}.
	   The maximum lag to compute the SF is set to be the half of the total length for each filament, as the method is only valid in this range \citep{2020NatAs...4.1064H}.
	   We present the results in Figure\ref{fig-structurefun}. 
	   We do not observe significant dip in the SF of F2 and F4, indicating that no evidence for periodicity is found in the data of these two filament; we visually identify dips in F1, F3, F5 and F6 for the column densities profile (their minimum locations are [20,30,11,8]pc), and in F3, F5 and F6 for the velocity profile (with minimum locations of [25,20,13]pc).The minimum potential characteristic lengths as revealed by these dips are approximately 8-30 pc, which are larger but still comparable with the average length of the segments($\lesssim 10$ pc). The flat slopes of the density structure functions imply that the \tht's signal-to-noise ratio is not high enough, which makes the periodicity less evident.

	  	\begin{deluxetable}{cccccccc}
		\tabletypesize{\small}
		\setlength{\tabcolsep}{0.11in}
		\tablecaption{Properties of the segments along the major filament \label{tb-segment}}
		\tablewidth{0pt} 
		\tablehead{\colhead{Name}& \colhead{($l,b, \mathrm{half-length}$)} & \colhead{segment}& \colhead{Length} & \colhead{Velocity difference} & \colhead{$\bigtriangledown v_{\shortparallel}$} & \colhead{Mass} 
		& \colhead{$\dot{M}$\tablenotemark{a}}\\
		\colhead{}&\colhead{(\degr, \degr,\degr)}&\colhead{(\degr, \degr)}&\colhead{($\mathrm{pc}$)} &\colhead{($\mathrm{\,km}\mathrm{\,s}^{-1}$)} & \colhead{$\mathrm{\,km} \mathrm{\,s}^{-1} \mathrm{\,pc}^{-1}$} & \colhead{($M_{\odot}$)} & \colhead{($M_{\odot} \mathrm{Myr}^{-1}$)} 
		} 
		\startdata                                                                                                                                         
F4+F1 &   (154.76, 2.43, 0.090 )&   (0.15 , 0.33 )	& 14.1   &   0.64 	&  0.05 	& 2261  	& 115    \\ 
      &   (154.53, 2.48, 0.060 )&   (0.43 , 0.55 )  & 9.4    &   2.56 	&  0.31 	& 842       & 262    \\ 
      &   (154.44, 2.54, 0.075 )&   (0.53 , 0.68 )	& 11.7 	 &   0.39 	&  0.04 	& 2920  	& 114    \\ 
      &   (154.37, 2.67, 0.065 )&   (0.75 , 0.88 )	& 10.2 	 &   1.27 	&  0.14 	& 5149  	& 720    \\ 
      &   (154.43, 2.86, 0.075 )&   (0.95 , 1.10 )  & 11.7 	 &   1.33 	&  0.13 	& 1499  	& 189    \\ 
      &   (154.30, 3.31, 0.050 )&   (1.48 , 1.58 )	& 7.8    &   0.77 	&  0.11 	& 337       & 39     \\ 
\hline                                                                        
F2    &   (154.47, 2.72, 0.070 )&   (0.10 , 0.24 ) 	& 11.0   &	 2.13   &	 0.22  	& 2105      & 462    \\ 
      &   (154.52, 2.66, 0.035 )&   (0.22 , 0.29 ) 	& 5.5  	 &   1.39   &	 0.29  	& 322   	& 95     \\ 
\hline                                                                        
F3    &   (154.54, 2.68, 0.040 )&   (0.04 , 0.12 ) 	& 6.3  	 &   0.71  	&  0.13  	& 226   	& 30     \\ 
      &   (154.57, 2.58, 0.055 )&   (0.13 , 0.24 ) 	& 8.6  	 &   1.34  	&  0.18  	& 355   	& 65     \\ 
      &   (154.75, 2.84, 0.040 )&   (0.65 , 0.73 ) 	& 6.3  	 &   1.71  	&  0.31  	& 802   	& 246    \\ 
      &   (154.72, 2.91, 0.035 )&   (0.74 , 0.81 ) 	& 5.5  	 &   1.28  	&  0.27  	& 295   	& 81     \\ 
      &   (154.79, 3.07, 0.075 )&   (0.90 , 1.05 ) 	& 11.7   &	 2.53  	&  0.24  	& 384   	& 92     \\ 
\hline                                                                        
F5    &   (155.28, 2.61, 0.050 )&   (0.35 , 0.45 ) 	& 7.8  	 &   2.22  	&  0.32  	& 390   	& 123    \\ 
      &   (155.40, 2.66, 0.050 )&   (0.50 , 0.60 ) 	& 7.8  	 &   1.42  	&  0.20  	& 756   	& 152    \\ 
      &   (155.83, 2.95, 0.050 )&   (1.10 , 1.20 )  & 7.8  	 &   1.41  	&  0.20  	& 284   	& 57     \\ 
      &   (155.99, 2.96, 0.050 )&   (1.30 , 1.40 )	& 7.8  	 &   2.90  	&  0.41  	& 303   	& 125    \\ 
\hline                                                                        
F6    &   (155.35, 2.36, 0.035 )&   (0.18 , 0.25 )  & 5.5  	 &   3.12  	&  0.66  	& 233       &  155    \\ 
		\enddata
		\tablenotetext{a}{The mass-flow rates ($\dot{M}$) is estimated using a simple cylindrical model from \cite{2013ApJ...766..115K}, $\dot{M} = \frac{\bigtriangledown v_{\shortparallel}M_{\rm gas}}{tan(\alpha)}$, where $\alpha$ is the angle of the inclination to the plane of the sky and $M_{\rm gas}$ is the mass of the filament segment, which is estimated from ${ }^{13} \mathrm{CO}$ in this work.}
	\end{deluxetable}	
 	   			
\subsection{Clumps}\label{clumps}
	
	The general models of infinite self-gravitating fluid cylinder \citep{1953ApJ...118..116C,1964ApJ...140.1056O,2010ApJ...719L.185J}
	suggest that
	the entire filament will collapse toward the short axis and then fragment into clumps or cores when the $M_{\text {line }}$ of an isothermal filament exceeds the critical value for equilibrium. 
	To better understand the evolutionary phases of the filaments and star formation therein, 
	we apply for the Gaussclumps algorithm in GILDAS \citep{1990ApJ...356..513S} to the \tht datacube to search for molecular clumps in the velocity interval of [-42.5, -30.0] $\mathrm{\,km} \mathrm{s}^{-1}$.
	The threshold is set to be $5\,\sigma$ to avoid false clumps. 
	The rest control parameters $s_{0}, s_{a}, s_{c}, w$ are set to be 1, 1, 1, 2, respectively, which are suggested by \cite{1998A&A...329..249K}.
	After removing the clumps with a short axis less than the spatial resolution 
	($50^{\prime \prime}$) and those located on the edge of the observed region, a total of $162$ clumps are identified. 
	The Gaussclumps fitting procedure gives the information of the clumps including the position in the Galactic coordinate system, angular sizes of the major ($\Theta_{\mathrm{maj}}$) and minor axis ($\Theta_{\mathrm{min}}$), 
	$v_{\mathrm{LSR}}$, peak temperature ($T_{\rm peak}$) and spectral FWHM ($\Delta V$). 
	Using the calculation methods 
	in \cite{2016A&A...588A.104G}, 
	the physical parameters including the radii after the beam deconvolution ($R_{\mathrm{eff}}$), excitation temperature ($T_{\rm ex}$), 
	optical depth ($\tau$), column density  ($N_{\mathrm{H}_{2}}$), volume density ($n_{\mathrm{H}_{2}}$), surface density ($\Sigma_{c}$),  
	LTE mass ($M_{\text {LTE }}$), virial mass ($M_{\mathrm{vir}}$), 
	and virial parameter ($\alpha_{\mathrm{vir}}=M_{\mathrm{vir}} / M_{\mathrm{LTE}}$) of these 
	clumps are derived. The results are tabulated in Table\ref{tb-clumps}.

	The effective radius after deconvolution of the clumps range from $0.3 \mathrm{\,pc}$ to $1.9 \mathrm{\,pc}$ with a median value of $0.8 \mathrm{\,pc}$. 
	The masses under local equilibrium range from $14 \,M_{\odot}$ to $2416 \,M_{\odot}$ with a median value of $78 \,M_{\odot}$.
	The excitation temperatures range from $4.6 \mathrm{\,K}$ to $22.8 \mathrm{\,K}$ with a median value of $9.5 \mathrm{\,K}$.
	The volume densities range from $70 \mathrm{\,cm}^{-3}$ to $6554 \mathrm{\,cm}^{-3}$ with a median value of $590 \mathrm{\,cm}^{-3}$.
	The virial parameters  span a range of $0.33 \sim 2.96$ with a median value  of $1.15$.

	As shown in Figure \ref{fig-dv}, we calculate the thermal and non-thermal velocity dispersions of the $162$ clumps following the methods of \cite{2017ApJ...838...49X}.   
	The thermal velocity dispersion of each species is given by $\sigma_{\rm T}(\mu_{\rm obs})=\sqrt{\frac{k_{\rm B}T_\mathrm{kin}}{\mu_{\rm obs}m_{\rm H}}}$, 
	where $k_{\mathrm{B}}$ is the Boltzmann constant, $\mu_{\rm obs}$ is atomic weight of the observed molecule ($\mu_{\rm obs}=29$ for \tht), $T_{\text {kin }}$ is the kinetic temperature that equals the excitation temperature and $m_{\mathrm{H}}$ is the mass of hydrogen atom.
	The non-thermal velocity dispersion is estimated by subtracting the thermal velocity dispersion 
	from the measured linewidth, $\sigma_{\rm NT}=\sqrt{{\sigma_{\rm obs}}^{2}-{\sigma_{\rm T}}^{2}(\mu_{\rm obs})}$,
	where $\sigma_{\rm obs}=\Delta V /\sqrt{8\ln2}$, and $\Delta V$ is the measured FWHM from the Gaussclumps procedure. 
	The three dimensional velocity dispersion ($\sigma_{\mathrm{3 D}}$) is estimated as $\sigma_{\mathrm{3 D}}=\sqrt{3} \sigma_{\mathrm{obs}}$.

\subsection{Young Stellar Objects}\label{YSO}

	Using data from 2MASS \citep[Two Micron All Sky Survey,][]{2006AJ....131.1163S} and WISE \citep[Wide-field Infrared Survey Explorer,][]{2010AJ....140.1868W} and a categorization scheme supplied by \cite{2014ApJ...791..131K}, we investigate the disk-bearing candidate YSOs in this region. To complement our catalogue, we use the YSO catalogues from \cite{2016MNRAS.458.3479M,2019MNRAS.487.2522M} who used machine learning classifiers to search for YSOs through WISE photometry and Gaia DR2 \citep[the second $Gaia$ Data Release,][]{2018A&A...616A...1G}.
	The disk-bearing YSOs can be categorized into three classes based on the \cite{1994ApJ...434..614G} concept: Class I ($\mathrm{ALPHA} \geqslant 0.3$), flat-spectrum ($0.3 \geqslant \mathrm{ALPHA} \geqslant-0.3$), and Class II ($-0.3 \geqslant \mathrm{ALPHA} \geqslant-1.6$).
	The infrared (IR) spectral index ALPHA is defined as $\frac{dlog(\lambda F_{\lambda})}{d log(\lambda)}$, 
	where $F\rm _{\lambda}$ is the flux as a function of wavelength $\lambda$.
	A total of $279$ disk-bearing candidate YSOs are identified in this region,
	of which 221 sources have matched Gaia objects and 203 sources have stellar parallax ($Plx$).
	To eliminate the foreground objects, we adopt a $30\%$ distance uncertainty of the cloud \citep{2019A&A...622A..52Z} and compare it with the parallax distance from Gaia ($1/Plx$). 
	Finally, a total of 103 sources are reserved as the YSO sample, including 26 (77) sources with (without) Gaia parallax distance information. 
	If $3 \,\sigma$ of the \tw emission is used to define the boundary of the cloud, $79$ sources are located within the cloud.
	The fraction of forground/backgroud contamination is estimated to be $\sim 6.6\%$ by the density of sources outside the cloud and the emission area of the cloud. 
	In the YSO sample, $39\%$ are from \cite{2016MNRAS.458.3479M} and $29\%$ are from \cite{2019MNRAS.487.2522M}. 
	The contamination level may be underestimated as the YSOs identified from  
	\cite{2019MNRAS.487.2522M} are only retained in the region where the dust opacity value is higher than $1.3\times10^{-5}$.
	Of the $79$ sources, $32$, $24$, and $23$ are classified as Class I, flat-spectrum,  and Class II, respectively.
	The IR photometric magnitudes, the ALPHA values and the origin of the YSOs sample are listed in Table\ref{tb-YSOs}.
	As shown in Figure \ref{fig-yso},  most of the candidate YSOs are distributed along the ridgelines of  the filaments.
	In addition, the Class I objects (magenta plus) appear to be assembled nearby the three \hii regions.

\section{Discussion}

\subsection{Large-scale Filaments}\label{Large-Scale Filaments}  

      Giant molecular filaments (GMF) can reach up to Galactic scale and play an important role as part of the spiral arms  \citep{2014A&A...568A..73R, 2015MNRAS.450.4043W, 2015ApJ...815...23Z,2017ApJS..229...24D}. However, there is no clear definition of the GMF. 
      According to the description of \cite{2019A&A...622A..52Z}, in position-position-velocity space, an elongated molecular cloud structure with length greater than 10\,pc and mass greater than $10^{3}M_{\odot}$ could be considered as GMF. In this MC, except F6, these filaments can be considered as GMF as their length range from $\sim$ 32\,pc to $\sim$ 90\,pc and mass range from $\sim 1725\,M_{\odot}$ to $\sim 13650 \,M_{\odot}$.
	  The fitted widths of these filaments are $1.3-3.9 \mathrm{\,pc}$, which are much larger than the typical value with of `$Herschel$ filaments' \citep[0.1$\mathrm{\,pc}$,][]{2011A&A...529L...6A}. The low-J CO only trace relatively diffuse gas and the angular resolution ($\sim 1.1 \mathrm{\,pc}$ at a distance of $\sim 4.5 \mathrm{\,kpc}$) is far from enough to resolve the widths of the filaments.

	  According to the criteria of hub-filament that a central body 
	  of low aspect ratio and high column density surrounded by branches with features of greater aspect ratio and lower 
	  column density \citep{2009ApJ...700.1609M}, and the criteria of ridge-nest structure  that a single dominating 
	  denser filament with a disorganised network of lower-density filaments \citep{2011A&A...533A..94H}, 
	  we suggest that the construction of the identified filaments is similar to the  `ridge-nest' rather than `hub-filament'.  
	  We consider F1 as the main filament and F2 to F6 as sub-filaments.
	  As shown in Table\ref{tb-filaments}, F1 $\sim$ F5 have larger $M_{\text {line }}$ than the critical mass per unit length 
	  ($M_{\text {line, crit}}$) in 
	  the isothermal cylinder model \citep{1964ApJ...140.1056O, 1997ApJ...480..681I},  here
	  $M_{\rm line, crit} = 2{c_{s}}^{2}/G \sim$ 16.4 ($T/10 \mathrm{K}$) $M_{\sun} \rm pc^{-1}$ 
	  ($c_{\rm s}$ is the isothermal sound speed $\sim 0.2 \mathrm{\,km}\mathrm{\,s}^{-1}$, and $G$ is the gravitational constant). 
	  The $M_{\text {line, crit}}$ is the critical value required for a filament to be gravitationally unstable to radial contraction and 
	  fragmentation along its length \citep{1997ApJ...480..681I}. 
	  The observed $M_{\text {line }}$ of F1 $\sim$ F5 exceed $M_{\text {line, crit}}$ except F6. 
	  It indicates that filaments F1 $\sim$ F5 are gravitationally unstable and should fragment into dense cores, 
	  while F6 may expand eventually.

\subsection{Kinematics of the Filaments}\label{Velocity Structures} 

	\begin{deluxetable*}{lccccccc}
		\tabletypesize{\small} 
		\setlength{\tabcolsep}{0.10in}
		\tablecaption{Characteristic fragmentation lengths of the filaments \label{tb-lengthscale}}
		\tablewidth{0pt}
		\tablehead{ \colhead{Name} &  \colhead{$l_{v}$\tablenotemark{a}}  &  \colhead{$l_{\mathrm{N_{H_{2}}}}$\tablenotemark{b}}
			& \colhead{$c_{s}$} & \colhead{$\sigma_{\mathrm{obs}}$}  & \colhead{$n_{c}$\tablenotemark{c}}
			& \colhead{$\lambda_{\mathrm{crit,th}}$}& \colhead{$\lambda_{\mathrm{crit,turb}}$}\\
			&   \colhead{pc} &   \colhead{pc} & \colhead{(km s$^{-1}$)} & \colhead{(km s$^{-1}$)}& \colhead{$\mathrm{~cm}^{-3}$} & \colhead{pc}& \colhead{pc}
			}
		\startdata
		F1 &      -   & 20  &  0.17 & 0.89   & 366   &1.7 &9.2 \\
		F3 &      25  & 30   &  0.16 & 0.54   & 67    &3.8 &13.0 \\
		F5 &      20  & 11  &  0.15 & 0.56   & 157   &2.3 &8.8 \\
		F6 &      13  & 8  &  0.14 & 0.93   & 12    &7.9 & 52.8\\
		\enddata 
		\tablenotetext{a}{Potential characteristic length scales estimated from \tw velocity structure function.}
		\tablenotetext{b}{Potential characteristic length scales estimated from \tht column density function.} 
		\tablenotetext{c}{$n_{c}= N_{\mathrm{c}}/\mathrm{width}$.}
	\end{deluxetable*}

According to the theoretical model of an infinite self-gravitating fluid cylinder \citep{1953ApJ...118..116C,1964ApJ...140.1056O,2010ApJ...719L.185J}, a hydrostatic filament can fragment if perturbations
are larger than the critical wavelength of $\lambda_{\mathrm{crit}} =11 H$,  and its gravitational instability would peak at $\lambda_{\max }=2\lambda_{\mathrm{crit}}$.
The isothermal scale height $H$ is expressed as $c_{s}\left(4 \pi G \rho_{c}\right)^{-1 / 2}$, where $c_{\mathrm{s}}$ is the isothermal sound speed ($\sqrt{\frac{k_{\mathrm{B}} T_{\mathrm{kin}}}{\mu_{\mathrm{obs}} m_{\mathrm{H}}}}$), $G$ is the gravitational constant, $\rho_{c}$ is an initial gas mass density at the center of the filament. If turbulent pressure dominates over thermal pressure, $c_{s}$ should be replaced by velocity dispersion $\sigma_{\mathrm{obs}}$.
We adopt the initial central density $\rho_{\mathrm{c}}=\mu m_{\mathrm{p}} n_{\mathrm{c}}= \mu m_{\mathrm{p}}N_{\mathrm{c}}/\mathrm{width}$, where $N_{\mathrm{c}}$ is the central column density along the filament, ``width" is the deconvolved FWHM from Gaussian fitting of the radial column density profiles (see, Figure \ref{fig-profile}), $m_{\mathrm{p}}$ is the proton mass, and $\mu=2.33$ is the molecular
weight of molecular gas. 
The typical observed velocity dispersion $\sigma_{\mathrm{obs}}$ is estimated by $\Delta V / \sqrt{8 \ln 2}$, where $\Delta V$ is the average value of FWHM (Gaussian fitting) derived from each \tht spectral line along the filaments. 
The theoretical predicted fragmentation characteristic lengths for the filaments with dips in their SF are listed in Table \ref{tb-lengthscale}. 
The observed fragmentation characteristic lengths of F1, F3, and F5 are comparable with the predicted critical wavelength ($\lambda_{\text {crit }}$) or maximum wavelength ($\lambda_{\max }$) of fragmentation of the self gravitating fluid cylinder.
As described in \cite{2011A&A...533A..34H}, based on the self-gravitating cylinder model, if the oscillatory pattern along filament is caused by core-forming motions, there will be a  phase-shift of $\lambda/4$ between density and velocity. However, using the method of cross-correlation as demonstrated by \cite{2020NatAs...4.1064H}, we have not found significant evidence of the co-oscillation at such phase-shift.

\begin{figure}[htb]
		\centering
		\includegraphics[width=0.88\textwidth]{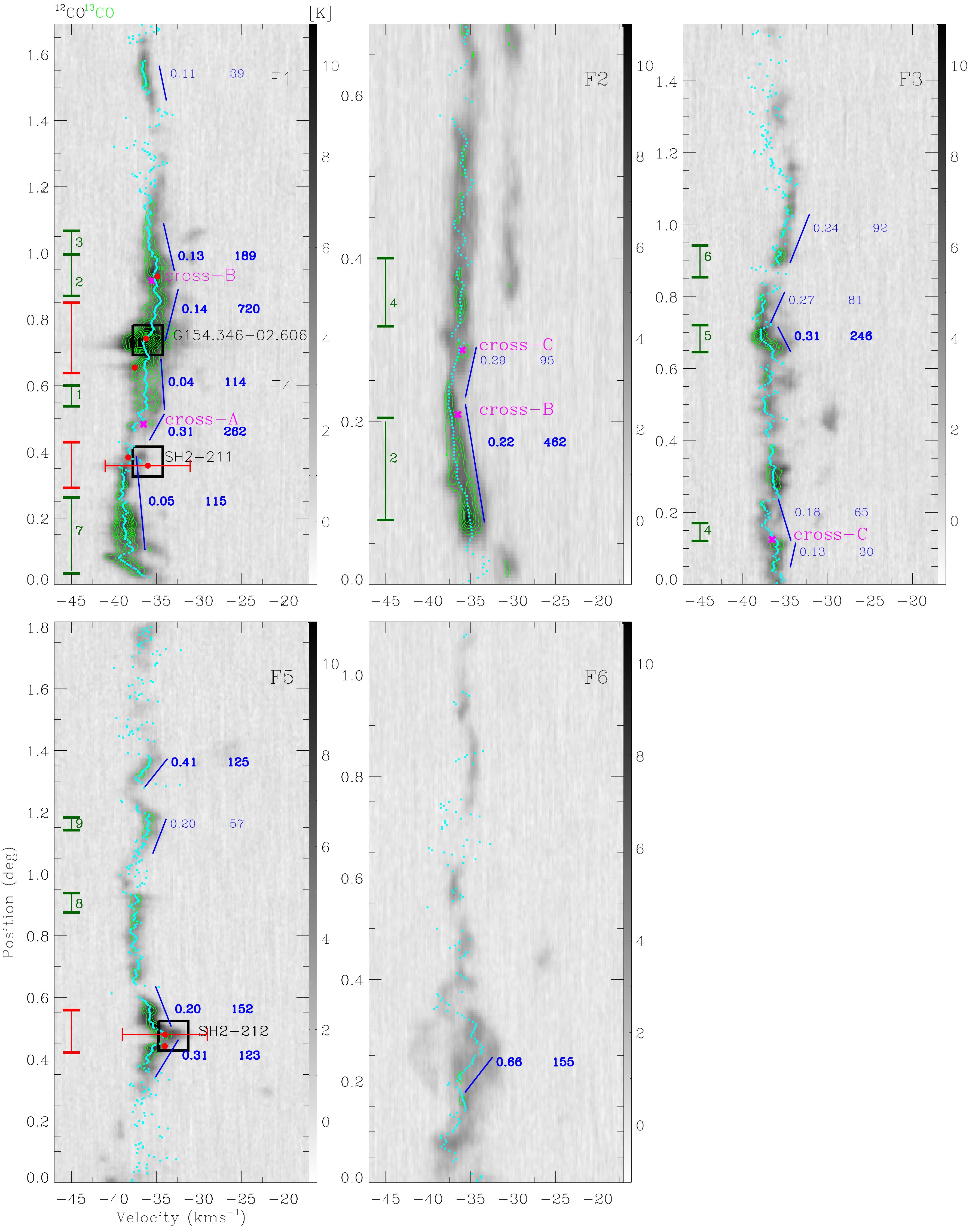} 
		\caption{The PV diagrams of \tw (grayscale) and \tht (green contours) along the filaments. 
		The green contours are set from 3$\,\sigma$ to peak value in step of 3$\,\sigma$. The colorbar is in unit of K. 
		In order to preserve the velocity continuity, pixels above 2$\,\sigma$ are kept to construct the \tht m1 profiles (cyan dotted lines). The blue lines are the linear fitting results of the corresponding segments. The corresponding velocity gradients in units of $\mathrm{km\,s}^{-1}\mathrm{\,pc}^{-1}$ and the mass-flow rates in units of  $M_{\sun}\mathrm{Myr}^{-1}$ are marked besides. The symbols of black squares, red solid circles, and magenta crosses are the star formation sets of \hii regions, IRAS sources, and intersection points between filaments.The red and green intervals mark out the zoomed-in areas in Figure\ref{fig-wise}}.
	\label{fig-pvfit}
    \end{figure}

\begin{figure}[htb]
		\centering
		\includegraphics[width=1.0\textwidth]{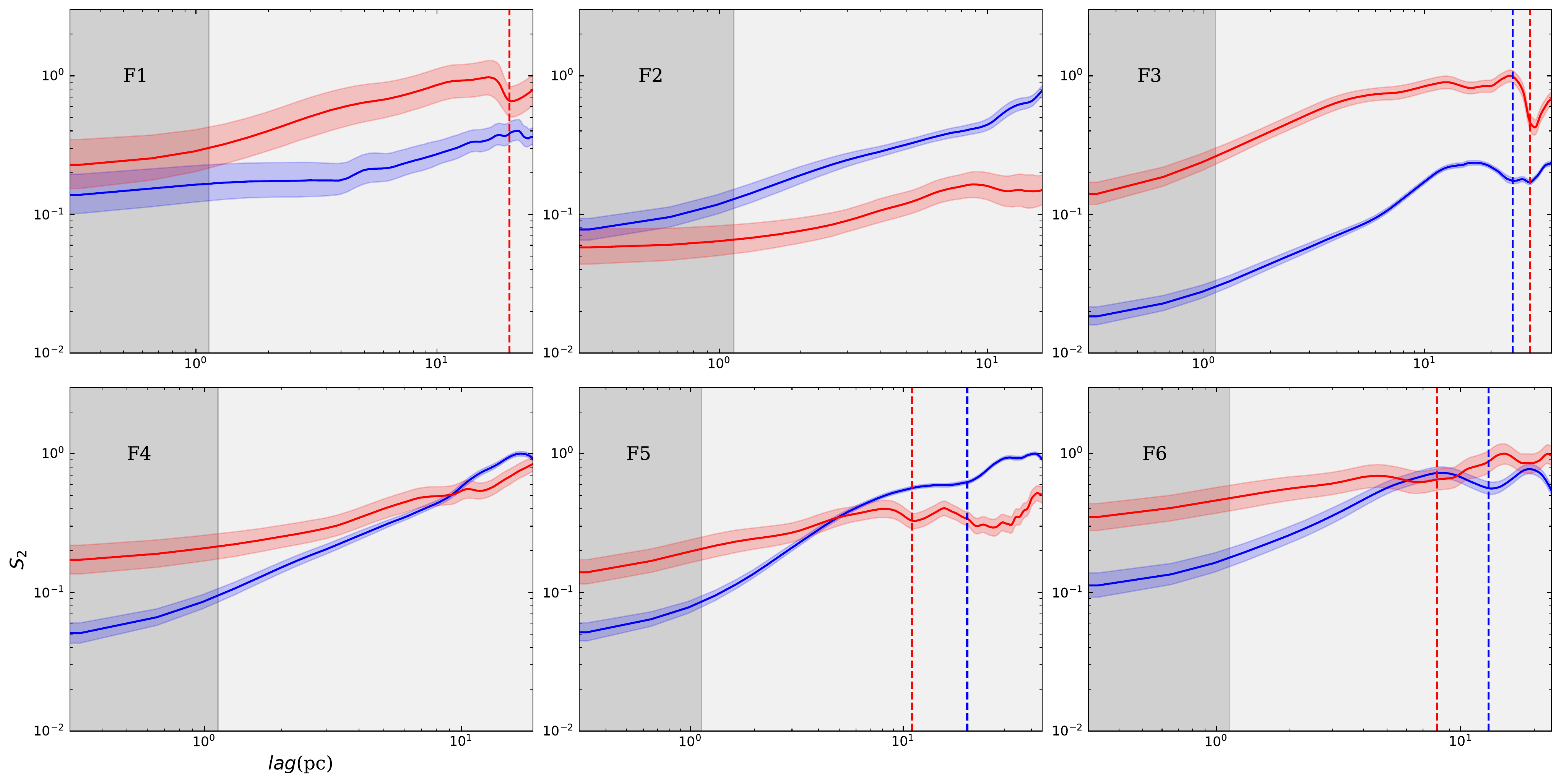} 
		\caption{The normalised second-order structure functions of velocity (blue solid lines) and column density (red solid lines). 
		In order to maintain the velocity consistent, 
		the intensity-weighted mean velocity of \tw along each filament is used to construct the velocity profiles. The column density estimated from \tht is used to construct the density profile. Pixels above $3\sigma$ are kept in the process. The blue and red shaded areas indicate the standard deviation ranges of  the structure function profiles. The vertical blue and red dashed lines indicate the potential characteristic length scales of periodic fluctuations in velocity and density.  
		The dark grey shaded area indicates the range of lags below spatial resolution of the data. 
		}
	\label{fig-structurefun}
    \end{figure}

\begin{figure}[htpb]
	    \centering
        \includegraphics[width=1.0\textwidth]{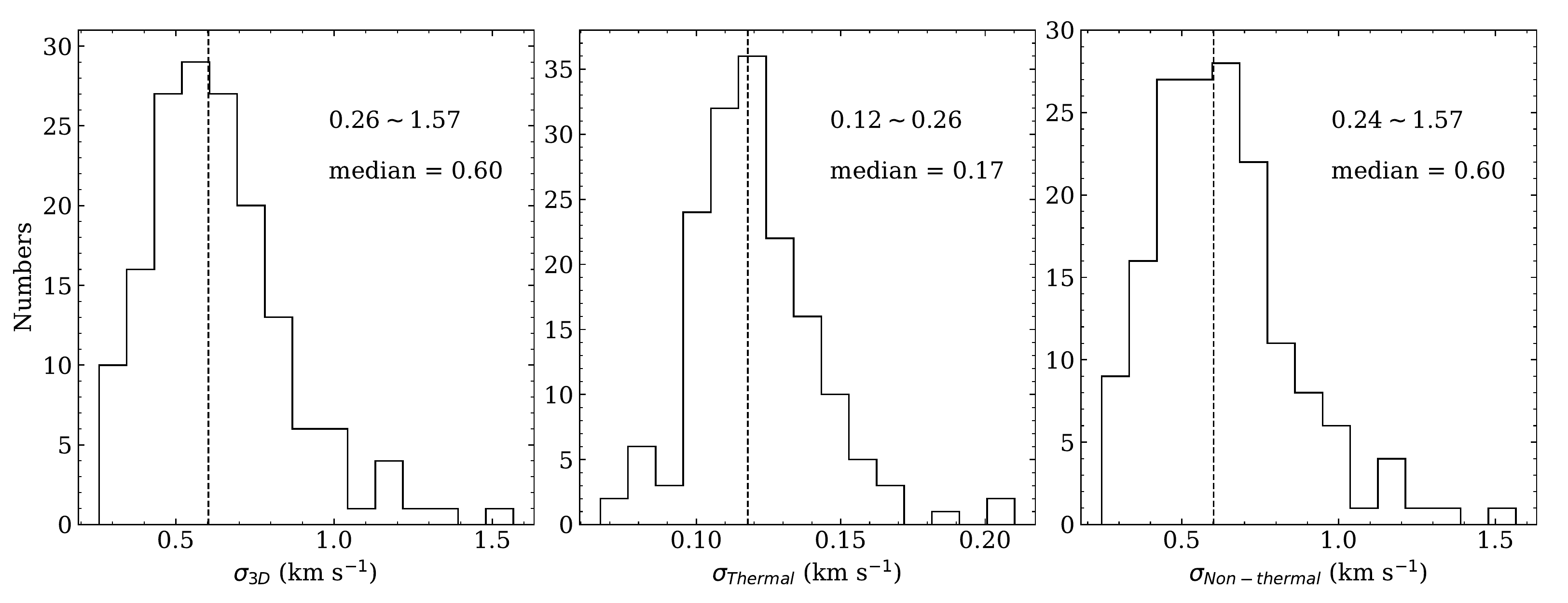}
	    \caption{Histogram of three-dimensional velocity dispersion (left panel), the one-dimensional thermal (middle panel) and non-thermal (right panel) velocity dispersions of the \tht clumps. The range and median values are marked on each panel.}
	      \label{fig-dv}
	\end{figure}

\begin{figure}[htpb]
	      \centering
	      \includegraphics[width=0.98\textwidth]{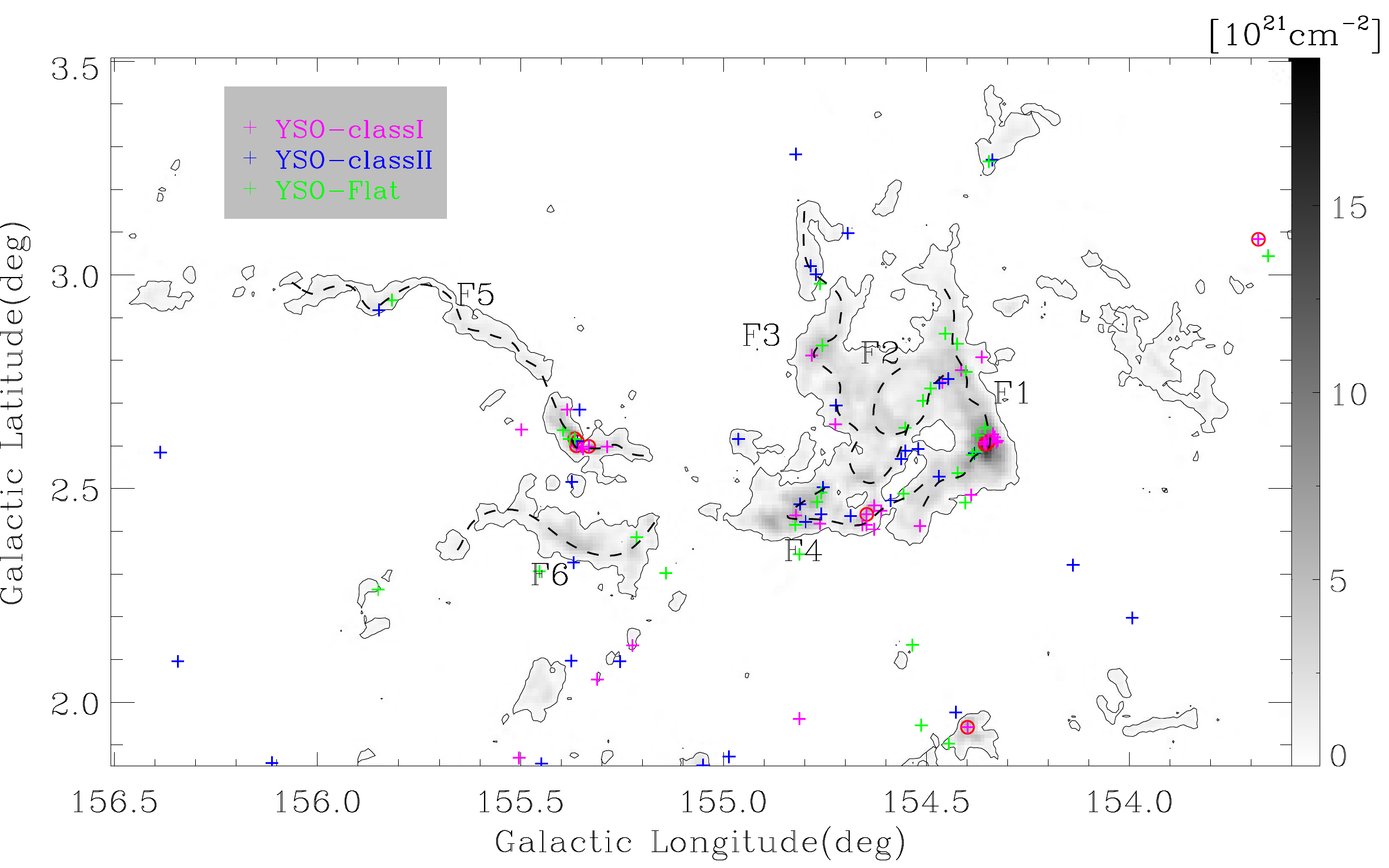}
	      \caption{The distribution of the candidate YSOs. 
	      The grayscale background is the $\mathrm{H}_{2}$ column density map estimated from \tw. 
	      The black dashed lines are the filament skeleton identified in Figure \ref{fig-cubemoment}.
	      The gray contour represents  the 3 $\sigma$ boundary of the \tw emission ( $0.2\times 10^{21} \mathrm{cm}^{-2}$).
	      The magenta, green, and blue crosses represent the candidate YSOs of Class $\mathrm{I}$, the Flat type, and 
	      Class $\mathrm{II}$, respectively. The red circle with plus indicates the candidate YSOs with luminosity greater than $100 L_{\odot}$. 
	      }
	      \label{fig-yso}
	\end{figure}

       The oscillation pattern along PV diagrams commonly appears in spectral line observations towards filamentary clouds \citep[see, e.g.,][]{2010ApJ...719L.185J,2011A&A...533A..34H,2013A&A...554A..55H,2021ApJ...921...23G}, which is generally interpreted as mass transport along the segment due to core accretion\citep[][]{2013ApJ...766..115K,2014MNRAS.440.2860H,2013ApJ...769..115H,2018ApJ...855....9L}.
	   We estimate mass-flow rates ($\dot{M}$) using a simple cylindrical model from \cite{2013ApJ...766..115K}, $\dot{M} = \frac{\bigtriangledown v_{\shortparallel}M_{\rm gas}}{tan(\alpha)}$, where $\alpha$ is the angle of the inclination to the plane of the sky, which is assumed to be $45^{\circ}$  \citep{2018ApJ...852...12Y,2021A&A...646A.137L}.
	   $M_{\rm gas}$ is the mass of the filament segment, which is estimated from ${ }^{13} \mathrm{CO}$ in this work.
	   We apply for this method to the regions where mass-flow may exist along the filaments (see, Figure\ref{fig-pvfit}). The results of the mass-flow rates ($\dot{M}$) along the filaments are listed in Table \ref{tb-segment}.

	  Around the \hii region G154.346+02.606, 
	  the estimated $\dot{M}$ are $\sim 720 \,M_{\odot} \mathrm{Myr}^{-1}$ and $\sim 114 \,M_{\odot} \mathrm{Myr}^{-1}$, respectively. 
	  This result is consistent with the infall rate of the high-mass star formation regions from high resolution observations,  e.g. the hub-filamentary cloud G22 \cite[$\sim 400\,M_{\odot}\mathrm{Myr}^{-1}$,][]{2018ApJ...852...12Y}, the infrared dark cloud G14.225-0.506 \cite[$\sim 100\,M_{\odot}\mathrm{Myr}^{-1}$,][]{2019ApJ...875...24C}. 
	  There may be deviations in mass estimation using different probes, e.g., 
	  the mass of G22 is derived from multiple mid-infrared extinction data, 
	  while the mass of G14.225-0.506 is estimated from N$_{2}$H$^{+}$(1-0) which has a higher critical density. 
	  Due to the assumption of LTE conditions throughout the cloud, optical thin of ${ }^{13} \mathrm{CO}$ and 
	  a constant \tht to H$_{2}$ abundance,
	  the mass maybe underestimated by 2-3 times  from \tht in this work \citep[see, e.g.,][]{2009ApJ...699.1092H,2008ApJ...680..428G}.
	  Even so, the estimated mass-flow rate is still higher than $100\,M_{\odot}\mathrm{Myr}^{-1}$ which is thought to 
	  be high enough to allow the formation of even O-type stars \citep{2018PASJ...70S..53I}.
	  For the region with a high mass flow rate of $\sim 720 M_{\odot} \mathrm{Myr}^{-1}$, we obtain a mean volume density ($\rho$) of $272 \mathrm{\,cm}^{-3}$ from \tht. The timescale of the free-fall time $t_{\mathrm{ff}}=\left(\frac{3 \pi}{32 G\rho}\right)^{0.5}$ is estimated to be about $3.1 \mathrm{\,Myr}$. 
	  Within one $t_{\mathrm{ff}}$, the mass would have accumulated to $2232\,M_{\odot}$, roughly 7.5\% of the total mass of the molecular cloud estimated from \tht.

	  There are about nine regions with mass-flow rate around them greater than $100\,M_{\odot}\mathrm{Myr}^{-1}$ (marked by bold numbers in Figure \ref{fig-pvfit}). These regions include the three \hii regions, the intersections of the filaments (cross-A and cross-B), and a dense region along F3. The \ei emission also appears in most of these areas correspondingly. 
	  All the systematic velocities at the intersections are relatively redshifted compared to that of the ambient gas, which implies that the intersection could play a role similar to a hub. The IRAS04335+5110 is overlaid with the cross-B and the velocity dispersion there seems relatively larger than its periphery, implying star formation activity is ongoing there.
	  In addition to the star forming zones and intersections, the most eye-catching region is the dense area along F3, which has a material flow rate of $246\,M_{\odot}\mathrm{Myr}^{-1}$. 
	  It is of interest to study 
	  star formation activity therein in the further observations.

\subsection{Distribution of the Clumps }\label{distribution of clumps}	

\begin{figure}[htpb]
	      \centering
	      \includegraphics[width=1.0\textwidth]{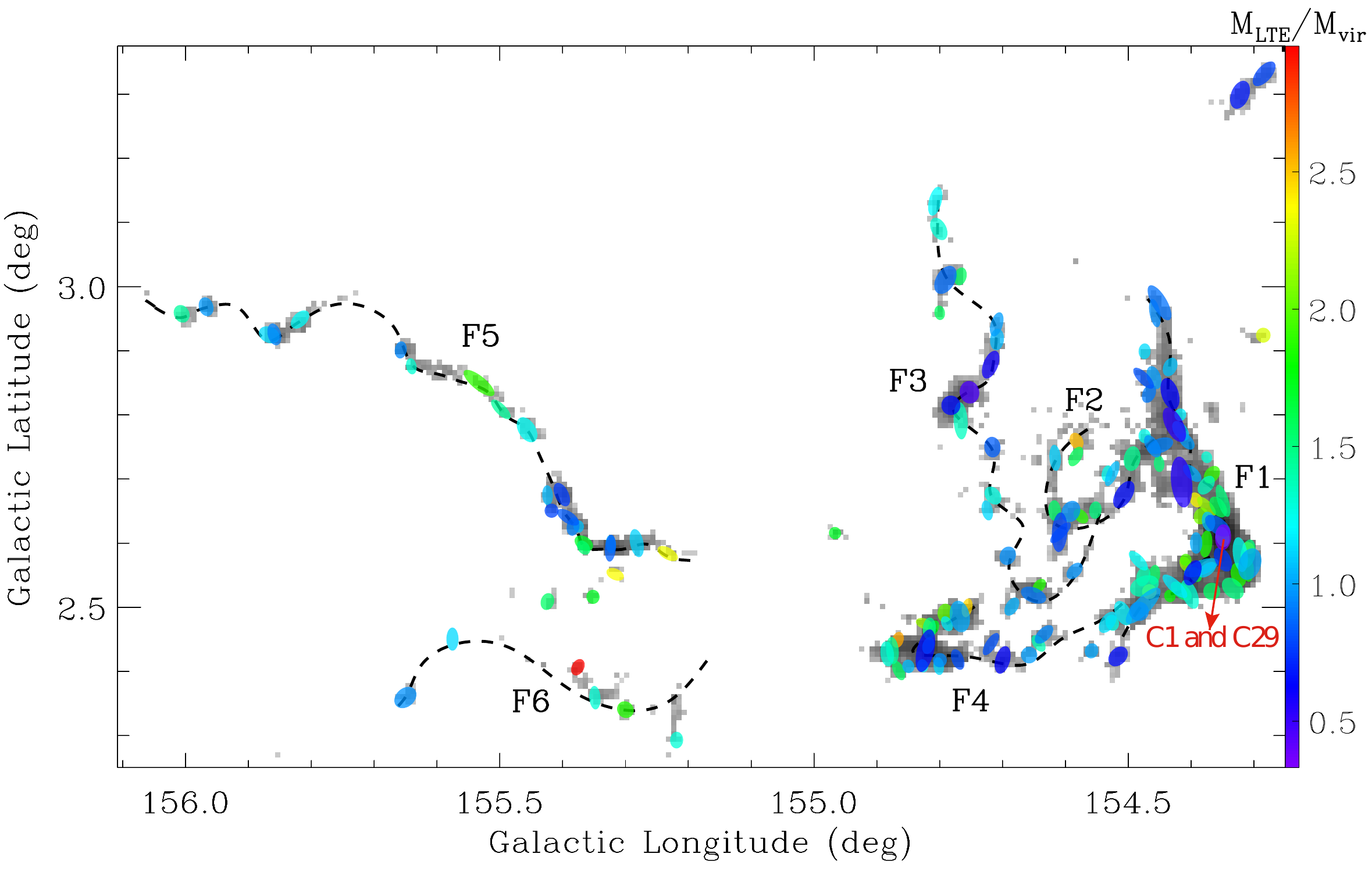}
	      \caption{The \tht velocity-integrated intensity map (moment 0) that overlapped with the clumps colorscaled by $\alpha_{\rm vir}$ . The black dashed lines 
	      are same as those in Figure \ref{fig-cubemoment}.
	      }
	      \label{fig-coredistr}
	\end{figure}

      As shown in Figure \ref{fig-coredistr}, most identified clumps are distributed along the filaments.  About $43 \%(66 / 153)$ clumps have a separation less than $2 \mathrm{\,pc}$ ($\sim$ half of the filaments width) from the nearby backbones of the filaments. The spatial association between filaments and clumps is remarkable, which suggests that clumps form primarily along filaments \citep[see, e.g.,][]{2013ASPC..476...95A}. The ratios of clumps with $\alpha_{\mathrm{vir}}<1$ and $\alpha_{\mathrm{vir}}<2$ are approximately $32 \%(52 / 162)$ and $90 \%(147 / 162)$, respectively. Most of the clumps are in gravitationally bound states, implying that the global contraction along filaments is efficient. As shown in Figure \ref{fig-dv}, similar distributions between the observed velocity dispersion and the calculated non-thermal velocity dispersion suggest that the clumps are dominated by non-thermal motions, e.g., hierarchical, global gravitational collapse, stellar feedbacks, or accretion-driven turbulence, etc.

\begin{figure}[htp]
	    \centering
        \includegraphics[width=1.0\textwidth]{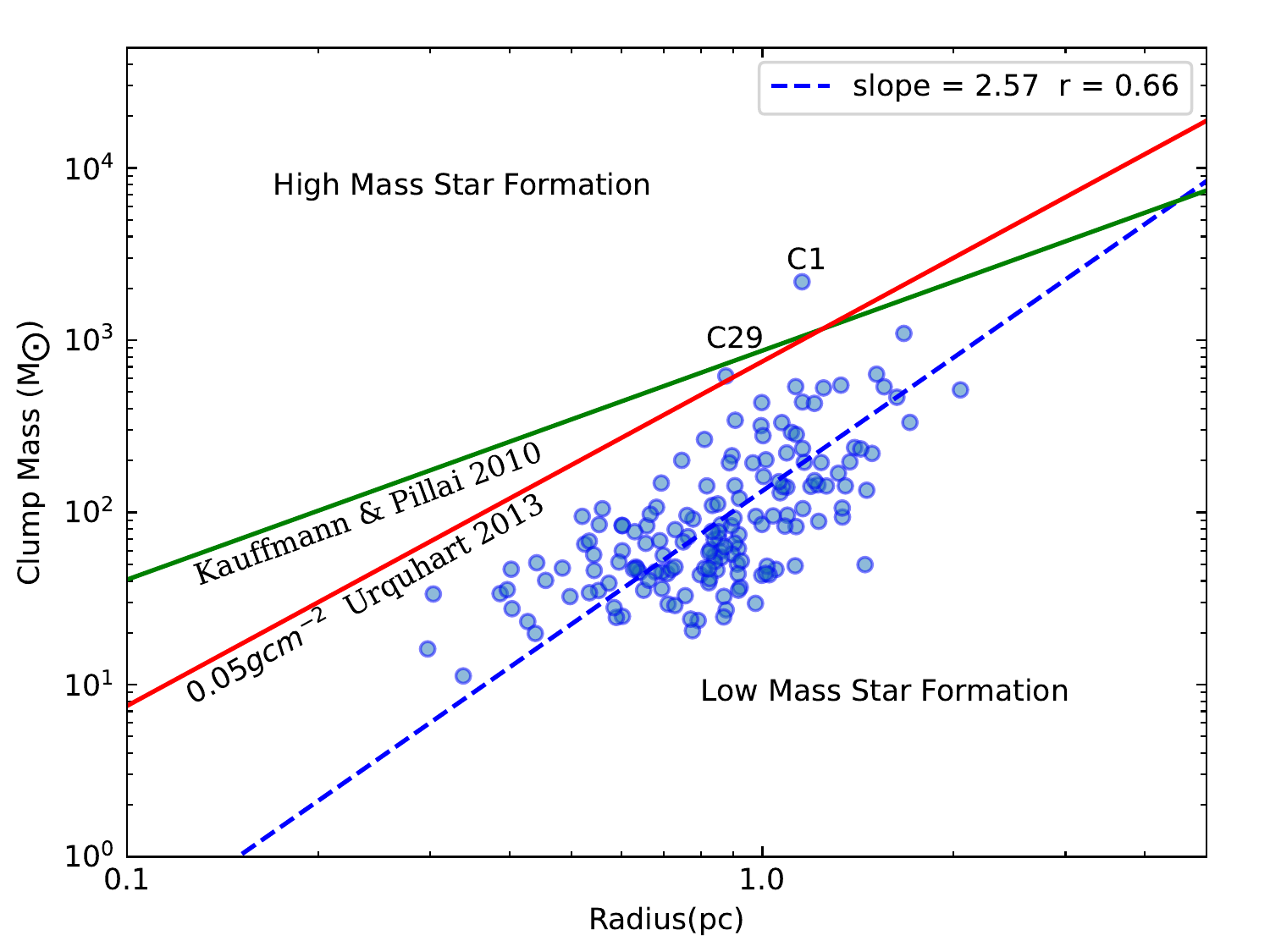}
        \caption{The mass versus radius diagram of \tht clumps. The green line and red line represent the lower limits of high-mass star formation ($M_{c}=870 M_{\odot}(R/ \mathrm{pc})^{1.33}$) from \cite{2010ApJ...716..433K} and $0.05 \mathrm{\,g\,cm}^{-2}$ from \cite{2013MNRAS.431.1752U}. The linear fitted slope and correlation coefficient of the clump sample are labled in the upper right corner.}
        \label{fig-Lar2}
	\end{figure}

	  As shown in Figure \ref{fig-Lar2}, the clump C1 (MWISP G154.349+2.609) is located above the demarcation line ($M_{\rm c}=870 \,M_{\odot}(\mathrm{R} / \mathrm{pc})^{1.33}$) between high- and low-mass star formation that suggested by \cite{2010ApJ...716..433K}. 
	  The surface densities of the clump C1 and C29 (MWISP G154.350+2.610) are higher than the lower limit ($0.05 \mathrm{\,g}\mathrm{\,cm}^{-2}$) of high-mass star formation suggested by \cite{2013MNRAS.431.1752U}.
	  It implies that the two clumps (C1 and C29) have potential to form high-mass stars. 
	  These two clumps are located in the apex of filament F1, which is similar with the scene revealed by $Herschel$ maps that massive stars tend to form at the junctions of supercritical filaments \citep[see,e.g.,][]{2010A&A...520A..49S,2012A&A...543L...3H}.
	  The most massive clump in this sample is Clump C1. It has a mass of $\sim 2416 \,M_{\odot}$ with a diameter of $\sim 1.1 \mathrm{\,pc}$ and a viral parameter of $\sim 0.33$.
	  In addition, we note that clumps with $\alpha_\mathrm{vir}\leqslant 1$ mainly distribute in the apexes of filaments, e.g., the apex of F1, F4, the north of F1, F3,  the south of F5. Clumps in these regions may go through rapid collapses to form stars.
	  Generally, a sample of star-forming regions with roughly constant column density should have a mass distribution $M \propto R^{\sigma}$ with $\sigma \simeq 2$. The linear fitted slope of this sample is $2.57$ with a correlation coefficient $r \simeq 0.66$, which agrees with the results of \cite{2015ApJ...805..157E} and \cite{2018MNRAS.477.2220T}, greater than the results $(\simeq 1.6-1.7)$ of \cite{2010ApJ...724..687L}, \cite{2010ApJ...716..433K}, and \cite{2014MNRAS.443.1555U}, and less than the results ($\geq 2.7$) from \cite{2009ApJ...698..324R}. The slope greater than 2 may be caused by the superposition of mass in the line of sight direction \citep{2020SSRv..216...76B}.

\subsection{Star Formation Rate and Efficiency}

	\begin{deluxetable}{lcccc}
		\tabletypesize{\scriptsize} 
		\setlength{\tabcolsep}{0.1in}
		\tablecaption{The 10 massive candidate YSOs in the observed field. \label{tb-massiveyso}}
		\tablewidth{0pt}
		\tablehead{ \colhead{allwise} & \colhead{$l$} & \colhead{$b$} & \colhead{from} & \colhead{Infrared luminosity}   \\
			& \colhead{($\degr$)} & \colhead{($\degr$)} &  & \colhead{($L_{\odot}$)}	}
		\startdata
        J043326.34+504330.2	&	154.3989685	&	1.941220484	&	SVMClassI/II	& 307            \\
J043543.73+520129.5	&	153.6823386	&	3.083409916	&	ClassI      	& 262             \\
J043620.88+511222.4	&	154.3536567	&	2.603973726	&	SVMClassI/II	& 793            \\
J043620.99+511254.1	&	154.3473066	&	2.610084698	&	SVMClassI/II	& 10895            \\
J043621.33+511218.1	&	154.3553349 &	2.604027456	&	SVMClassI/II	& 232             \\
J043649.54+505242.8	&	154.647162 	&	2.439575961	&	SVMClassIII 	& 235             \\
J044027.19+502828.8	&	155.331961  &	2.59873844  &	SVMClassI/II	& 2944             \\
J044035.03+502710.5	&	155.3620685	&	2.599936929	&	SVMClassI/II	& 102             \\
J044037.27+502740.7	&	155.3597037	&	2.609947118	&	SVMClassIII 	& 138             \\
J044040.84+502738.5	&	155.3664467	&	2.616627731	&	SVMClassIII 	& 105            \\
                                                                                    
		\enddata
		\tablenotetext{a}{Integral luminosity estimated from WISE.}
	\end{deluxetable}

\begin{figure}[htpb]
	      \centering
	      \includegraphics[width=0.48\textwidth]{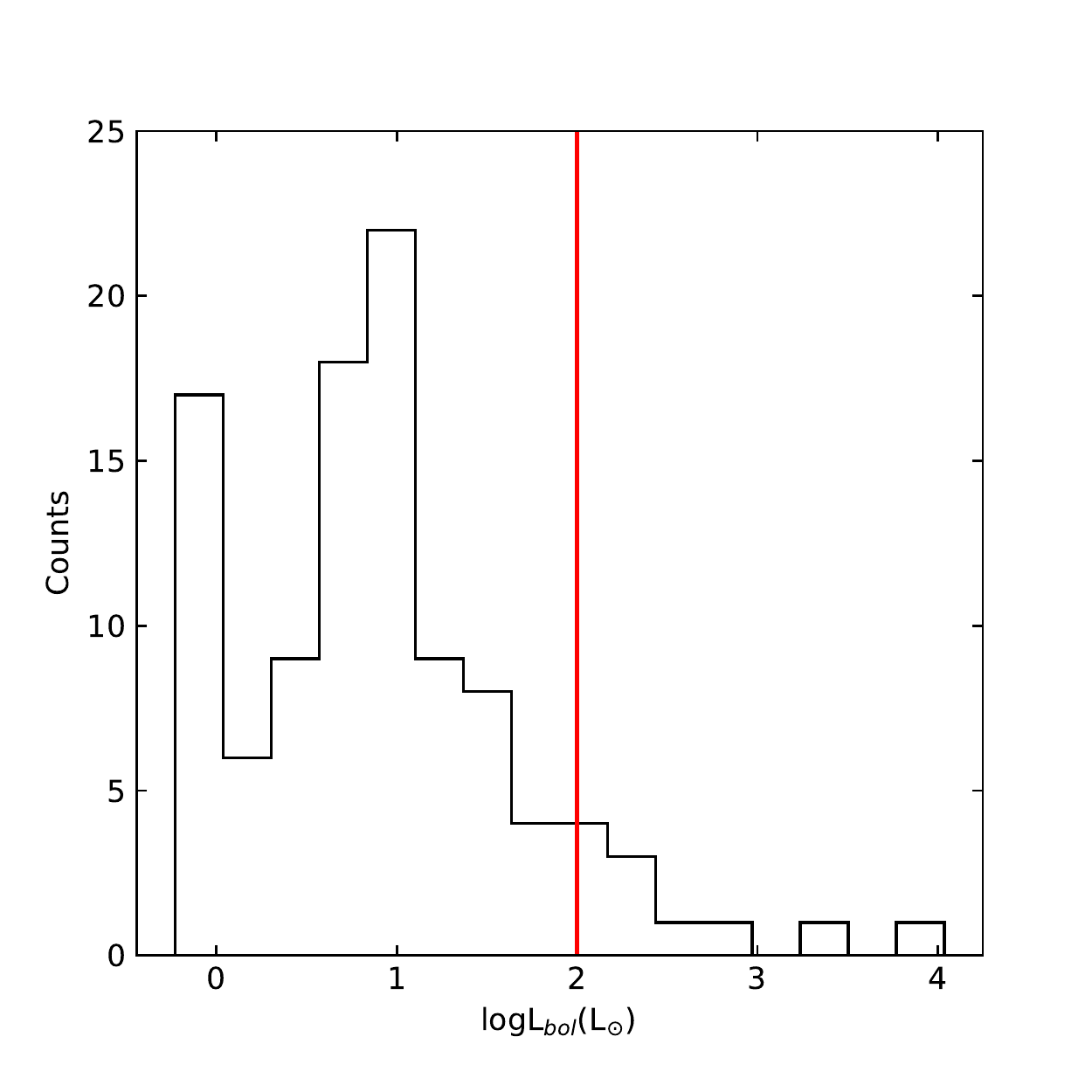}
	      \includegraphics[width=0.48\textwidth]{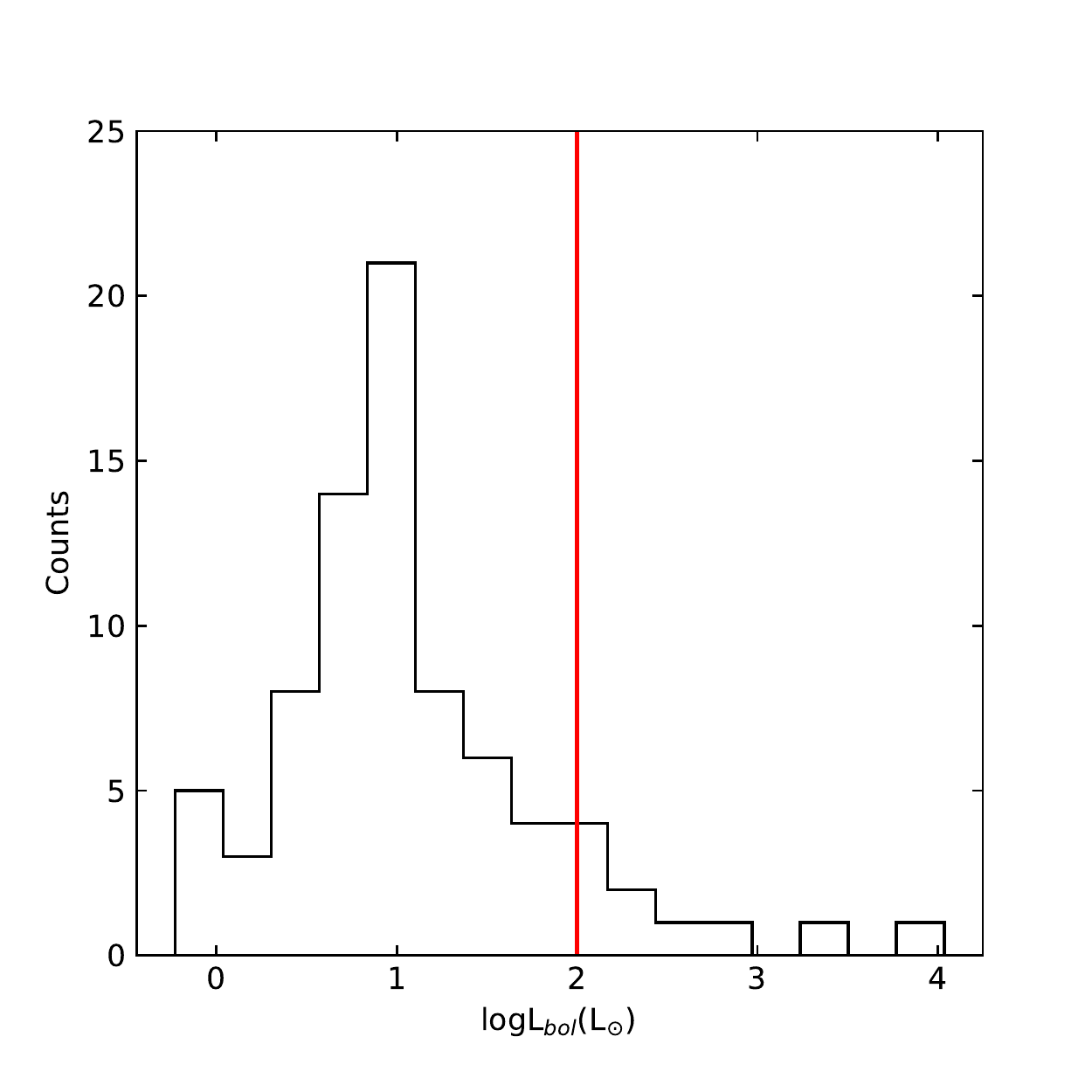}
	      \caption{Luminosity function of the YSOs sample (left panel) and the sources that located within the cloud (right panel). The red vertical line is the position $100 L_{\odot}$.}
	      \label{fig-Lbol}
	\end{figure}

      The star formation efficiency is defined as the ratio of the mass of the YSOs to the total mass of YSOs and gas: $\mathrm{SFE}=\frac{M_{\mathrm{YSO}}}{M_{\mathrm{YSO}}+M_{\mathrm{gas}}}$. 
      We use the trapezoid rule \citep{2008ApJS..179..249D,2015ApJS..220...11D} to integrate over the SEDs of the YSOs sample and the sources that located within the cloud, respectively.
      As shown in Figure\ref{fig-Lbol}, the peak position of histogram is $\sim 10 L_{\odot}$, indicating that the YSOs sample is complete down to $\sim 10 L_{\odot}$. 
      There are 9 out of 10 sources with luminosities above $100 L_{\odot}$ located within the $3\,\sigma$ boundary of \tw emission (see Figure\ref{fig-yso} and Table\ref{tb-massiveyso}), 
      three of which are distributed around \hii region G154.346+02.606, one in \hii region SH2-211, four in \hii region SH2-212, and one to the south of the filamentary molecular cloud. The contamination from foreground/background on the candidate massive YSOs is estimated to be $\sim 2.3\%$, which is neglected in our following calculation.  
      A total of $8$ sources overlap with the filamentary cloud.
      According to \cite{2012AJ....144...31K}, the luminosity function (LF) of high-mass star forming clouds (HLF) have a characteristic tail extending toward luminosities above $100 L_{\odot}$. Thus, the candidate YSOs with infrared luminosities above $100 L_{\odot}$ could be candidate massive YSOs.
      However, the number of candidate massive YSOs is incomplete due to the saturation problem of WISE survey. We refer to the massive young stellar catalog form the Red MSX Source Survey\citep{2013ApJS..208...11L} for supplement. 
      Only two \hii regions (G155.3319+02.5989 and G154.3472+02.6099) with bolometric luminosity up to $10^{4} L_{\odot}$ are found and they are coincides with the candidate massive YSOs in our sample (J044027.19+502828.8 and J043620.99+511254.1, see Table\ref{tb-massiveyso}).
      Based on the Kroupa IMF, the total number of YSO within the filamentary cloud is estimated to be $1255$ assuming that the $8$ sources would evolve into high-mass protostellars with mass greater than $8 M_{\odot}$.
      The total mass of YSOs is estimated to be $\sim 627\,M_{\odot}$ provided that the mean mass of a single candidate YSO is $0.5 \,M_{\odot}$. 
      Given that the mass of molecular cloud ($2.96 \times10^{4} M_{\odot}$, from \tht), the SFE will be $\sim 2.1\%$ in this case.

      The \tht emission area is $\sim 506$ arcmin$^{2}$ which is corresponded to $\sim 867 \mathrm{\,pc}^{2}$ at a distance of $4.5 \mathrm{\,kpc}$. Taking 2 Myr \citep{2009ApJS..181..321E}  and $0.54 \mathrm{\,Myr}$ \citep{2015ApJ...806..231H} as the Class $\mathrm{I}$ + $\mathrm{II}$ and Class $\mathrm{I}$ lifetimes,  
      the estimated star formation rate ($\mathrm{SFR}=\frac{M\rm_{YSOs}}{\tau}$, $\tau$ is the average lifetime of YSOs), 
      the gas mass surface density ($\Sigma_{\mathrm{^{13}CO}}$) 
      and the SFR surface density $\left(\Sigma_{\mathrm{SFR}}\right)$ 
      are $\sim 313 \,M_{\odot} \mathrm{Myr}^{-1}$, 
      $\sim 34 \,M_{\odot} \mathrm{pc}^{-2}$, 
      and  $\sim 0.36 \,M_{\odot} \mathrm{Myr}^{-1} \mathrm{pc}^{-2}$, respectively. These results are comparable with that of \cite{2019A&A...622A..52Z}, who investigated systematically the star-forming content of a sample of $57$ GMFs and suggested that the SFRs of the GMFs scale similarly with dense gas as those of nearby molecular clouds.
      However, limited by the resolution, the seemingly individual object could potentially  be an unresolved cluster at such a great distance. Considering the influence of YSO clustering, the estimated SFR 
      can only be regarded as a lower limit which 
      may be underestimated by a factor of $1.5-7$\citep{2019A&A...622A..52Z}.

\subsection{WISE Three-Color Diagram}\label{wise}     

\begin{figure}[htpb]
	      \centering
	      \includegraphics[width=0.98\textwidth]{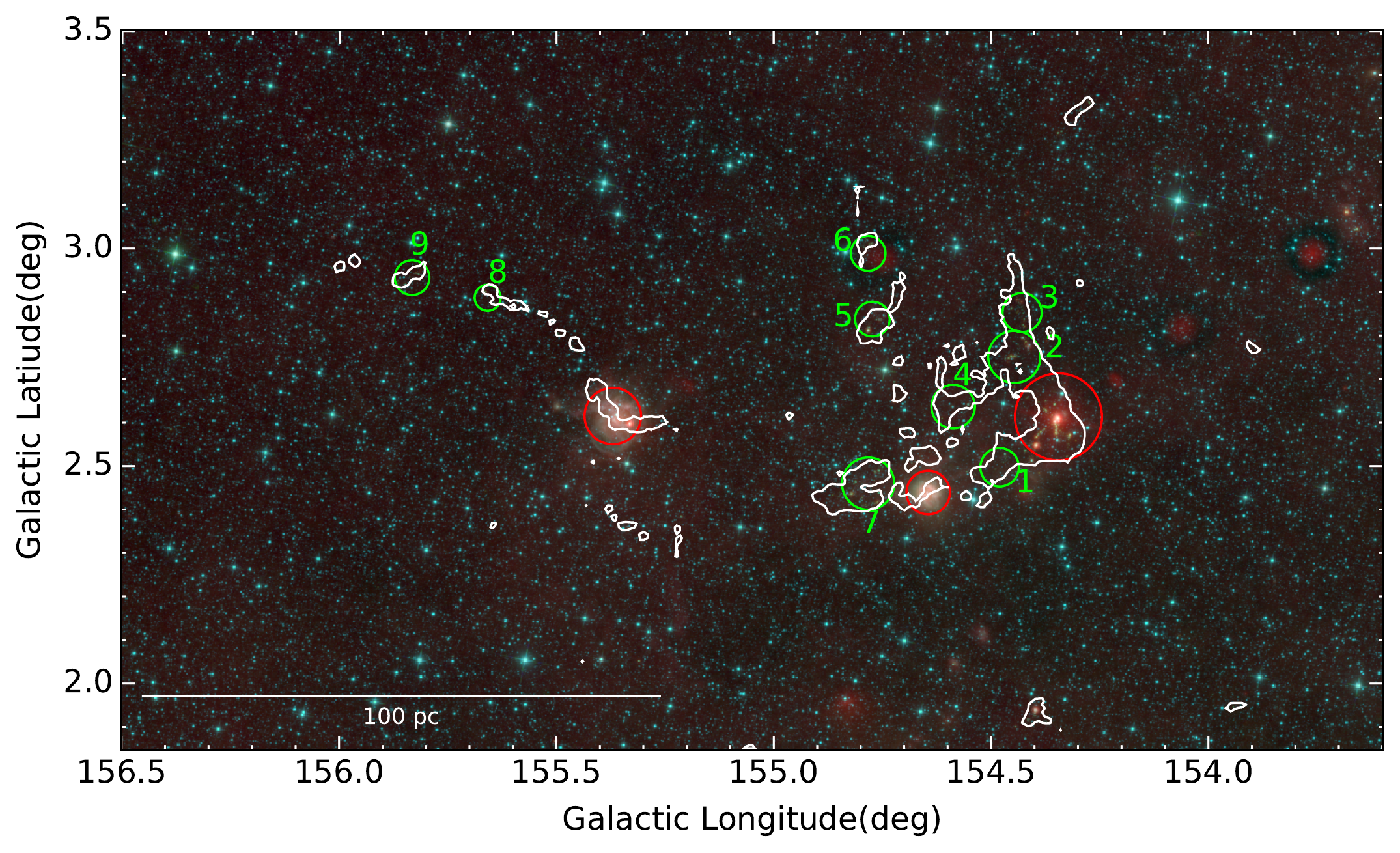}
	      \caption{The composite three-color image of three WISE bands ($22\,\mu$m in red, $4.6\,\mu$m in green, $3.4\,\mu$m in blue). The white contour represents  the $3\,\sigma$ boundary of the \tht emission. The red circles are the locations of the three \hii regions. The nine green circles indicate the regions where mass-flow activities are detected. 
	      }
	      \label{fig-wise}
	\end{figure}

      Figure \ref{fig-wise} shows the multi-color WISE image ($22 \,\mu \mathrm{m}$ in red, $4.6 \,\mu \mathrm{m}$ in green, and $3.4 \,\mu \mathrm{m}$ in blue) in the observed region. The $22 \,\mu \mathrm{m}$ emission is used to trace the small warm dust grains heated by ionized gases. The $4.6 \,\mu \mathrm{m}$ emission mostly originates from stars associated with the \hii regions and is considered as a promising diagnostic in the search for massive candidate YSOs \citep{2008AJ....136.2391C,2011ApJ...743...56C}. The $3.4 \,\mu \mathrm{m}$ band contains polycyclic aromatic hydrocarbon (PAH) emission at $3.3 \,\mu \mathrm{m}$ as well as a prominent molecular hydrogen feature at $3.234 \,\mu \mathrm{m}$. 
      As shown in Figure \ref{fig-wise}, the infrared emission in the observed region is dominated by the three \hii regions and we describe them below.

\subsubsection{\hii  Regions in the Filaments}\label{HII regions}     
\paragraph{G154.346+02.606}\label{G154.346+02.606}  

\begin{figure}[htpb]
	      \centering
	      \includegraphics[width=1\textwidth]{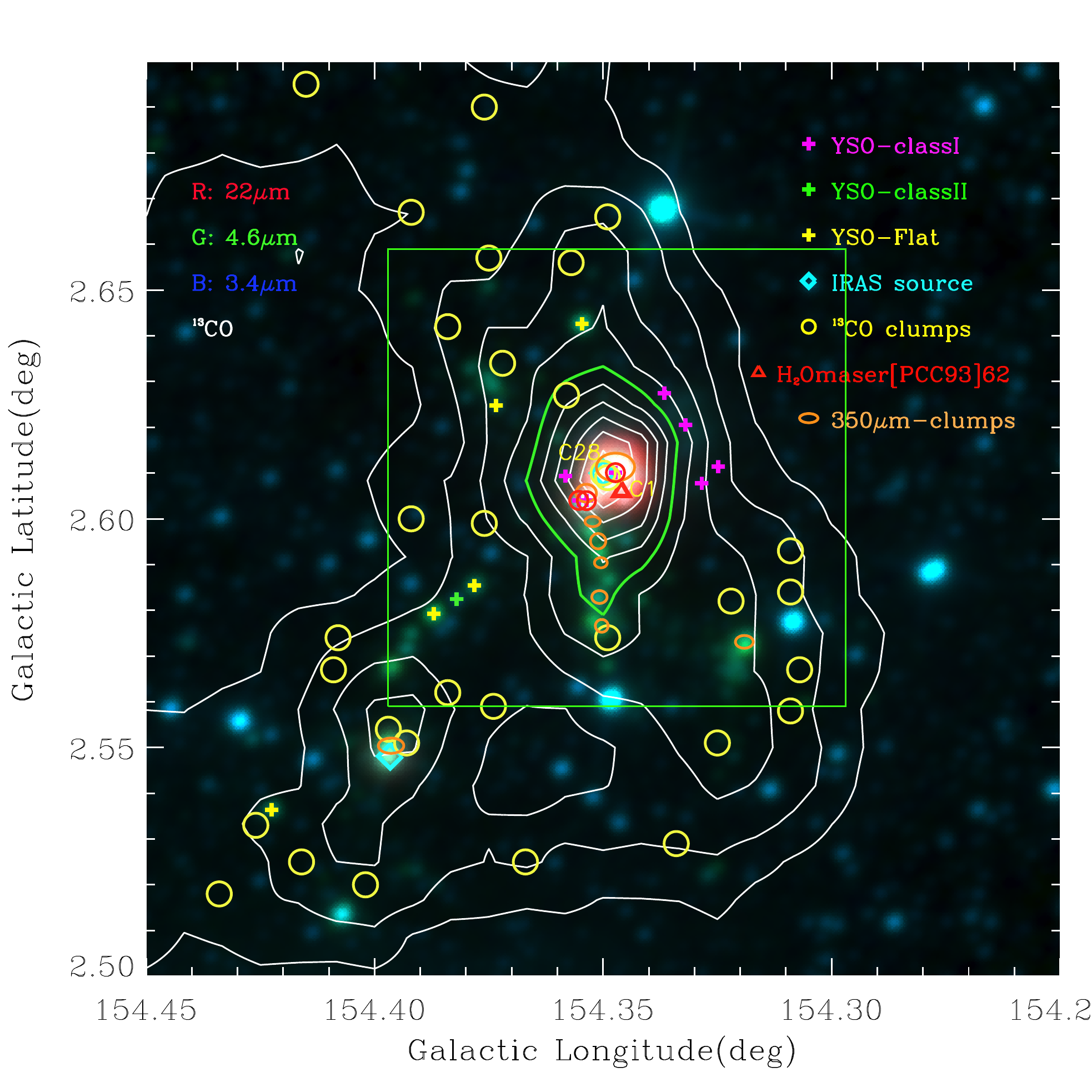}
	      \caption{The background is the composite three-color image of three WISE bands (22 $\mu$m in red, 4.6$\mu$m in green, 3.4$\mu$m in blue) 
	      toward the zoom-in region of G154.346+02.606.
	      The white contours show the \tht intensity integrated in the velocity interval of $[-39.5, -33.5] \mathrm{\,km}\mathrm{\,s}^{-1}$, 
	      which are set from $10\,\sigma$ to the peak value in this region by an $10 \%$ average step.
	      The green box is cantered at $(l,b)\sim$ ($154\fdg347, 2\fdg609$) with a size of $0\fdg1 \times 0\fdg1$. 
	      The green contour indicates the region with intensity greater than $12.65 \mathrm{\,K km} \mathrm{s}^{-1}$, or  roughly $N_{\mathrm{H}_{2}} \gtrsim 6.1 \times 10^{21} \mathrm{\,cm}^{-2}$.
	      The magenta, yellow, and green plus signs represent the candidate YSOs of Class $\mathrm{I}$, Class $\mathrm{II}$, and Flat type, respectively. 
	      The red circle with plus signs represent the candidate massive YSOs.
	      The yellow circles, cyan diamond, red triangle, and orange ellipses represent the \tht clumps,  IRAS sources, H$_{2}$O maser, and the 350 $\mu$m clumps found out by BGPS, respectively.
	      }
	      \label{fig-IR-region1}
	\end{figure}

	  As shown in Figure \ref{fig-IR-region1}, the infrared emission is enclosed by the \tht emission.
	  In addition, the velocity of hydrogen radio recombination lines (HRRLs) is consistent with that of the CO emission (see Table\ref{tb-IRAS}). Thus the \hii region G154.346+02.606 is confirmed to be associated with filament F1. 
	  IRAS04324+5106 was identified as high-mass protostellar object due to its low FIR luminosity to virial mass ratio  \citep{2018AJ....156..210A}, which is spatially coincident with the \tht peak emission. 
	  One H$_{2}$O maser towards IRAS04324+5106 was detected by \cite{1988A&A...191..323W} using the $100 \mathrm{\,m}$ Effelsberg telescope. The peak velocity of the maser is $-41 \mathrm{\,km}\mathrm{\,s}^{-1}$. \cite{2007PASJ...59.1185S} confirmed this maser in the $\mathrm{H}_{2} \mathrm{O}$ survey with the Nobeyama $45 \mathrm{\,m}$ telescope. 
	  The $\mathrm{H}_{2} \mathrm{O}$ maser is suggested to be one of the best tracers to the early phases of high-mass star formation \citep{1989ApJ...346..983E,2001ApJ...559L.143F} as it is often associated with outflow or accretion processes \citep{2007PASJ...59.1185S}.
	  A chain of $350 \,\mu \mathrm{m}$ clumps (orange ellipse) were identified by Bolocam Galactic Plane Survey \citep[BGPS,][]{2015ApJS..218....1M}.
	  
	  In our observations, about 10 \tht clumps are located around the \hii region. Among them, C1 has potential to form high-mass star (see above) and undergoes rapid collapse as it has the largest mass and the smallest virial parameter (0.39). 
	  In the sub-region (154\fdg30$\leqslant$ l $\leqslant$ 154\fdg40, 2\fdg55$\leqslant$ b $\leqslant$ 2\fdg65, $\sim 61 \mathrm{\,pc}^{2}$), 
	  the gas mass estimated from \tht is $\sim 7803 \,M_{\odot}$. 
	  Three Class $\mathrm{I}$ sources are found with luminosity greater than 100 $L_{\odot}$. 
	  Assuming that the three Class $\mathrm{I}$ sources would evolve into high-mass protostellars ($>8 M_{\odot}$), the total number of YSOs is estimated to be $\sim 470$ based on the Kroupa IMF.
	  Thus, the SFE, SFR, $\Sigma_{\mathrm{^{13}CO}}$, and $\Sigma_{\mathrm{SFR}}$ are estimated to be about $2.9\%$, 117 $M_{\odot}$My$^{-1}$, 124 $M_{\odot}\mathrm{pc}^{-2}$, and 1.88 $M_{\odot}$My$^{-1}\mathrm{pc}^{-2}$, respectively. 
      Within the green contour ($\sim 8.1\,\mathrm{pc}^{-2}$, the \tht integrated intensity greater $\gtrsim  12.65 \mathrm{\,K} \mathrm{\,kms}^{-1}$
      , equivalent to $N_{\mathrm{H}_{2}} \gtrsim 6.1 \times 10^{21} \mathrm{\,cm}^{-2}$
      ), the gas mass is estimated to be $\sim 2832 \,M_{\odot}$.      
      The SFE, SFR, $\Sigma_{\mathrm{^{13}CO}}$, and $\Sigma_{\mathrm{SFR}}$ in this area is estimated to be $\sim 7.6 \%$,  $\sim 117 \,M_{\odot} \mathrm{My}^{-1}$,  $\sim 388 \,M_{\odot} \mathrm{pc}^{-2}$,  $\sim 16 \,M_{\odot} \mathrm{My}^{-1} \mathrm{pc}^{-2}$. According to the starburst certification standard of $\Sigma_{\mathrm{SFR}}>1 M_{\odot} \mathrm{yr}^{-1} \mathrm{kpc}^{-2}$ \citep{2016ApJ...833...23N}, G154.346+02.606 would be a promising candidate for mini-starburst.

\paragraph{SH2-211}\label{SH2-211}  

\begin{figure}[htpb]
	      \centering
	      \includegraphics[width=1\textwidth]{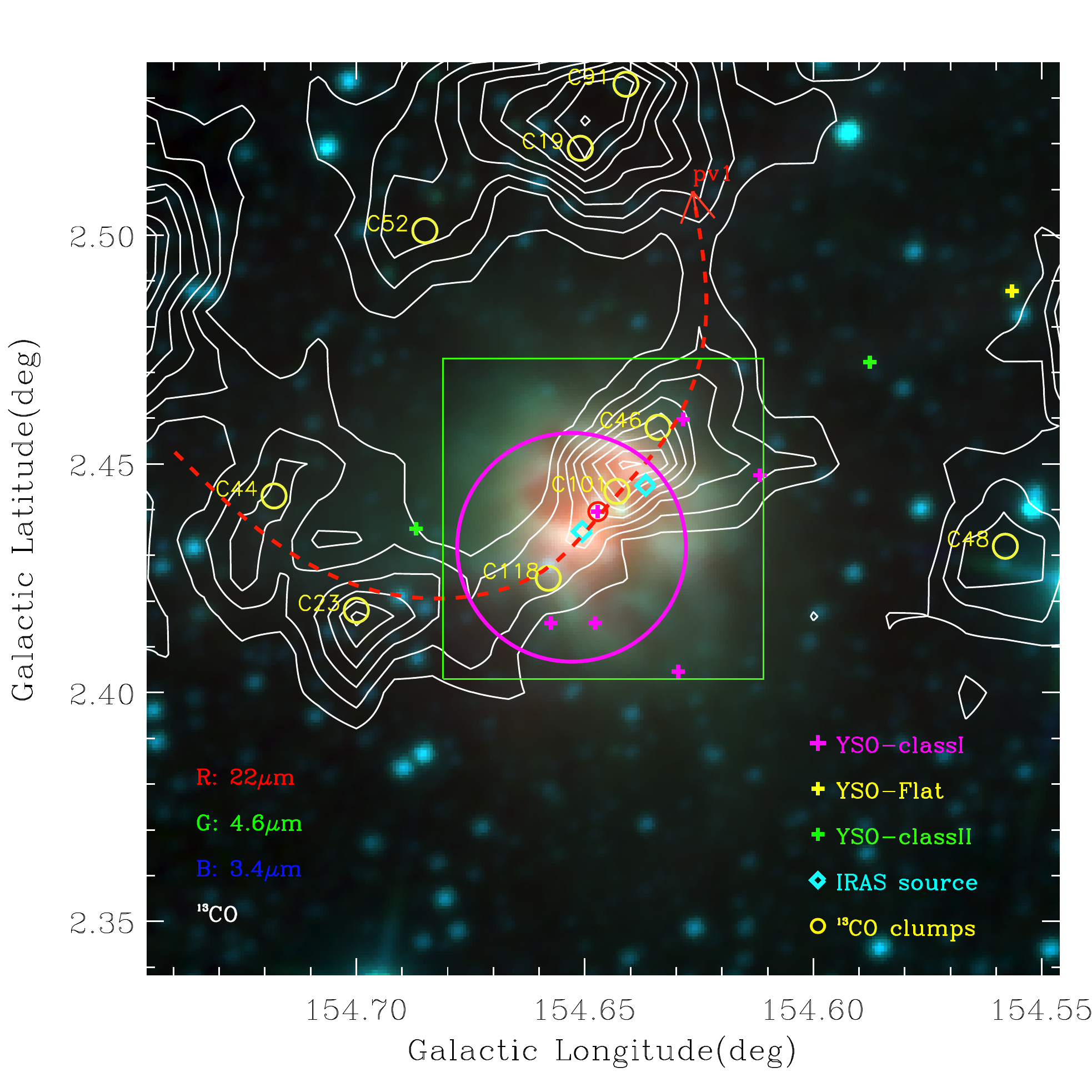}
	      \caption{Same as Figure \ref{fig-IR-region1} but for the region of SH2-211. The white contours show the \tht intensity integrated in the velocity interval of $[-39.5, -33.5] \mathrm{\,km}\mathrm{\,s}^{-1}$, 
	      which are set from $5\,\sigma$ to the peak value in this region by an $10 \%$ average step. 
	      The green box is cantered at$(l, b) \sim$ ($154\fdg646, 2\fdg438$) with a size of $0\fdg07 \times 0\fdg07$.
	      The red dotted arrow line indicate the PV slice path (width $\sim$ 2 pixels). The IR bubble is indicated by a magenta circle with a center of (l, b) $\sim$ (154.653$\degr$, 2.432$\degr$) and a radius of 0.025$\degr$. 
	      }
	      \label{fig-IR-region2}
	\end{figure}

	 As shown in Figure \ref{fig-IR-region2}, in the zoom-in region, the \tht morphology displays a ringlike structure and is spatially consistent with the infrared bubble in the northwestward. 
	 The rough consistency of the $v_{\rm LSR}$s ranges between HRRLs and CO suggests that the CO emission is associated with SH2-211 (see Table\ref{tb-IRAS}). 
	 \cite{1986BAAS...18Q.921M} revealed three gas components of HII, HI and $\mathrm{CO}$ associated with SH2-211 and SH2-212, and suggested that SH2-211 may be located in a more complex molecular environment according to the estimated parameters, e.g, sizes, masses. 
	 Two IRAS sources \citep[IRAS04329+5047 and IRAS04329+5045, ][]{2005ApJS..161..361K} are located on the opposite sides of the IR bubble separately. 
 	 Peak emission of \tht appears around IRAS04329+5047. 
 	 Three early-type stars (BOV, V=15.78; O9V, V=15.23; and O9Ib, V=13.54) are suggested as the ionization sources of SH2-211 \citep{1984A&A...139L...5C}. 
	 Three clumps (C118, C101, C46) are located around the bubble.
	 The velocities of C118 ($-39.2 \mathrm{\,km}\mathrm{\,s}^{-1}$) and C46 ($-36.7 \mathrm{\,km}\mathrm{\,s}^{-1}$) are consistent with F4 and F3 respectively (see Figure \ref{fig-fil-spectrum}), while C101 has the moderate velocity among the three.
	 The gas mass calculated from \tht within the green box ($0\fdg07 \times 0\fdg07$) is $\sim 275 \,M_{\odot}$. The \tht emission area is $\sim 14 \mathrm{\,pc}^{2}$. 
 	 Compared with G154.346+02.606 and SH2-212, the mass-flow rate along the filament around SH2-211 is the lowest. 
	 There are 1 candidate massive YSOs (J043649.54+505242.8) at the center of the IR bubble.
	 Based on the Kroupa IMF, the SFE, SFR, $\Sigma_{\mathrm{^{13}CO}}$, and $\Sigma_{\mathrm{SFR}}$ within the green box are estimated to be about $22\%$,  $39 M_{\odot}$My$^{-1}$,  $19 M_{\odot}\mathrm{pc}^{-2}$, and $2.8 M_{\odot}$My$^{-1}\mathrm{pc}^{-2}$, respectively. The \hii region SH2-211 
	 may host ongoing starburst event\citep{2016ApJ...833...23N}.
	 As mentioned before, F3 and F4 has a distinct $v_{\rm LSR}$ (F3 at $-36 \mathrm{\,km}\mathrm{\,s}^{-1}$, F4 at $-38 \mathrm{\,km}\mathrm{\,s}^{-1}$).  
	 Based on the velocity continuity of \tw, the northwest part is likely to be associated with F3 and the southeast part is likely to be associated with F4. The bubble appears in the intersection of F3 and F4. It is likely that the collision of the two filaments accounts for the star formation and then the \hii region.

\begin{figure}[htpb]
	      \centering
              \includegraphics[width=0.5\textwidth]{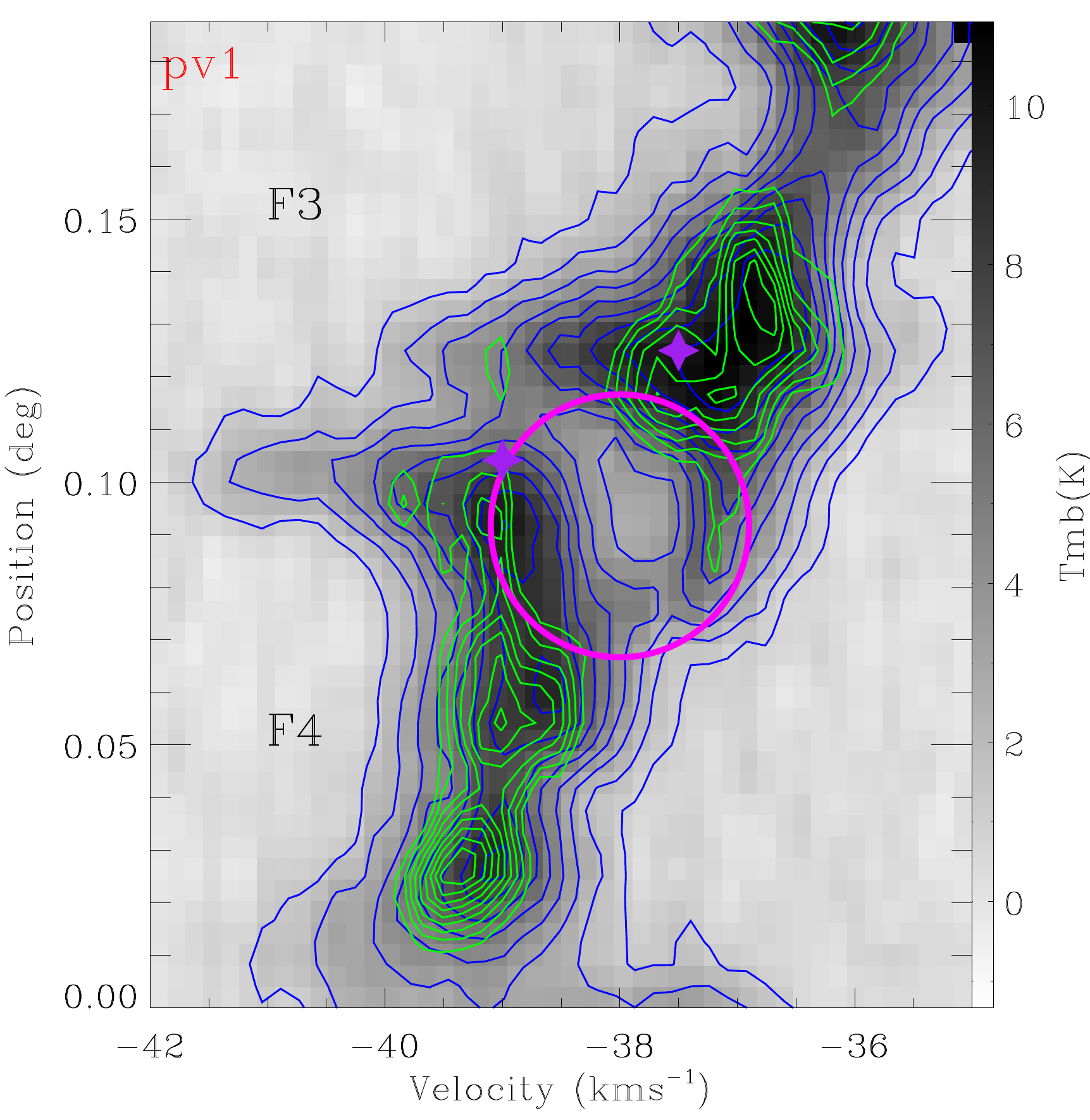}
	      \caption{The PV diagram of \tw (grayscale and blue contours) and \tht (green contours) along the PV slice path in Figure \ref{fig-IR-region2}. 
	      The colorbar is in unit of K. 
	      The contours are set from 3 $\sigma$ to the peak emission by an average step of $10\%$. 
	      The magenta ellipse is centered at -38 km s$^{-1}$ with velocity radius of $1.5 \mathrm{\,km}\mathrm{\,s}^{-1}$ and position radius of $0\fdg025$, which is corresponding to the magenta circle in Figure \ref{fig-IR-region2}.
	      }
	      \label{fig-IR-region2_pv}
	\end{figure}

	The heated bubble has a relatively circular morphology. 
	As shown in Figure \ref{fig-IR-region2_pv}, the CO PV diagram along the bubble also shows a circle morphology, indicating the expansion of the bubble. Assuming the bubble is produced by stellar wind, 
	the mechanical luminosity of the stellar wind can be estimated by 
	$L_{\text {wind }} \approx \frac{1}{3}\left(\frac{n_{\mathrm{gas}}}{\mathrm{cm}^{-3}}\right)\left(\frac{R_{\mathrm{c}}}{\mathrm{pc}}\right)^{2}\left(\frac{V_{\mathrm{c}}}{\mathrm{km} \mathrm{s}^{-1}}\right)^{3} \times 10^{30} \text { erg } \mathrm{s}^{-1}$ \citep{1977ApJ...218..377W}, where $R_{\mathrm{c}}$ is the radius of the shell, $V_{\mathrm{c}}$ is the shell expansion velocity, $n_{\mathrm{gas}}$  is the density of the cavity within the MC. 
	The $R_{\mathrm{c}}$ ($\sim 2.1 \mathrm{pc}$) and $V_{\mathrm{c}}$ ($\sim 1.5 \,\mathrm{km\,s}^{-1}$) of the shell are
	obtained from Figure \ref{fig-IR-region2_pv}. 
	The $n_{\mathrm{gas}}$ ($n_{\mathrm{gas}}=\frac{3 N_{\text {shell }}}{R_{\mathrm{c}}}$) estimated from the \tw emission within the shell
	is $\sim 1.7\times 10^{3} \mathrm{\,cm}^{-3}$. Then the $L_{\text {wind }}$ is estimated to be $\sim 8.4 \times 10^{33} \mathrm{\,erg} \mathrm{\,s}^{-1}$.
	The kinetic time-scale ($t_{\text {kin}} \approx \frac{16}{27}\frac{R_{\rm c}}{\rm pc}\frac{\rm kms^{-1}}{v_{\rm c}}$) 
	for opening such a shell is estimated to be about $\sim 8.2 \times 10^{5} \mathrm{\,yr}$.

\paragraph{SH2-212}\label{SH2-212}  	

\begin{figure}[htpb]
	      \centering
              \includegraphics[width=1\textwidth]{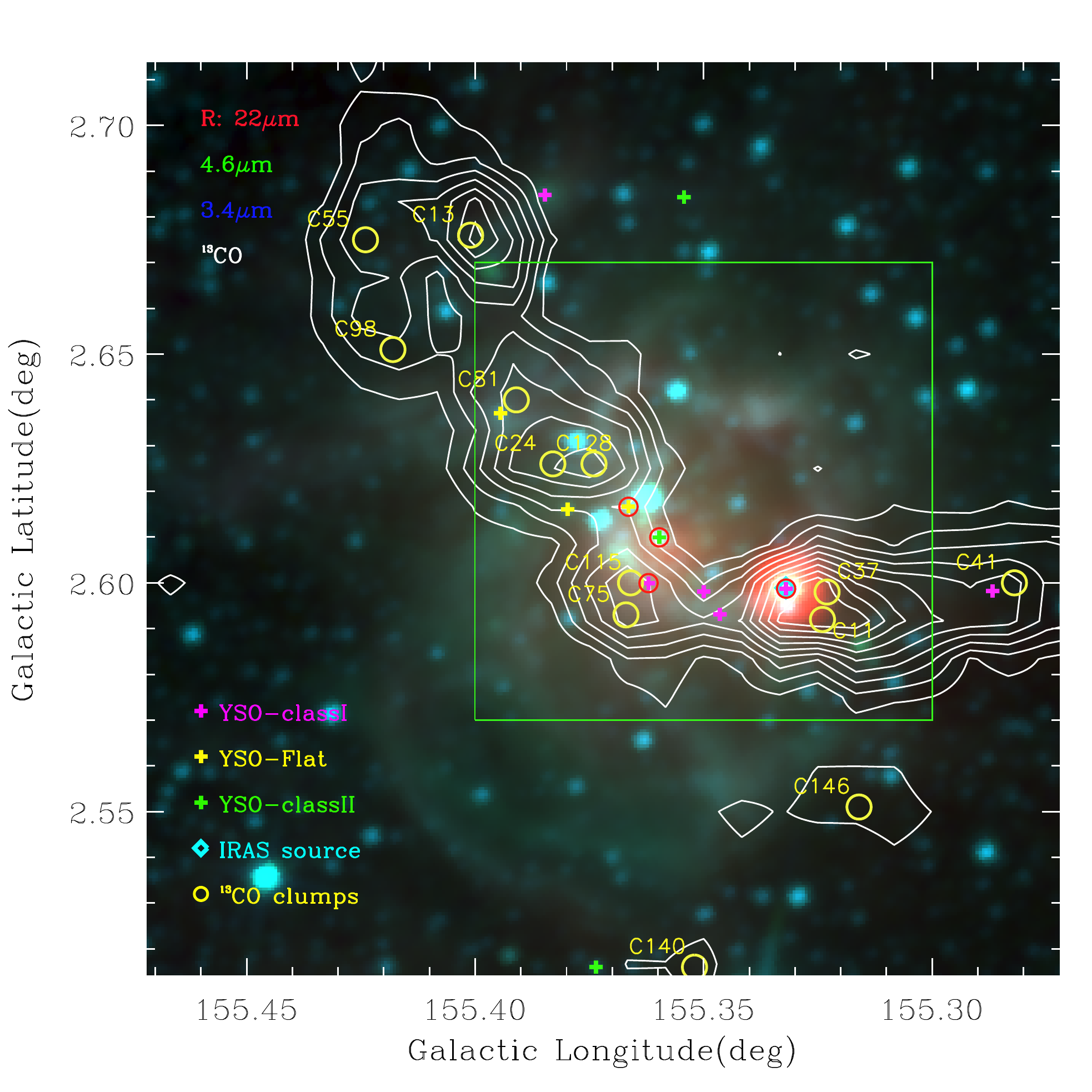}
	      \caption{Same as Figure \ref{fig-IR-region1} but for the region of SH2-212. The white contours show the \tht intensity integrated in the velocity interval of $[-39.5, -33.5] \mathrm{\,km}\mathrm{\,s}^{-1}$, 
	      which are set from $10\,\sigma$ to the peak value in this region by an $10 \%$ average step. 
	      The green box is cantered at$(l, b) \sim$ ($155\fdg350, 2\fdg620$) with a size of $0\fdg1 \times 0\fdg1$.
	      }
	      \label{fig-IR-region3}
	\end{figure}

	As shown in Figure \ref{fig-IR-region3}, the \tht morphology displays an arc-like structure encircling the infrared bubble. 
	The velocities of the HRRLs (see Table\ref{tb-IRAS})
	differ from the CO molecular gas with an offset of $\sim 5  \mathrm{\,km}\mathrm{\,s}^{-1}$. 
	\cite{2008A&A...482..585D} attributed this velocity difference to a scenario of ``champagne flow'' \citep{1979A&A....71...59T}, the ionized gas flows away from the cloud. About six \tht clumps are semicircling the ionized gas.
	C37 and C128 have a velocity at $-36 \mathrm{\,km}\mathrm{\,s}^{-1}$, which is roughly same as that of F5 (see, Figure \ref{fig-fil-spectrum}).
	C11, C24, and  C75 have velocities at $-35 \mathrm{\,km}\mathrm{\,s}^{-1}$, differ from F5 with $\sim 1 \mathrm{\,km}\mathrm{\,s}^{-1}$. C115 is the closest clumps to the infrared bubble and has velocity at $ \sim -34 \mathrm{\,km}\mathrm{\,s}^{-1}$, differ from F5 with $\sim 2 \mathrm{\,km}\mathrm{\,s}^{-1}$. 
	\cite{2008A&A...482..585D} suggested the origin of the fragmented molecular gas around the infrared bubble through the collect and collapse process.
	Near C11 and C37, where the infrared cluster IRAS 04366+5022 is located, a possible second-generation UC \hii region is forming \citep{2008A&A...482..585D}.
	The velocity difference between C11 and C37 may be caused by the formation and ionization of the secondary UC \hii region. 
	The mass of the exciting star 
	\citep[No. 228 in][]{2008A&A...482..585D} of the second-generation UC \hii region estimated from SED between $1.25 \,\mu \mathrm{m}$ and $21.3 \,\mu \mathrm{m}$ is $\sim 14 \,M_{\odot}$, and the spectral class is $\sim$ B0.5V \citep{2011MNRAS.411.2530J}.
	The gas mass calculated from \tht within the green box ($0\fdg1 \times 0\fdg1$) is $\sim 905 \,M_{\odot}$. 
	The \tht emission area is $\sim 24 \mathrm{\,pc}^{2}$. 
	Four candidate massive YSOs are found at the boundary of IR bubble, and the locations of two candidates (J044027.19+502828.8 and J044037.27+502740.7) are consistent with the region of 22 $\mu \mathrm{m}$ emission. 
	Based on the Kroupa IMF, the lower limits of SFE, SFR, $\Sigma_{\mathrm{^{13}CO}}$, and $\Sigma_{\mathrm{SFR}}$ within the green box are estimated to be about $15\%$,  $78 M_{\odot}$My$^{-1}$,  $36 M_{\odot}\mathrm{pc}^{-2}$, and $3.1 M_{\odot}$My$^{-1}\mathrm{pc}^{-2}$, respectively. Therefor, the \hii region SH2-212
	 may also host ongoing starburst event.

\subsubsection{Other Regions}\label{HII}   	
	
\begin{figure}[htpb]
	      \centering
	      \includegraphics[width=0.98\textwidth]{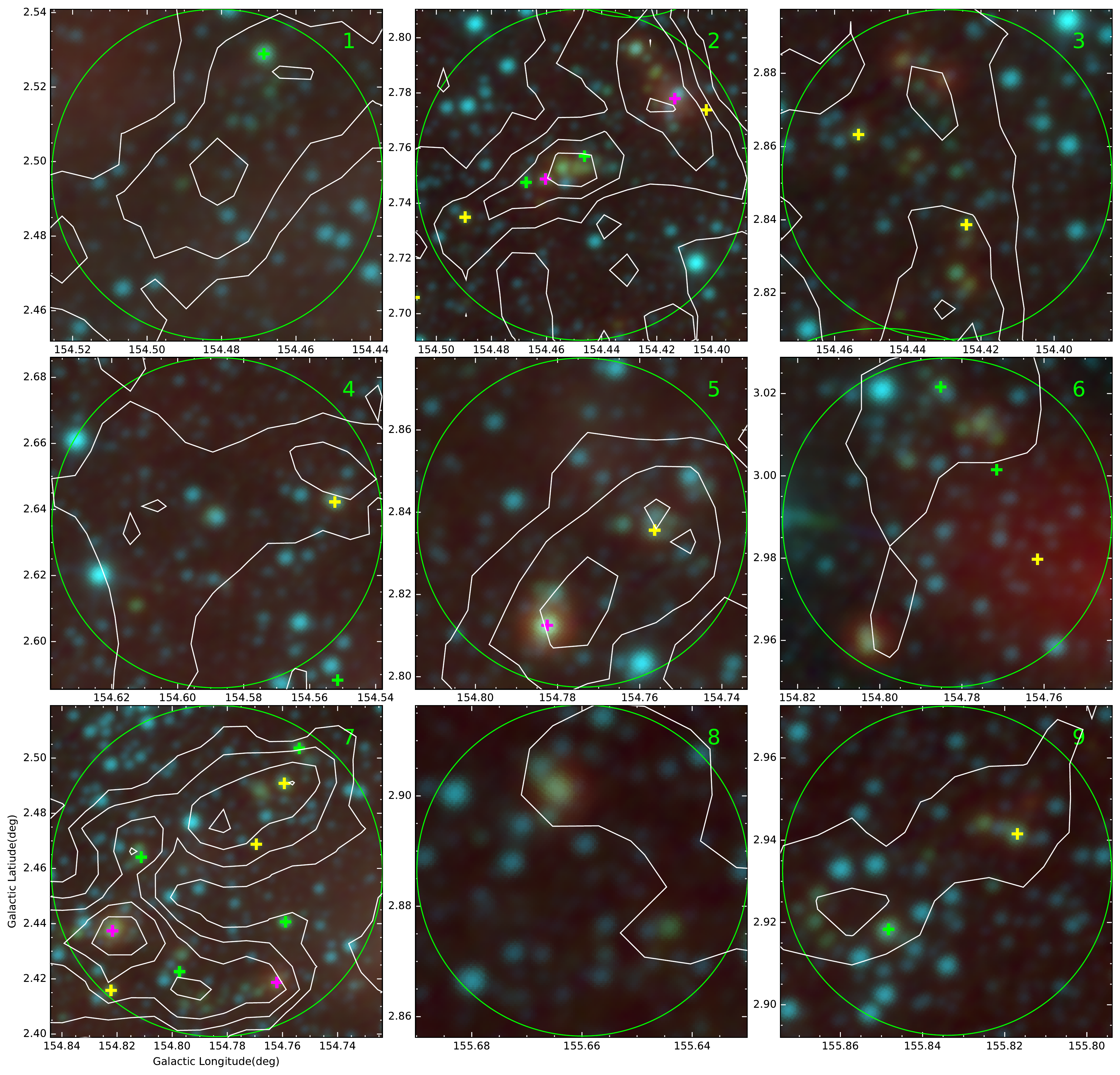}
	      \caption{The zoom-in of the nine green circles indicated in Figure \ref{fig-wise}. The red, yellow, and green crosses represent the candidate YSOs of Class $\mathrm{I}$, Class $\mathrm{II}$, and Flat type, respectively. 
	      The red circle with plus signs represent the candidate massive YSOs.
	      The white contours are the \tht intensity integrated in the velocity interval of $[-39.5, -33.5] \mathrm{\,km}\mathrm{\,s}^{-1}$, 
	      which are set from $3\,\sigma$ (1 $\sigma \sim 0.28\mathrm{\,K} \mathrm{\,km}\mathrm{\,s}^{-1}$) to  peak value ($27 \mathrm{\,K} \mathrm{\,km}\mathrm{\,s}^{-1}$) in an $6 \%$ average step.
	      }
	      \label{fig-wise-LMC}
	\end{figure}

	As shown in Figure \ref{fig-wise-LMC}, based on the analysis of CO molecular dynamics structure (see Figure \ref{fig-pvfit}), we zoom in to nine regions where mass-flow activities exist, including cross-A, cross-B, and cross-C (corresponding to the zoom-in areas 1, 2, and 4). The WISE point sources appear in clusters in these regions and the emission at $4.6 \,\mu \mathrm{m}$ and $22 \,\mu \mathrm{m}$ bands is obvious. 
	The $4.6 \mu \mathrm{m}$  emission is consistent with the strong \tht emission, which implies that YSOs are embedded in the CO clouds.
	The estimated accretion rates around these regions are $\sim 100 \,M_{\odot} \mathrm{Myr}^{-1}$, several times smaller than those around the \hii regions. 
	Among the nine regions, the area 2 (cross-B) is probably the most active one with star formation as it has the largest accretion rate. 
	Actually, when examining the entire WISE image, we consider that the entire MC is undergoing active star formation.

\section{Conclusion}	
	 We present a larg-field survey of the $J=1-0$ transition lines of $^{12}$CO, $^{13}$CO, and C$^{18}$O towards the Galactic region of l=[153\fdg6, 156\fdg5] and b=[1\fdg85, 3\fdg5], using the PMO 13.7 m millimeter telescope. The main results are summarized as follows:
	
	 1,  A network-shaped cloud is found in the velocity interval of $[-42.5,-30.0] \mathrm{\,km}\mathrm{\,s}^{-1}$. The distance is estimated to be $\sim$ 4.5 kpc. The basic physical parameters including excitation temperatures, H$_2$ column densities, and masses are estimated from ${ }^{12} \mathrm{CO}$ and ${ }^{13} \mathrm{CO}$, respectively. 
	
	 2.  Six large-scale filaments are identified from the \tht emission. By comparing the observed average $M_{\text {line}}$ and the $M_{\text {line, crit}}$, filaments F1-F5 are suggested to be under the gravitationally unstable conditions and will fragment into dense cores.

 	 3. The PV diagrams of the filaments (e.g. F1, F5) show oscillation patterns, which is explained by the mass-flow caused by accretion activity. 
 	 The material seems to transport along the filaments to feed the \hii regions G154.346+2.606, SH2-212 and cross-B, with high mass-flow rates larger than $100 \,M_{\odot} \mathrm{Myr}^{-1}$. The mass-flow rates in the rest parts of filaments are relatively lower, at a level of a few tens of $M_{\odot} \mathrm{Myr}^{-1}$.

 	 4. A total of $162$ CO clumps are extracted from the \tht datacube, of which $32 \%$ have $\alpha_\mathrm{vir}$ less than 1 and $90 \%$ have $\alpha_\mathrm{vir}$ less than 2. Rapid collapse seems to take place in the apexs of the filaments as most clumps with $\alpha_\mathrm{vir}$ less than 1  are located therein. Global contraction along filaments is efficient. We find $163$ candidate YSOs within the cloud traced by \tw. The SFE along the filaments is $\sim 2.1 \%$, which is comparable with that of other GMFs.

  	5. Three \hii regions, including G154.346+02.606, SH2-211 and SH2-212, are associated with the filaments.  According to the estimated SFR densities ($> 1 M_{\odot}$My$^{-1}\mathrm{pc}^{-2}$), 
  	  the three \hii regions are considered to host ongoing mini starburst events. 
  	  In addition, other 9 regions with material flow around show signature of star forming activity in the WISE three-color diagram. The results from both the CO and infrared emission suggest that the entire filamentary cloud is undergoing intense star formation.

\begin{acknowledgments}
This work was supported by the National key R\&D Program of China (grant No.2017YFA0402702), the National Natural Science Foundation of China (grant No. 12041305), and CAS International Cooperation Program (grant No. 114332KYSB20190009). M.Zhang was supported by the National Natural Science Foundation of China (grants No. 12073079).
This research made use of the data from the Milky Way Imaging Scroll Painting (MWISP) project, which is a multi-line survey in ${ }^{12} \mathrm{CO}$/${ }^{13} \mathrm{CO}$/$\mathrm{C}{ }^{12} \mathrm{O}$ along the northern galactic plane with PMO-13.7m telescope. We are grateful to all the members of the MWISP working group, particularly the staff members at PMO-13.7m telescope, for their long-term support. MWISP was sponsored by National Key R\&D Program of China with grant 2017YFA0402701 and  CAS Key Research Program of Frontier Sciences with grant QYZDJ-SSW-SLH047. 
\end{acknowledgments}

\vspace{5mm}
\facilities{PMO-13.7m}

\software{GILDAS/CLASS \citep{2005sf2a.conf..721P}}

\bibliography{export-bibtex}{}

\begin{thebibliography}{}
\expandafter\ifx\csname natexlab\endcsname\relax\def\natexlab#1{#1}\fi
\providecommand{\url}[1]{\href{#1}{#1}}
\providecommand{\dodoi}[1]{doi:~\href{http://doi.org/#1}{\nolinkurl{#1}}}
\providecommand{\doeprint}[1]{\href{http://ascl.net/#1}{\nolinkurl{http://ascl.net/#1}}}
\providecommand{\doarXiv}[1]{\href{https://arxiv.org/abs/#1}{\nolinkurl{https://arxiv.org/abs/#1}}}

\bibitem[{{Anderson} {et~al.}(2015){Anderson}, {Armentrout}, {Johnstone},
  {Bania}, {Balser}, {Wenger}, \& {Cunningham}}]{2015ApJS..221...26A}
{Anderson}, L.~D., {Armentrout}, W.~P., {Johnstone}, B.~M., {et~al.} 2015,
  \apjs, 221, 26, \dodoi{10.1088/0067-0049/221/2/26}

\bibitem[{{Andr{\'e}} {et~al.}(2014){Andr{\'e}}, {Di Francesco},
  {Ward-Thompson}, {Inutsuka}, {Pudritz}, \& {Pineda}}]{2014prpl.conf...27A}
{Andr{\'e}}, P., {Di Francesco}, J., {Ward-Thompson}, D., {et~al.} 2014, in
  Protostars and Planets VI, ed. H.~{Beuther}, R.~S. {Klessen}, C.~P.
  {Dullemond}, \& T.~{Henning}, 27,
  \dodoi{10.2458/azu\_uapress\_9780816531240-ch002}

\bibitem[{{Andr{\'e}} {et~al.}(2013){Andr{\'e}}, {K{\"o}nyves}, {Arzoumanian},
  {Palmeirim}, \& {Peretto}}]{2013ASPC..476...95A}
{Andr{\'e}}, P., {K{\"o}nyves}, V., {Arzoumanian}, D., {Palmeirim}, P., \&
  {Peretto}, N. 2013, in Astronomical Society of the Pacific Conference Series,
  Vol. 476, New Trends in Radio Astronomy in the ALMA Era: The 30th Anniversary
  of Nobeyama Radio Observatory, ed. R.~{Kawabe}, N.~{Kuno}, \& S.~{Yamamoto},
  95

\bibitem[{{Andr{\'e}} {et~al.}(2010){Andr{\'e}}, {Men'shchikov}, {Bontemps},
  {K{\"o}nyves}, {Motte}, {Schneider}, {Didelon}, {Minier}, {Saraceno},
  {Ward-Thompson}, {di Francesco}, {White}, {Molinari}, {Testi}, {Abergel},
  {Griffin}, {Henning}, {Royer}, {Mer{\'\i}n}, {Vavrek}, {Attard},
  {Arzoumanian}, {Wilson}, {Ade}, {Aussel}, {Baluteau}, {Benedettini},
  {Bernard}, {Blommaert}, {Cambr{\'e}sy}, {Cox}, {di Giorgio}, {Hargrave},
  {Hennemann}, {Huang}, {Kirk}, {Krause}, {Launhardt}, {Leeks}, {Le Pennec},
  {Li}, {Martin}, {Maury}, {Olofsson}, {Omont}, {Peretto}, {Pezzuto}, {Prusti},
  {Roussel}, {Russeil}, {Sauvage}, {Sibthorpe}, {Sicilia-Aguilar}, {Spinoglio},
  {Waelkens}, {Woodcraft}, \& {Zavagno}}]{2010A&A...518L.102A}
{Andr{\'e}}, P., {Men'shchikov}, A., {Bontemps}, S., {et~al.} 2010, \aap, 518,
  L102, \dodoi{10.1051/0004-6361/201014666}

\bibitem[{{Ao} {et~al.}(2018){Ao}, {Yang}, {Tatematsu}, {Henkel}, {Sunada}, \&
  {Nguyen-Luong}}]{2018AJ....156..210A}
{Ao}, Y., {Yang}, J., {Tatematsu}, K., {et~al.} 2018, \aj, 156, 210,
  \dodoi{10.3847/1538-3881/aae259}

\bibitem[{{Arzoumanian} {et~al.}(2011){Arzoumanian}, {Andr{\'e}}, {Didelon},
  {K{\"o}nyves}, {Schneider}, {Men'shchikov}, {Sousbie}, {Zavagno}, {Bontemps},
  {di Francesco}, {Griffin}, {Hennemann}, {Hill}, {Kirk}, {Martin}, {Minier},
  {Molinari}, {Motte}, {Peretto}, {Pezzuto}, {Spinoglio}, {Ward-Thompson},
  {White}, \& {Wilson}}]{2011A&A...529L...6A}
{Arzoumanian}, D., {Andr{\'e}}, P., {Didelon}, P., {et~al.} 2011, \aap, 529,
  L6, \dodoi{10.1051/0004-6361/201116596}

\bibitem[{{Ballesteros-Paredes} {et~al.}(2020){Ballesteros-Paredes},
  {Andr{\'e}}, {Hennebelle}, {Klessen}, {Kruijssen}, {Chevance}, {Nakamura},
  {Adamo}, \& {V{\'a}zquez-Semadeni}}]{2020SSRv..216...76B}
{Ballesteros-Paredes}, J., {Andr{\'e}}, P., {Hennebelle}, P., {et~al.} 2020,
  \ssr, 216, 76, \dodoi{10.1007/s11214-020-00698-3}

\bibitem[{{Bally}(1981)}]{1981PhDT.........3B}
{Bally}, J. 1981, PhD thesis, Massachusetts Univ., Amherst.

\bibitem[{{Balser} {et~al.}(2011){Balser}, {Rood}, {Bania}, \&
  {Anderson}}]{2011ApJ...738...27B}
{Balser}, D.~S., {Rood}, R.~T., {Bania}, T.~M., \& {Anderson}, L.~D. 2011,
  \apj, 738, 27, \dodoi{10.1088/0004-637X/738/1/27}

\bibitem[{{Cantat-Gaudin} \& {Anders}(2020)}]{2020A&A...633A..99C}
{Cantat-Gaudin}, T., \& {Anders}, F. 2020, \aap, 633, A99,
  \dodoi{10.1051/0004-6361/201936691}

\bibitem[{{Carpenter} {et~al.}(1993){Carpenter}, {Snell}, {Schloerb}, \&
  {Skrutskie}}]{1993ApJ...407..657C}
{Carpenter}, J.~M., {Snell}, R.~L., {Schloerb}, F.~P., \& {Skrutskie}, M.~F.
  1993, \apj, 407, 657, \dodoi{10.1086/172548}

\bibitem[{{Castets} \& {Langer}(1995)}]{1995A&A...294..835C}
{Castets}, A., \& {Langer}, W.~D. 1995, \aap, 294, 835

\bibitem[{{Chandrasekhar} \& {Fermi}(1953)}]{1953ApJ...118..116C}
{Chandrasekhar}, S., \& {Fermi}, E. 1953, \apj, 118, 116,
  \dodoi{10.1086/145732}

\bibitem[{{Chen} {et~al.}(2019){Chen}, {Zhang}, {Wright}, {Busquet}, {Lin},
  {Liu}, {Olguin}, {Sanhueza}, {Nakamura}, {Palau}, {Ohashi}, {Tatematsu}, \&
  {Liao}}]{2019ApJ...875...24C}
{Chen}, H.-R.~V., {Zhang}, Q., {Wright}, M.~C.~H., {et~al.} 2019, \apj, 875,
  24, \dodoi{10.3847/1538-4357/ab0f3e}

\bibitem[{{Chini} \& {Wink}(1984)}]{1984A&A...139L...5C}
{Chini}, R., \& {Wink}, J.~E. 1984, \aap, 139, L5

\bibitem[{{Colombo} {et~al.}(2021){Colombo}, {K{\"o}nig}, {Urquhart},
  {Wyrowski}, {Mattern}, {Menten}, {Lee}, {Brand}, {Wienen}, {Mazumdar},
  {Schuller}, \& {Leurini}}]{2021A&A...655L...2C}
{Colombo}, D., {K{\"o}nig}, C., {Urquhart}, J.~S., {et~al.} 2021, \aap, 655,
  L2, \dodoi{10.1051/0004-6361/202142182}

\bibitem[{{Cyganowski} {et~al.}(2011){Cyganowski}, {Brogan}, {Hunter}, \&
  {Churchwell}}]{2011ApJ...743...56C}
{Cyganowski}, C.~J., {Brogan}, C.~L., {Hunter}, T.~R., \& {Churchwell}, E.
  2011, \apj, 743, 56, \dodoi{10.1088/0004-637X/743/1/56}

\bibitem[{{Cyganowski} {et~al.}(2008){Cyganowski}, {Whitney}, {Holden},
  {Braden}, {Brogan}, {Churchwell}, {Indebetouw}, {Watson}, {Babler},
  {Benjamin}, {Gomez}, {Meade}, {Povich}, {Robitaille}, \&
  {Watson}}]{2008AJ....136.2391C}
{Cyganowski}, C.~J., {Whitney}, B.~A., {Holden}, E., {et~al.} 2008, \aj, 136,
  2391, \dodoi{10.1088/0004-6256/136/6/2391}

\bibitem[{{Dame} {et~al.}(2001){Dame}, {Hartmann}, \&
  {Thaddeus}}]{2001ApJ...547..792D}
{Dame}, T.~M., {Hartmann}, D., \& {Thaddeus}, P. 2001, \apj, 547, 792,
  \dodoi{10.1086/318388}

\bibitem[{{Deharveng} {et~al.}(2008){Deharveng}, {Lefloch}, {Kurtz}, {Nadeau},
  {Pomar{\`e}s}, {Caplan}, \& {Zavagno}}]{2008A&A...482..585D}
{Deharveng}, L., {Lefloch}, B., {Kurtz}, S., {et~al.} 2008, \aap, 482, 585,
  \dodoi{10.1051/0004-6361:20079233}

\bibitem[{{Du} {et~al.}(2017){Du}, {Xu}, {Yang}, \&
  {Sun}}]{2017ApJS..229...24D}
{Du}, X., {Xu}, Y., {Yang}, J., \& {Sun}, Y. 2017, \apjs, 229, 24,
  \dodoi{10.3847/1538-4365/aa5d9d}

\bibitem[{{Du} {et~al.}(2016){Du}, {Xu}, {Yang}, {Sun}, {Li}, {Zhang}, \&
  {Zhou}}]{2016ApJS..224....7D}
{Du}, X., {Xu}, Y., {Yang}, J., {et~al.} 2016, \apjs, 224, 7,
  \dodoi{10.3847/0067-0049/224/1/7}

\bibitem[{{Dunham} {et~al.}(2008){Dunham}, {Crapsi}, {Evans}, {Bourke},
  {Huard}, {Myers}, \& {Kauffmann}}]{2008ApJS..179..249D}
{Dunham}, M.~M., {Crapsi}, A., {Evans}, Neal~J., I., {et~al.} 2008, \apjs, 179,
  249, \dodoi{10.1086/591085}

\bibitem[{{Dunham} {et~al.}(2015){Dunham}, {Allen}, {Evans},
  {Broekhoven-Fiene}, {Cieza}, {Di Francesco}, {Gutermuth}, {Harvey},
  {Hatchell}, {Heiderman}, {Huard}, {Johnstone}, {Kirk}, {Matthews}, {Miller},
  {Peterson}, \& {Young}}]{2015ApJS..220...11D}
{Dunham}, M.~M., {Allen}, L.~E., {Evans}, Neal~J., I., {et~al.} 2015, \apjs,
  220, 11, \dodoi{10.1088/0067-0049/220/1/11}

\bibitem[{{Elitzur} {et~al.}(1989){Elitzur}, {Hollenbach}, \&
  {McKee}}]{1989ApJ...346..983E}
{Elitzur}, M., {Hollenbach}, D.~J., \& {McKee}, C.~F. 1989, \apj, 346, 983,
  \dodoi{10.1086/168080}

\bibitem[{{Ellsworth-Bowers} {et~al.}(2015){Ellsworth-Bowers}, {Glenn},
  {Riley}, {Rosolowsky}, {Ginsburg}, {Evans}, {Bally}, {Battersby}, {Shirley},
  \& {Merello}}]{2015ApJ...805..157E}
{Ellsworth-Bowers}, T.~P., {Glenn}, J., {Riley}, A., {et~al.} 2015, \apj, 805,
  157, \dodoi{10.1088/0004-637X/805/2/157}

\bibitem[{{Evans} {et~al.}(2009){Evans}, {Dunham}, {J{\o}rgensen}, {Enoch},
  {Mer{\'\i}n}, {van Dishoeck}, {Alcal{\'a}}, {Myers}, {Stapelfeldt}, {Huard},
  {Allen}, {Harvey}, {van Kempen}, {Blake}, {Koerner}, {Mundy}, {Padgett}, \&
  {Sargent}}]{2009ApJS..181..321E}
{Evans}, Neal~J., I., {Dunham}, M.~M., {J{\o}rgensen}, J.~K., {et~al.} 2009,
  \apjs, 181, 321, \dodoi{10.1088/0067-0049/181/2/321}

\bibitem[{{Foster} \& {Brunt}(2015)}]{2015AJ....150..147F}
{Foster}, T., \& {Brunt}, C.~M. 2015, \aj, 150, 147,
  \dodoi{10.1088/0004-6256/150/5/147}

\bibitem[{{Frerking} {et~al.}(1982){Frerking}, {Langer}, \&
  {Wilson}}]{1982ApJ...262..590F}
{Frerking}, M.~A., {Langer}, W.~D., \& {Wilson}, R.~W. 1982, \apj, 262, 590,
  \dodoi{10.1086/160451}

\bibitem[{{Fukui} {et~al.}(2018){Fukui}, {Kohno}, {Yokoyama}, {Torii},
  {Hattori}, {Sano}, {Nishimura}, {Ohama}, {Yamamoto}, \&
  {Tachihara}}]{2018PASJ...70S..41F}
{Fukui}, Y., {Kohno}, M., {Yokoyama}, K., {et~al.} 2018, \pasj, 70, S41,
  \dodoi{10.1093/pasj/psy017}

\bibitem[{{Furuya} {et~al.}(2001){Furuya}, {Kitamura}, {Wootten}, {Claussen},
  \& {Kawabe}}]{2001ApJ...559L.143F}
{Furuya}, R.~S., {Kitamura}, Y., {Wootten}, H.~A., {Claussen}, M.~J., \&
  {Kawabe}, R. 2001, \apjl, 559, L143, \dodoi{10.1086/324012}

\bibitem[{{Gaia Collaboration} {et~al.}(2018){Gaia Collaboration}, {Brown},
  {Vallenari}, {Prusti}, {de Bruijne}, {Babusiaux}, {Bailer-Jones}, {Biermann},
  {Evans}, {Eyer}, {Jansen}, {Jordi}, {Klioner}, {Lammers}, {Lindegren},
  {Luri}, {Mignard}, {Panem}, {Pourbaix}, {Randich}, {Sartoretti}, {Siddiqui},
  {Soubiran}, {van Leeuwen}, {Walton}, {Arenou}, {Bastian}, {Cropper},
  {Drimmel}, {Katz}, {Lattanzi}, {Bakker}, {Cacciari}, {Casta{\~n}eda},
  {Chaoul}, {Cheek}, {De Angeli}, {Fabricius}, {Guerra}, {Holl}, {Masana},
  {Messineo}, {Mowlavi}, {Nienartowicz}, {Panuzzo}, {Portell}, {Riello},
  {Seabroke}, {Tanga}, {Th{\'e}venin}, {Gracia-Abril}, {Comoretto},
  {Garcia-Reinaldos}, {Teyssier}, {Altmann}, {Andrae}, {Audard},
  {Bellas-Velidis}, {Benson}, {Berthier}, {Blomme}, {Burgess}, {Busso},
  {Carry}, {Cellino}, {Clementini}, {Clotet}, {Creevey}, {Davidson}, {De
  Ridder}, {Delchambre}, {Dell'Oro}, {Ducourant},
  {Fern{\'a}ndez-Hern{\'a}ndez}, {Fouesneau}, {Fr{\'e}mat}, {Galluccio},
  {Garc{\'\i}a-Torres}, {Gonz{\'a}lez-N{\'u}{\~n}ez}, {Gonz{\'a}lez-Vidal},
  {Gosset}, {Guy}, {Halbwachs}, {Hambly}, {Harrison}, {Hern{\'a}ndez},
  {Hestroffer}, {Hodgkin}, {Hutton}, {Jasniewicz}, {Jean-Antoine-Piccolo},
  {Jordan}, {Korn}, {Krone-Martins}, {Lanzafame}, {Lebzelter}, {L{\"o}ffler},
  {Manteiga}, {Marrese}, {Mart{\'\i}n-Fleitas}, {Moitinho}, {Mora}, {Muinonen},
  {Osinde}, {Pancino}, {Pauwels}, {Petit}, {Recio-Blanco}, {Richards},
  {Rimoldini}, {Robin}, {Sarro}, {Siopis}, {Smith}, {Sozzetti}, {S{\"u}veges},
  {Torra}, {van Reeven}, {Abbas}, {Abreu Aramburu}, {Accart}, {Aerts},
  {Altavilla}, {{\'A}lvarez}, {Alvarez}, {Alves}, {Anderson}, {Andrei},
  {Anglada Varela}, {Antiche}, {Antoja}, {Arcay}, {Astraatmadja}, {Bach},
  {Baker}, {Balaguer-N{\'u}{\~n}ez}, {Balm}, {Barache}, {Barata}, {Barbato},
  {Barblan}, {Barklem}, {Barrado}, {Barros}, {Barstow}, {Bartholom{\'e}
  Mu{\~n}oz}, {Bassilana}, {Becciani}, {Bellazzini}, {Berihuete}, {Bertone},
  {Bianchi}, {Bienaym{\'e}}, {Blanco-Cuaresma}, {Boch}, {Boeche}, {Bombrun},
  {Borrachero}, {Bossini}, {Bouquillon}, {Bourda}, {Bragaglia}, {Bramante},
  {Breddels}, {Bressan}, {Brouillet}, {Br{\"u}semeister}, {Brugaletta},
  {Bucciarelli}, {Burlacu}, {Busonero}, {Butkevich}, {Buzzi}, {Caffau},
  {Cancelliere}, {Cannizzaro}, {Cantat-Gaudin}, {Carballo}, {Carlucci},
  {Carrasco}, {Casamiquela}, {Castellani}, {Castro-Ginard}, {Charlot},
  {Chemin}, {Chiavassa}, {Cocozza}, {Costigan}, {Cowell}, {Crifo}, {Crosta},
  {Crowley}, {Cuypers}, {Dafonte}, {Damerdji}, {Dapergolas}, {David}, {David},
  {de Laverny}, {De Luise}, {De March}, {de Martino}, {de Souza}, {de Torres},
  {Debosscher}, {del Pozo}, {Delbo}, {Delgado}, {Delgado}, {Di Matteo},
  {Diakite}, {Diener}, {Distefano}, {Dolding}, {Drazinos}, {Dur{\'a}n},
  {Edvardsson}, {Enke}, {Eriksson}, {Esquej}, {Eynard Bontemps}, {Fabre},
  {Fabrizio}, {Faigler}, {Falc{\~a}o}, {Farr{\`a}s Casas}, {Federici},
  {Fedorets}, {Fernique}, {Figueras}, {Filippi}, {Findeisen}, {Fonti},
  {Fraile}, {Fraser}, {Fr{\'e}zouls}, {Gai}, {Galleti}, {Garabato},
  {Garc{\'\i}a-Sedano}, {Garofalo}, {Garralda}, {Gavel}, {Gavras}, {Gerssen},
  {Geyer}, {Giacobbe}, {Gilmore}, {Girona}, {Giuffrida}, {Glass}, {Gomes},
  {Granvik}, {Gueguen}, {Guerrier}, {Guiraud}, {Guti{\'e}rrez-S{\'a}nchez},
  {Haigron}, {Hatzidimitriou}, {Hauser}, {Haywood}, {Heiter}, {Helmi}, {Heu},
  {Hilger}, {Hobbs}, {Hofmann}, {Holland}, {Huckle}, {Hypki}, {Icardi},
  {Jan{\ss}en}, {Jevardat de Fombelle}, {Jonker}, {Juh{\'a}sz}, {Julbe},
  {Karampelas}, {Kewley}, {Klar}, {Kochoska}, {Kohley}, {Kolenberg},
  {Kontizas}, {Kontizas}, {Koposov}, {Kordopatis}, {Kostrzewa-Rutkowska},
  {Koubsky}, {Lambert}, {Lanza}, {Lasne}, {Lavigne}, {Le Fustec}, {Le
  Poncin-Lafitte}, {Lebreton}, {Leccia}, {Leclerc}, {Lecoeur-Taibi},
  {Lenhardt}, {Leroux}, {Liao}, {Licata}, {Lindstr{\o}m}, {Lister}, {Livanou},
  {Lobel}, {L{\'o}pez}, {Managau}, {Mann}, {Mantelet}, {Marchal}, {Marchant},
  {Marconi}, {Marinoni}, {Marschalk{\'o}}, {Marshall}, {Martino}, {Marton},
  {Mary}, {Massari}, {Matijevi{\v{c}}}, {Mazeh}, {McMillan}, {Messina},
  {Michalik}, {Millar}, {Molina}, {Molinaro}, {Moln{\'a}r}, {Montegriffo},
  {Mor}, {Morbidelli}, {Morel}, {Morris}, {Mulone}, {Muraveva}, {Musella},
  {Nelemans}, {Nicastro}, {Noval}, {O'Mullane}, {Ord{\'e}novic},
  {Ord{\'o}{\~n}ez-Blanco}, {Osborne}, {Pagani}, {Pagano}, {Pailler},
  {Palacin}, {Palaversa}, {Panahi}, {Pawlak}, {Piersimoni}, {Pineau}, {Plachy},
  {Plum}, {Poggio}, {Poujoulet}, {Pr{\v{s}}a}, {Pulone}, {Racero}, {Ragaini},
  {Rambaux}, {Ramos-Lerate}, {Regibo}, {Reyl{\'e}}, {Riclet}, {Ripepi}, {Riva},
  {Rivard}, {Rixon}, {Roegiers}, {Roelens}, {Romero-G{\'o}mez}, {Rowell},
  {Royer}, {Ruiz-Dern}, {Sadowski}, {Sagrist{\`a} Sell{\'e}s}, {Sahlmann},
  {Salgado}, {Salguero}, {Sanna}, {Santana-Ros}, {Sarasso}, {Savietto},
  {Schultheis}, {Sciacca}, {Segol}, {Segovia}, {S{\'e}gransan}, {Shih},
  {Siltala}, {Silva}, {Smart}, {Smith}, {Solano}, {Solitro}, {Sordo}, {Soria
  Nieto}, {Souchay}, {Spagna}, {Spoto}, {Stampa}, {Steele},
  {Steidelm{\"u}ller}, {Stephenson}, {Stoev}, {Suess}, {Surdej}, {Szabados},
  {Szegedi-Elek}, {Tapiador}, {Taris}, {Tauran}, {Taylor}, {Teixeira},
  {Terrett}, {Teyssandier}, {Thuillot}, {Titarenko}, {Torra Clotet}, {Turon},
  {Ulla}, {Utrilla}, {Uzzi}, {Vaillant}, {Valentini}, {Valette}, {van Elteren},
  {Van Hemelryck}, {van Leeuwen}, {Vaschetto}, {Vecchiato}, {Veljanoski},
  {Viala}, {Vicente}, {Vogt}, {von Essen}, {Voss}, {Votruba}, {Voutsinas},
  {Walmsley}, {Weiler}, {Wertz}, {Wevers}, {Wyrzykowski}, {Yoldas},
  {{\v{Z}}erjal}, {Ziaeepour}, {Zorec}, {Zschocke}, {Zucker}, {Zurbach}, \&
  {Zwitter}}]{2018A&A...616A...1G}
{Gaia Collaboration}, {Brown}, A.~G.~A., {Vallenari}, A., {et~al.} 2018, \aap,
  616, A1, \dodoi{10.1051/0004-6361/201833051}

\bibitem[{{Garden} {et~al.}(1991){Garden}, {Hayashi}, {Gatley}, {Hasegawa}, \&
  {Kaifu}}]{1991ApJ...374..540G}
{Garden}, R.~P., {Hayashi}, M., {Gatley}, I., {Hasegawa}, T., \& {Kaifu}, N.
  1991, \apj, 374, 540, \dodoi{10.1086/170143}

\bibitem[{{Goldsmith} {et~al.}(2008){Goldsmith}, {Heyer}, {Narayanan}, {Snell},
  {Li}, \& {Brunt}}]{2008ApJ...680..428G}
{Goldsmith}, P.~F., {Heyer}, M., {Narayanan}, G., {et~al.} 2008, \apj, 680,
  428, \dodoi{10.1086/587166}

\bibitem[{{Gong} {et~al.}(2016){Gong}, {Mao}, {Fang}, {Zhang}, {Su}, {Yang},
  {Jiang}, {Xu}, {Wang}, {Wang}, {Lu}, \& {Sun}}]{2016A&A...588A.104G}
{Gong}, Y., {Mao}, R.~Q., {Fang}, M., {et~al.} 2016, \aap, 588, A104,
  \dodoi{10.1051/0004-6361/201527334}

\bibitem[{{Greene} {et~al.}(1994){Greene}, {Wilking}, {Andre}, {Young}, \&
  {Lada}}]{1994ApJ...434..614G}
{Greene}, T.~P., {Wilking}, B.~A., {Andre}, P., {Young}, E.~T., \& {Lada},
  C.~J. 1994, \apj, 434, 614, \dodoi{10.1086/174763}

\bibitem[{{Guo} {et~al.}(2021){Guo}, {Chen}, {Feng}, {Sun}, {Wang}, {Su},
  {Sun}, {Ao}, {Zhang}, {Zhou}, {Yuan}, \& {Yang}}]{2021ApJ...921...23G}
{Guo}, W., {Chen}, X., {Feng}, J., {et~al.} 2021, \apj, 921, 23,
  \dodoi{10.3847/1538-4357/ac15fe}

\bibitem[{{Hacar} \& {Tafalla}(2011)}]{2011A&A...533A..34H}
{Hacar}, A., \& {Tafalla}, M. 2011, \aap, 533, A34,
  \dodoi{10.1051/0004-6361/201117039}

\bibitem[{{Hacar} {et~al.}(2018){Hacar}, {Tafalla}, {Forbrich}, {Alves},
  {Meingast}, {Grossschedl}, \& {Teixeira}}]{2018A&A...610A..77H}
{Hacar}, A., {Tafalla}, M., {Forbrich}, J., {et~al.} 2018, \aap, 610, A77,
  \dodoi{10.1051/0004-6361/201731894}

\bibitem[{{Hacar} {et~al.}(2013){Hacar}, {Tafalla}, {Kauffmann}, \&
  {Kov{\'a}cs}}]{2013A&A...554A..55H}
{Hacar}, A., {Tafalla}, M., {Kauffmann}, J., \& {Kov{\'a}cs}, A. 2013, \aap,
  554, A55, \dodoi{10.1051/0004-6361/201220090}

\bibitem[{{Heiderman} \& {Evans}(2015)}]{2015ApJ...806..231H}
{Heiderman}, A., \& {Evans}, Neal~J., I. 2015, \apj, 806, 231,
  \dodoi{10.1088/0004-637X/806/2/231}

\bibitem[{{Heitsch}(2013)}]{2013ApJ...769..115H}
{Heitsch}, F. 2013, \apj, 769, 115, \dodoi{10.1088/0004-637X/769/2/115}

\bibitem[{{Hennemann} {et~al.}(2012){Hennemann}, {Motte}, {Schneider},
  {Didelon}, {Hill}, {Arzoumanian}, {Bontemps}, {Csengeri}, {Andr{\'e}},
  {Konyves}, {Louvet}, {Marston}, {Men'shchikov}, {Minier}, {Nguyen Luong},
  {Palmeirim}, {Peretto}, {Sauvage}, {Zavagno}, {Anderson}, {Bernard}, {Di
  Francesco}, {Elia}, {Li}, {Martin}, {Molinari}, {Pezzuto}, {Russeil}, {Rygl},
  {Schisano}, {Spinoglio}, {Sousbie}, {Ward-Thompson}, \&
  {White}}]{2012A&A...543L...3H}
{Hennemann}, M., {Motte}, F., {Schneider}, N., {et~al.} 2012, \aap, 543, L3,
  \dodoi{10.1051/0004-6361/201219429}

\bibitem[{{Henshaw} {et~al.}(2014){Henshaw}, {Caselli}, {Fontani},
  {Jim{\'e}nez-Serra}, \& {Tan}}]{2014MNRAS.440.2860H}
{Henshaw}, J.~D., {Caselli}, P., {Fontani}, F., {Jim{\'e}nez-Serra}, I., \&
  {Tan}, J.~C. 2014, \mnras, 440, 2860, \dodoi{10.1093/mnras/stu446}

\bibitem[{{Henshaw} {et~al.}(2020){Henshaw}, {Kruijssen}, {Longmore}, {Riener},
  {Leroy}, {Rosolowsky}, {Ginsburg}, {Battersby}, {Chevance}, {Meidt},
  {Glover}, {Hughes}, {Kainulainen}, {Klessen}, {Schinnerer}, {Schruba},
  {Beuther}, {Bigiel}, {Blanc}, {Emsellem}, {Henning}, {Herrera}, {Koch},
  {Pety}, {Ragan}, \& {Sun}}]{2020NatAs...4.1064H}
{Henshaw}, J.~D., {Kruijssen}, J.~M.~D., {Longmore}, S.~N., {et~al.} 2020,
  Nature Astronomy, 4, 1064, \dodoi{10.1038/s41550-020-1126-z}

\bibitem[{{Heyer} \& {Dame}(2015)}]{2015ARA&A..53..583H}
{Heyer}, M., \& {Dame}, T.~M. 2015, \araa, 53, 583,
  \dodoi{10.1146/annurev-astro-082214-122324}

\bibitem[{{Heyer} {et~al.}(2009){Heyer}, {Krawczyk}, {Duval}, \&
  {Jackson}}]{2009ApJ...699.1092H}
{Heyer}, M., {Krawczyk}, C., {Duval}, J., \& {Jackson}, J.~M. 2009, \apj, 699,
  1092, \dodoi{10.1088/0004-637X/699/2/1092}

\bibitem[{{Hill} {et~al.}(2011){Hill}, {Motte}, {Didelon}, {Bontemps},
  {Minier}, {Hennemann}, {Schneider}, {Andr{\'e}}, {Men'shchikov}, {Anderson},
  {Arzoumanian}, {Bernard}, {di Francesco}, {Elia}, {Giannini}, {Griffin},
  {K{\"o}nyves}, {Kirk}, {Marston}, {Martin}, {Molinari}, {Nguyen Luong},
  {Peretto}, {Pezzuto}, {Roussel}, {Sauvage}, {Sousbie}, {Testi},
  {Ward-Thompson}, {White}, {Wilson}, \& {Zavagno}}]{2011A&A...533A..94H}
{Hill}, T., {Motte}, F., {Didelon}, P., {et~al.} 2011, \aap, 533, A94,
  \dodoi{10.1051/0004-6361/201117315}

\bibitem[{{Hunter} \& {Massey}(1990)}]{1990AJ.....99..846H}
{Hunter}, D.~A., \& {Massey}, P. 1990, \aj, 99, 846, \dodoi{10.1086/115378}

\bibitem[{{Inoue} {et~al.}(2001){Inoue}, {Hirashita}, \&
  {Kamaya}}]{2001ApJ...555..613I}
{Inoue}, A.~K., {Hirashita}, H., \& {Kamaya}, H. 2001, \apj, 555, 613,
  \dodoi{10.1086/321499}

\bibitem[{{Inoue} {et~al.}(2018){Inoue}, {Hennebelle}, {Fukui}, {Matsumoto},
  {Iwasaki}, \& {Inutsuka}}]{2018PASJ...70S..53I}
{Inoue}, T., {Hennebelle}, P., {Fukui}, Y., {et~al.} 2018, \pasj, 70, S53,
  \dodoi{10.1093/pasj/psx089}

\bibitem[{{Inutsuka} \& {Miyama}(1997)}]{1997ApJ...480..681I}
{Inutsuka}, S.-i., \& {Miyama}, S.~M. 1997, \apj, 480, 681,
  \dodoi{10.1086/303982}

\bibitem[{{Jackson} {et~al.}(2019){Jackson}, {Allingham}, {Patterson},
  {Killerby-Smith}, {Whitaker}, \& {Hogge}}]{2019AAS...23312706J}
{Jackson}, J.~M., {Allingham}, D., {Patterson}, P., {et~al.} 2019, in American
  Astronomical Society Meeting Abstracts, Vol. 233, American Astronomical
  Society Meeting Abstracts \#233, 127.06

\bibitem[{{Jackson} {et~al.}(2010){Jackson}, {Finn}, {Chambers}, {Rathborne},
  \& {Simon}}]{2010ApJ...719L.185J}
{Jackson}, J.~M., {Finn}, S.~C., {Chambers}, E.~T., {Rathborne}, J.~M., \&
  {Simon}, R. 2010, \apjl, 719, L185, \dodoi{10.1088/2041-8205/719/2/L185}

\bibitem[{{Johnstone} \& {Bally}(1999)}]{1999ApJ...510L..49J}
{Johnstone}, D., \& {Bally}, J. 1999, \apjl, 510, L49, \dodoi{10.1086/311792}

\bibitem[{{Jose} {et~al.}(2011){Jose}, {Pandey}, {Ogura}, {Ojha}, {Bhatt},
  {Samal}, {Chauhan}, {Sahu}, \& {Rawat}}]{2011MNRAS.411.2530J}
{Jose}, J., {Pandey}, A.~K., {Ogura}, K., {et~al.} 2011, \mnras, 411, 2530,
  \dodoi{10.1111/j.1365-2966.2010.17860.x}

\bibitem[{{Kauffmann} {et~al.}(2008){Kauffmann}, {Bertoldi}, {Bourke}, {Evans},
  \& {Lee}}]{2008A&A...487..993K}
{Kauffmann}, J., {Bertoldi}, F., {Bourke}, T.~L., {Evans}, N.~J., I., \& {Lee},
  C.~W. 2008, \aap, 487, 993, \dodoi{10.1051/0004-6361:200809481}

\bibitem[{{Kauffmann} {et~al.}(2010){Kauffmann}, {Pillai}, {Shetty}, {Myers},
  \& {Goodman}}]{2010ApJ...716..433K}
{Kauffmann}, J., {Pillai}, T., {Shetty}, R., {Myers}, P.~C., \& {Goodman},
  A.~A. 2010, \apj, 716, 433, \dodoi{10.1088/0004-637X/716/1/433}

\bibitem[{{Kirk} {et~al.}(2013){Kirk}, {Myers}, {Bourke}, {Gutermuth},
  {Hedden}, \& {Wilson}}]{2013ApJ...766..115K}
{Kirk}, H., {Myers}, P.~C., {Bourke}, T.~L., {et~al.} 2013, \apj, 766, 115,
  \dodoi{10.1088/0004-637X/766/2/115}

\bibitem[{{Klein} {et~al.}(2005){Klein}, {Posselt}, {Schreyer}, {Forbrich}, \&
  {Henning}}]{2005ApJS..161..361K}
{Klein}, R., {Posselt}, B., {Schreyer}, K., {Forbrich}, J., \& {Henning}, T.
  2005, \apjs, 161, 361, \dodoi{10.1086/496962}

\bibitem[{{Koenig} \& {Leisawitz}(2014)}]{2014ApJ...791..131K}
{Koenig}, X.~P., \& {Leisawitz}, D.~T. 2014, \apj, 791, 131,
  \dodoi{10.1088/0004-637X/791/2/131}

\bibitem[{{K{\"o}nyves} {et~al.}(2015){K{\"o}nyves}, {Andr{\'e}},
  {Men'shchikov}, {Palmeirim}, {Arzoumanian}, {Schneider}, {Roy}, {Didelon},
  {Maury}, {Shimajiri}, {Di Francesco}, {Bontemps}, {Peretto}, {Benedettini},
  {Bernard}, {Elia}, {Griffin}, {Hill}, {Kirk}, {Ladjelate}, {Marsh}, {Martin},
  {Motte}, {Nguy{\^e}n Luong}, {Pezzuto}, {Roussel}, {Rygl}, {Sadavoy},
  {Schisano}, {Spinoglio}, {Ward-Thompson}, \& {White}}]{2015A&A...584A..91K}
{K{\"o}nyves}, V., {Andr{\'e}}, P., {Men'shchikov}, A., {et~al.} 2015, \aap,
  584, A91, \dodoi{10.1051/0004-6361/201525861}

\bibitem[{{Kramer} {et~al.}(1998){Kramer}, {Stutzki}, {Rohrig}, \&
  {Corneliussen}}]{1998A&A...329..249K}
{Kramer}, C., {Stutzki}, J., {Rohrig}, R., \& {Corneliussen}, U. 1998, \aap,
  329, 249

\bibitem[{{Kryukova} {et~al.}(2012){Kryukova}, {Megeath}, {Gutermuth},
  {Pipher}, {Allen}, {Allen}, {Myers}, \& {Muzerolle}}]{2012AJ....144...31K}
{Kryukova}, E., {Megeath}, S.~T., {Gutermuth}, R.~A., {et~al.} 2012, \aj, 144,
  31, \dodoi{10.1088/0004-6256/144/2/31}

\bibitem[{{Lada} {et~al.}(2017){Lada}, {Lewis}, {Lombardi}, \&
  {Alves}}]{2017A&A...606A.100L}
{Lada}, C.~J., {Lewis}, J.~A., {Lombardi}, M., \& {Alves}, J. 2017, \aap, 606,
  A100, \dodoi{10.1051/0004-6361/201731221}

\bibitem[{{Lada} {et~al.}(2009){Lada}, {Lombardi}, \&
  {Alves}}]{2009ApJ...703...52L}
{Lada}, C.~J., {Lombardi}, M., \& {Alves}, J.~F. 2009, \apj, 703, 52,
  \dodoi{10.1088/0004-637X/703/1/52}

\bibitem[{{Lada} {et~al.}(2010){Lada}, {Lombardi}, \&
  {Alves}}]{2010ApJ...724..687L}
---. 2010, \apj, 724, 687, \dodoi{10.1088/0004-637X/724/1/687}

\bibitem[{{Lim} {et~al.}(2015){Lim}, {Sung}, {Bessell}, {Kim}, {Hur}, \&
  {Park}}]{2015AJ....149..127L}
{Lim}, B., {Sung}, H., {Bessell}, M.~S., {et~al.} 2015, \aj, 149, 127,
  \dodoi{10.1088/0004-6256/149/4/127}

\bibitem[{{Liu} {et~al.}(2021){Liu}, {Xu}, {Wang}, {Yu}, {Zhang}, {Li}, \&
  {Zhang}}]{2021A&A...646A.137L}
{Liu}, X.-L., {Xu}, J.-L., {Wang}, J.-J., {et~al.} 2021, \aap, 646, A137,
  \dodoi{10.1051/0004-6361/201935035}

\bibitem[{{Lockman}(1989)}]{1989ApJS...71..469L}
{Lockman}, F.~J. 1989, \apjs, 71, 469, \dodoi{10.1086/191383}

\bibitem[{{Louvet} {et~al.}(2014){Louvet}, {Motte}, {Hennebelle}, {Maury},
  {Bonnell}, {Bontemps}, {Gusdorf}, {Hill}, {Gueth}, {Peretto},
  {Duarte-Cabral}, {Stephan}, {Schilke}, {Csengeri}, {Nguyen Luong}, \&
  {Lis}}]{2014A&A...570A..15L}
{Louvet}, F., {Motte}, F., {Hennebelle}, P., {et~al.} 2014, \aap, 570, A15,
  \dodoi{10.1051/0004-6361/201423603}

\bibitem[{{Lu} {et~al.}(2018){Lu}, {Zhang}, {Liu}, {Sanhueza}, {Tatematsu},
  {Feng}, {Smith}, {Myers}, {Sridharan}, \& {Gu}}]{2018ApJ...855....9L}
{Lu}, X., {Zhang}, Q., {Liu}, H.~B., {et~al.} 2018, \apj, 855, 9,
  \dodoi{10.3847/1538-4357/aaad11}

\bibitem[{{Lumsden} {et~al.}(2013){Lumsden}, {Hoare}, {Urquhart}, {Oudmaijer},
  {Davies}, {Mottram}, {Cooper}, \& {Moore}}]{2013ApJS..208...11L}
{Lumsden}, S.~L., {Hoare}, M.~G., {Urquhart}, J.~S., {et~al.} 2013, \apjs, 208,
  11, \dodoi{10.1088/0067-0049/208/1/11}

\bibitem[{{Marton} {et~al.}(2016){Marton}, {T{\'o}th}, {Paladini}, {Kun},
  {Zahorecz}, {McGehee}, \& {Kiss}}]{2016MNRAS.458.3479M}
{Marton}, G., {T{\'o}th}, L.~V., {Paladini}, R., {et~al.} 2016, \mnras, 458,
  3479, \dodoi{10.1093/mnras/stw398}

\bibitem[{{Marton} {et~al.}(2019){Marton}, {{\'A}brah{\'a}m}, {Szegedi-Elek},
  {Varga}, {Kun}, {K{\'o}sp{\'a}l}, {Varga-Vereb{\'e}lyi}, {Hodgkin},
  {Szabados}, {Beck}, \& {Kiss}}]{2019MNRAS.487.2522M}
{Marton}, G., {{\'A}brah{\'a}m}, P., {Szegedi-Elek}, E., {et~al.} 2019, \mnras,
  487, 2522, \dodoi{10.1093/mnras/stz1301}

\bibitem[{{Mattern} {et~al.}(2018){Mattern}, {Kauffmann}, {Csengeri},
  {Urquhart}, {Leurini}, {Wyrowski}, {Giannetti}, {Barnes}, {Beuther},
  {Bronfman}, {Duarte-Cabral}, {Henning}, {Kainulainen}, {Menten}, {Schisano},
  \& {Schuller}}]{2018A&A...619A.166M}
{Mattern}, M., {Kauffmann}, J., {Csengeri}, T., {et~al.} 2018, \aap, 619, A166,
  \dodoi{10.1051/0004-6361/201833406}

\bibitem[{{McCutcheon} {et~al.}(1986){McCutcheon}, {Dewdney}, {Purton}, {Wall},
  \& {Sato}}]{1986BAAS...18Q.921M}
{McCutcheon}, W.~H., {Dewdney}, P.~E., {Purton}, C.~R., {Wall}, W.~F., \&
  {Sato}, T. 1986, in Bulletin of the American Astronomical Society, Vol.~18,
  921

\bibitem[{{Merello} {et~al.}(2015){Merello}, {Evans}, {Shirley}, {Rosolowsky},
  {Ginsburg}, {Bally}, {Battersby}, \& {Dunham}}]{2015ApJS..218....1M}
{Merello}, M., {Evans}, Neal~J., I., {Shirley}, Y.~L., {et~al.} 2015, \apjs,
  218, 1, \dodoi{10.1088/0067-0049/218/1/1}

\bibitem[{{Moffat} {et~al.}(1979){Moffat}, {Fitzgerald}, \&
  {Jackson}}]{1979A&AS...38..197M}
{Moffat}, A.~F.~J., {Fitzgerald}, M.~P., \& {Jackson}, P.~D. 1979, \aaps, 38,
  197

\bibitem[{{Motte} {et~al.}(2018){Motte}, {Bontemps}, \&
  {Louvet}}]{2018ARA&A..56...41M}
{Motte}, F., {Bontemps}, S., \& {Louvet}, F. 2018, \araa, 56, 41,
  \dodoi{10.1146/annurev-astro-091916-055235}

\bibitem[{{Myers}(2009)}]{2009ApJ...700.1609M}
{Myers}, P.~C. 2009, \apj, 700, 1609, \dodoi{10.1088/0004-637X/700/2/1609}

\bibitem[{{Nguyen Luong} {et~al.}(2011){Nguyen Luong}, {Motte}, {Hennemann},
  {Hill}, {Rygl}, {Schneider}, {Bontemps}, {Men'shchikov}, {Andr{\'e}},
  {Peretto}, {Anderson}, {Arzoumanian}, {Deharveng}, {Didelon}, {di Francesco},
  {Griffin}, {Kirk}, {K{\"o}nyves}, {Martin}, {Maury}, {Minier}, {Molinari},
  {Pestalozzi}, {Pezzuto}, {Reid}, {Roussel}, {Sauvage}, {Schuller}, {Testi},
  {Ward-Thompson}, {White}, \& {Zavagno}}]{2011A&A...535A..76N}
{Nguyen Luong}, Q., {Motte}, F., {Hennemann}, M., {et~al.} 2011, \aap, 535,
  A76, \dodoi{10.1051/0004-6361/201117831}

\bibitem[{{Nguyen-Luong} {et~al.}(2016){Nguyen-Luong}, {Nguyen}, {Motte},
  {Schneider}, {Fujii}, {Louvet}, {Hill}, {Sanhueza}, {Chibueze}, \&
  {Didelon}}]{2016ApJ...833...23N}
{Nguyen-Luong}, Q., {Nguyen}, H. V.~V., {Motte}, F., {et~al.} 2016, \apj, 833,
  23, \dodoi{10.3847/0004-637X/833/1/23}

\bibitem[{{Ostriker}(1964)}]{1964ApJ...140.1056O}
{Ostriker}, J. 1964, \apj, 140, 1056, \dodoi{10.1086/148005}

\bibitem[{{Palmeirim} {et~al.}(2013){Palmeirim}, {Andr{\'e}}, {Kirk},
  {Ward-Thompson}, {Arzoumanian}, {K{\"o}nyves}, {Didelon}, {Schneider},
  {Benedettini}, {Bontemps}, {Di Francesco}, {Elia}, {Griffin}, {Hennemann},
  {Hill}, {Martin}, {Men'shchikov}, {Molinari}, {Motte}, {Nguyen Luong},
  {Nutter}, {Peretto}, {Pezzuto}, {Roy}, {Rygl}, {Spinoglio}, \&
  {White}}]{2013A&A...550A..38P}
{Palmeirim}, P., {Andr{\'e}}, P., {Kirk}, J., {et~al.} 2013, \aap, 550, A38,
  \dodoi{10.1051/0004-6361/201220500}

\bibitem[{{Pety}(2005)}]{2005sf2a.conf..721P}
{Pety}, J. 2005, in SF2A-2005: Semaine de l'Astrophysique Francaise, ed.
  F.~{Casoli}, T.~{Contini}, J.~M. {Hameury}, \& L.~{Pagani}, 721

\bibitem[{{Pirogov}(1999)}]{Pirogov1999}
{Pirogov}, L. 1999, \aap, 348, 600

\bibitem[{{Pismis} {et~al.}(1991){Pismis}, {Hasse}, \&
  {Quintero}}]{1991PASP..103..843P}
{Pismis}, P., {Hasse}, I., \& {Quintero}, A. 1991, \pasp, 103, 843,
  \dodoi{10.1086/132891}

\bibitem[{{Ragan} {et~al.}(2009){Ragan}, {Bergin}, \&
  {Gutermuth}}]{2009ApJ...698..324R}
{Ragan}, S.~E., {Bergin}, E.~A., \& {Gutermuth}, R.~A. 2009, \apj, 698, 324,
  \dodoi{10.1088/0004-637X/698/1/324}

\bibitem[{{Ragan} {et~al.}(2014){Ragan}, {Henning}, {Tackenberg}, {Beuther},
  {Johnston}, {Kainulainen}, \& {Linz}}]{2014A&A...568A..73R}
{Ragan}, S.~E., {Henning}, T., {Tackenberg}, J., {et~al.} 2014, \aap, 568, A73,
  \dodoi{10.1051/0004-6361/201423401}

\bibitem[{{Reid} {et~al.}(2016){Reid}, {Dame}, {Menten}, \&
  {Brunthaler}}]{2016ApJ...823...77R}
{Reid}, M.~J., {Dame}, T.~M., {Menten}, K.~M., \& {Brunthaler}, A. 2016, \apj,
  823, 77, \dodoi{10.3847/0004-637X/823/2/77}

\bibitem[{{Reid} {et~al.}(2019){Reid}, {Menten}, {Brunthaler}, {Zheng}, {Dame},
  {Xu}, {Li}, {Sakai}, {Wu}, {Immer}, {Zhang}, {Sanna}, {Moscadelli}, {Rygl},
  {Bartkiewicz}, {Hu}, {Quiroga-Nu{\~n}ez}, \& {van
  Langevelde}}]{2019ApJ...885..131R}
{Reid}, M.~J., {Menten}, K.~M., {Brunthaler}, A., {et~al.} 2019, \apj, 885,
  131, \dodoi{10.3847/1538-4357/ab4a11}

\bibitem[{{Russeil} {et~al.}(2013){Russeil}, {Schneider}, {Anderson},
  {Zavagno}, {Molinari}, {Persi}, {Bontemps}, {Motte}, {Ossenkopf},
  {Andr{\'e}}, {Arzoumanian}, {Bernard}, {Deharveng}, {Didelon}, {Di
  Francesco}, {Elia}, {Hennemann}, {Hill}, {K{\"o}nyves}, {Li}, {Martin},
  {Nguyen Luong}, {Peretto}, {Pezzuto}, {Polychroni}, {Roussel}, {Rygl},
  {Spinoglio}, {Testi}, {Tig{\'e}}, {Vavrek}, {Ward-Thompson}, \&
  {White}}]{2013A&A...554A..42R}
{Russeil}, D., {Schneider}, N., {Anderson}, L.~D., {et~al.} 2013, \aap, 554,
  A42, \dodoi{10.1051/0004-6361/201219971}

\bibitem[{{Schneider} {et~al.}(2010){Schneider}, {Csengeri}, {Bontemps},
  {Motte}, {Simon}, {Hennebelle}, {Federrath}, \&
  {Klessen}}]{2010A&A...520A..49S}
{Schneider}, N., {Csengeri}, T., {Bontemps}, S., {et~al.} 2010, \aap, 520, A49,
  \dodoi{10.1051/0004-6361/201014481}

\bibitem[{{Shan} {et~al.}(2012){Shan}, {Yang}, {Shi}, {Yao}, {Zuo}, {Lin},
  {Chen}, {Zhang}, {Duan}, {Cao}, {Li}, {Li}, {Liu}, \&
  {Zhong}}]{2012ITTST...2..593S}
{Shan}, W., {Yang}, J., {Shi}, S., {et~al.} 2012, IEEE Transactions on
  Terahertz Science and Technology, 2, 593, \dodoi{10.1109/TTHZ.2012.2213818}

\bibitem[{{Skrutskie} {et~al.}(2006){Skrutskie}, {Cutri}, {Stiening},
  {Weinberg}, {Schneider}, {Carpenter}, {Beichman}, {Capps}, {Chester},
  {Elias}, {Huchra}, {Liebert}, {Lonsdale}, {Monet}, {Price}, {Seitzer},
  {Jarrett}, {Kirkpatrick}, {Gizis}, {Howard}, {Evans}, {Fowler}, {Fullmer},
  {Hurt}, {Light}, {Kopan}, {Marsh}, {McCallon}, {Tam}, {Van Dyk}, \&
  {Wheelock}}]{2006AJ....131.1163S}
{Skrutskie}, M.~F., {Cutri}, R.~M., {Stiening}, R., {et~al.} 2006, \aj, 131,
  1163, \dodoi{10.1086/498708}

\bibitem[{{Snell} {et~al.}(1990){Snell}, {Dickman}, \&
  {Huang}}]{1990ApJ...352..139S}
{Snell}, R.~L., {Dickman}, R.~L., \& {Huang}, Y.~L. 1990, \apj, 352, 139,
  \dodoi{10.1086/168521}

\bibitem[{{Soler} {et~al.}(2020){Soler}, {Beuther}, {Syed}, {Wang}, {Anderson},
  {Glover}, {Hennebelle}, {Heyer}, {Henning}, {Izquierdo}, {Klessen}, {Linz},
  {McClure-Griffiths}, {Ott}, {Ragan}, {Rugel}, {Schneider}, {Smith},
  {Sormani}, {Stil}, {Tre{\ss}}, \& {Urquhart}}]{2020A&A...642A.163S}
{Soler}, J.~D., {Beuther}, H., {Syed}, J., {et~al.} 2020, \aap, 642, A163,
  \dodoi{10.1051/0004-6361/202038882}

\bibitem[{{Sousbie}(2011)}]{2011MNRAS.414..350S}
{Sousbie}, T. 2011, \mnras, 414, 350, \dodoi{10.1111/j.1365-2966.2011.18394.x}

\bibitem[{{Stutzki} \& {Guesten}(1990)}]{1990ApJ...356..513S}
{Stutzki}, J., \& {Guesten}, R. 1990, \apj, 356, 513, \dodoi{10.1086/168859}

\bibitem[{{Su} {et~al.}(2019){Su}, {Yang}, {Zhang}, {Gong}, {Wang}, {Zhou},
  {Wang}, {Chen}, {Sun}, {Chen}, {Xu}, \& {Jiang}}]{2019ApJS..240....9S}
{Su}, Y., {Yang}, J., {Zhang}, S., {et~al.} 2019, \apjs, 240, 9,
  \dodoi{10.3847/1538-4365/aaf1c8}

\bibitem[{{Sun} {et~al.}(2015){Sun}, {Xu}, {Yang}, {Li}, {Du}, {Zhang}, \&
  {Zhou}}]{2015ApJ...798L..27S}
{Sun}, Y., {Xu}, Y., {Yang}, J., {et~al.} 2015, \apjl, 798, L27,
  \dodoi{10.1088/2041-8205/798/2/L27}

\bibitem[{{Sun} {et~al.}(2021){Sun}, {Yang}, {Yan}, {Lin}, {Zhang}, {Su}, {Xu},
  {Chen}, {Wang}, \& {Zhou}}]{2021ApJS..256...32S}
{Sun}, Y., {Yang}, J., {Yan}, Q.-Z., {et~al.} 2021, \apjs, 256, 32,
  \dodoi{10.3847/1538-4365/ac11fe}

\bibitem[{{Sunada} {et~al.}(2007){Sunada}, {Nakazato}, {Ikeda}, {Hongo},
  {Kitamura}, \& {Yang}}]{2007PASJ...59.1185S}
{Sunada}, K., {Nakazato}, T., {Ikeda}, N., {et~al.} 2007, \pasj, 59, 1185,
  \dodoi{10.1093/pasj/59.6.1185}

\bibitem[{{Syed} {et~al.}(2021){Syed}, {Soler}, {Beuther}, {Wang}, {Suri},
  {Henshaw}, {Riener}, {Bialy}, {Rezaei Kh.}, {Stil}, {Goldsmith}, {Rugel},
  {Glover}, {Klessen}, {Kerp}, {Urquhart}, {Ott}, {Roy}, {Schneider}, {Smith},
  {Longmore}, \& {Linz}}]{2021arXiv211101057S}
{Syed}, J., {Soler}, J.~D., {Beuther}, H., {et~al.} 2021, arXiv e-prints,
  arXiv:2111.01057.
\newblock \doarXiv{2111.01057}

\bibitem[{{Tatematsu} {et~al.}(2008){Tatematsu}, {Kandori}, {Umemoto}, \&
  {Sekimoto}}]{2008PASJ...60..407T}
{Tatematsu}, K., {Kandori}, R., {Umemoto}, T., \& {Sekimoto}, Y. 2008, \pasj,
  60, 407, \dodoi{10.1093/pasj/60.3.407}

\bibitem[{{Tenorio-Tagle}(1979)}]{1979A&A....71...59T}
{Tenorio-Tagle}, G. 1979, \aap, 71, 59

\bibitem[{{Traficante} {et~al.}(2018){Traficante}, {Duarte-Cabral}, {Elia},
  {Fuller}, {Merello}, {Molinari}, {Peretto}, {Schisano}, \& {Di
  Giorgio}}]{2018MNRAS.477.2220T}
{Traficante}, A., {Duarte-Cabral}, A., {Elia}, D., {et~al.} 2018, \mnras, 477,
  2220, \dodoi{10.1093/mnras/sty798}

\bibitem[{{Trevi{\~n}o-Morales} {et~al.}(2019){Trevi{\~n}o-Morales}, {Fuente},
  {S{\'a}nchez-Monge}, {Kainulainen}, {Didelon}, {Suri}, {Schneider},
  {Ballesteros-Paredes}, {Lee}, {Hennebelle}, {Pilleri},
  {Gonz{\'a}lez-Garc{\'\i}a}, {Kramer}, {Garc{\'\i}a-Burillo}, {Luna},
  {Goicoechea}, {Tremblin}, \& {Geen}}]{2019A&A...629A..81T}
{Trevi{\~n}o-Morales}, S.~P., {Fuente}, A., {S{\'a}nchez-Monge}, {\'A}.,
  {et~al.} 2019, \aap, 629, A81, \dodoi{10.1051/0004-6361/201935260}

\bibitem[{{Urquhart} {et~al.}(2013){Urquhart}, {Moore}, {Schuller}, {Wyrowski},
  {Menten}, {Thompson}, {Csengeri}, {Walmsley}, {Bronfman}, \&
  {K{\"o}nig}}]{2013MNRAS.431.1752U}
{Urquhart}, J.~S., {Moore}, T.~J.~T., {Schuller}, F., {et~al.} 2013, \mnras,
  431, 1752, \dodoi{10.1093/mnras/stt287}

\bibitem[{{Urquhart} {et~al.}(2014){Urquhart}, {Moore}, {Csengeri}, {Wyrowski},
  {Schuller}, {Hoare}, {Lumsden}, {Mottram}, {Thompson}, {Menten}, {Walmsley},
  {Bronfman}, {Pfalzner}, {K{\"o}nig}, \& {Wienen}}]{2014MNRAS.443.1555U}
{Urquhart}, J.~S., {Moore}, T.~J.~T., {Csengeri}, T., {et~al.} 2014, \mnras,
  443, 1555, \dodoi{10.1093/mnras/stu1207}

\bibitem[{{Vall{\'e}e}(2008)}]{2008AJ....135.1301V}
{Vall{\'e}e}, J.~P. 2008, \aj, 135, 1301, \dodoi{10.1088/0004-6256/135/4/1301}

\bibitem[{{Wang} {et~al.}(2015){Wang}, {Testi}, {Ginsburg}, {Walmsley},
  {Molinari}, \& {Schisano}}]{2015MNRAS.450.4043W}
{Wang}, K., {Testi}, L., {Ginsburg}, A., {et~al.} 2015, \mnras, 450, 4043,
  \dodoi{10.1093/mnras/stv735}

\bibitem[{{Weaver} {et~al.}(1977){Weaver}, {McCray}, {Castor}, {Shapiro}, \&
  {Moore}}]{1977ApJ...218..377W}
{Weaver}, R., {McCray}, R., {Castor}, J., {Shapiro}, P., \& {Moore}, R. 1977,
  \apj, 218, 377, \dodoi{10.1086/155692}

\bibitem[{{Wouterloot} {et~al.}(1988){Wouterloot}, {Brand}, \&
  {Henkel}}]{1988A&A...191..323W}
{Wouterloot}, J.~G.~A., {Brand}, J., \& {Henkel}, C. 1988, \aap, 191, 323

\bibitem[{{Wright} {et~al.}(2010){Wright}, {Eisenhardt}, {Mainzer}, {Ressler},
  {Cutri}, {Jarrett}, {Kirkpatrick}, {Padgett}, {McMillan}, {Skrutskie},
  {Stanford}, {Cohen}, {Walker}, {Mather}, {Leisawitz}, {Gautier}, {McLean},
  {Benford}, {Lonsdale}, {Blain}, {Mendez}, {Irace}, {Duval}, {Liu}, {Royer},
  {Heinrichsen}, {Howard}, {Shannon}, {Kendall}, {Walsh}, {Larsen}, {Cardon},
  {Schick}, {Schwalm}, {Abid}, {Fabinsky}, {Naes}, \&
  {Tsai}}]{2010AJ....140.1868W}
{Wright}, E.~L., {Eisenhardt}, P. R.~M., {Mainzer}, A.~K., {et~al.} 2010, \aj,
  140, 1868, \dodoi{10.1088/0004-6256/140/6/1868}

\bibitem[{{Xiong} {et~al.}(2017){Xiong}, {Chen}, {Yang}, {Fang}, {Zhang},
  {Zhang}, {Du}, \& {Long}}]{2017ApJ...838...49X}
{Xiong}, F., {Chen}, X., {Yang}, J., {et~al.} 2017, \apj, 838, 49,
  \dodoi{10.3847/1538-4357/aa6443}

\bibitem[{{Yan} {et~al.}(2020){Yan}, {Yang}, {Su}, {Sun}, \&
  {Wang}}]{2020ApJ...898...80Y}
{Yan}, Q.-Z., {Yang}, J., {Su}, Y., {Sun}, Y., \& {Wang}, C. 2020, \apj, 898,
  80, \dodoi{10.3847/1538-4357/ab9f9c}

\bibitem[{{Yang} {et~al.}(2002){Yang}, {Jiang}, {Wang}, {Ju}, \&
  {Wang}}]{2002ApJS..141..157Y}
{Yang}, J., {Jiang}, Z., {Wang}, M., {Ju}, B., \& {Wang}, H. 2002, \apjs, 141,
  157, \dodoi{10.1086/340038}

\bibitem[{{Yuan} {et~al.}(2018){Yuan}, {Li}, {Wu}, {Ellingsen}, {Henkel},
  {Wang}, {Liu}, {Liu}, {Zavagno}, {Ren}, \& {Huang}}]{2018ApJ...852...12Y}
{Yuan}, J., {Li}, J.-Z., {Wu}, Y., {et~al.} 2018, \apj, 852, 12,
  \dodoi{10.3847/1538-4357/aa9d40}

\bibitem[{{Yuan} {et~al.}(2021){Yuan}, {Yang}, {Du}, {Liu}, {Zhang}, {Lin},
  {Sun}, {Yan}, {Ma}, {Su}, {Sun}, \& {Zhou}}]{2021ApJS..257...51Y}
{Yuan}, L., {Yang}, J., {Du}, F., {et~al.} 2021, \apjs, 257, 51,
  \dodoi{10.3847/1538-4365/ac242a}

\bibitem[{{Zhang} {et~al.}(2019){Zhang}, {Kainulainen}, {Mattern}, {Fang}, \&
  {Henning}}]{2019A&A...622A..52Z}
{Zhang}, M., {Kainulainen}, J., {Mattern}, M., {Fang}, M., \& {Henning}, T.
  2019, \aap, 622, A52, \dodoi{10.1051/0004-6361/201732400}

\bibitem[{{Zucker} {et~al.}(2015){Zucker}, {Battersby}, \&
  {Goodman}}]{2015ApJ...815...23Z}
{Zucker}, C., {Battersby}, C., \& {Goodman}, A. 2015, \apj, 815, 23,
  \dodoi{10.1088/0004-637X/815/1/23}

\end{thebibliography}


\begin{longrotatetable}
\begin{deluxetable*}{cccccccccccccccccc}
		\setlength{\tabcolsep}{0.020in}
		\tablecaption{Properties of the identified dense clumps \label{tb-clumps}}
		\tablewidth{0pt}
		\tablehead{ \colhead{NO.} & \colhead{Clump}  & \colhead{$ \Theta_\mathrm{maj} $} & \colhead{$ \Theta_\mathrm{min} $} & \colhead{$PA$} 
		& \colhead{$v$}  & \colhead{$T_{\rm peak}$} & \colhead{$\Delta V$} & \colhead{$R_{\rm eff}$\tablenotemark{a}} 
		& \colhead{$T_{\rm ex}$} & \colhead{$\tau$} & \colhead{$N_{\mathrm{H}_{2}}$} & \colhead{$ \Sigma_{c} $}
		& \colhead{$n_{\mathrm{H}_{2}}$}  & \colhead{$ M_\mathrm{LTE} $}  & \colhead{$M_\mathrm{vir}$}  & \colhead{$\alpha_{\text {vir}}$}\\
		&   & \colhead{($ \arcmin $)} & \colhead{($ \arcmin $)} & \colhead{($\degr$)} 
		& \colhead{($\mathrm{km} \mathrm{s}^{-1}$)} & \colhead{(K)} &\colhead{($\mathrm{km} \mathrm{s}^{-1}$)} & \colhead{( pc)} 
		& \colhead{(K)} &  & \colhead{($10^{21} \mathrm{\,cm}^{-2}$)} & \colhead{($\mathrm{M}_{\odot} \mathrm{pc}^{-2}$)} 
		& \colhead{($ 10^{3}\mathrm{cm}^{-3}$)}  & \colhead{($\mathrm{M}_{\odot}$)}  & \colhead{($\mathrm{M}_{\odot}$)}  & \\
		\colhead{$(1)$} & \colhead{$(2)$} & \colhead{$(3)$} & \colhead{$(4)$} & \colhead{$(5)$} & \colhead{$(6)$} 
		& \colhead{$(7)$}  & \colhead{$(8)$} & \colhead{$(9)$} & \colhead{$(10)$} 
		& \colhead{$(11)$} & \colhead{$(12)$} & \colhead{$(13)$} 
		& \colhead{$(14)$}  & \colhead{$(15)$}  & \colhead{$(16)$} & \colhead{$(17)$}
		}
        \startdata                 
        C1    &   MWISP G154.349 +2.609   &    1.53    &    2.40     &    87.0       &    -36.36    &    7.4     &   1.88     &    1.08     &     22.8      &     0.48       &    12.10     &    656.4    &    6.55      &   2416     &     800     &    0.33    \\                                                                                                                                            
C2    &   MWISP G154.397 +2.554   &    1.47    &    2.49     &    64.4       &    -36.52    &    4.0     &   1.24     &    1.08     &     11.4      &     0.68       &    2.42      &    143.5    &    1.43      &   530      &     348     &    0.66    \\                                                                                                                                            
C3    &   MWISP G154.823 +2.432   &    4.07    &    1.63     &    -14.1      &    -38.69    &    4.0     &   1.58     &    1.57     &     12.9      &     0.53       &    3.35      &    161.2    &    1.11      &   1241     &     817     &    0.66    \\                                                                                                                                            
C4    &   MWISP G154.753 +2.835   &    2.12    &    1.86     &    11.5       &    -38.19    &    4.1     &   0.97     &    1.13     &     13.5      &     0.52       &    2.23      &    120.3    &    1.15      &   484      &     223     &    0.46    \\                                                                                                                                            
C5    &   MWISP G154.427 +2.785   &    1.67    &    3.53     &    117.8      &    -35.5     &    3.4     &   1        &    1.46     &     13.2      &     0.42       &    1.85      &    86.5     &    0.64      &   577      &     305     &    0.53    \\
\enddata
		\tablenotetext{a}{Deconvolution radius: $R_{eff} = \frac{1}{2}\times d \times \frac{\pi}{2} \sqrt{\Theta_\mathrm{maj}\Theta_\mathrm{min}-\theta_\mathrm{beam}^2}$.}
\end{deluxetable*}
\end{longrotatetable}

\begin{longrotatetable}
\begin{deluxetable*}{lcccccccccccccc}
		\tabletypesize{\tiny} 
		\setlength{\tabcolsep}{0.015in}
		\tablecaption{Infrared Photometric Magnitudes of Candidate YSOs \label{tb-YSOs}}
		\tablewidth{0pt}
		\tablehead{\colhead{allwise} & \colhead{$l$} & \colhead{$b$}  
		& \colhead{$W1$ (3.4 $\mu m)$}  & \colhead{$W2$ (4.6 $\mu m)$} &\colhead{$W3$ (12 $\mu m)$} & \colhead{$W4$ (22 $\mu m)$} & \colhead{$J$ (1.25 $\mu m)$} & \colhead{$H$ (1.65 $\mu m)$} & \colhead{$K_{\rm s}$ (2.16 $\mu m)$} &\colhead{FROM\tablenotemark{a}} &  \colhead{ GAIA$\_$SOURCE$\_$ID }
		&\colhead{ALPHA}  &\colhead{$plx$}& \colhead{distance} \\ 
		& \colhead{($\degr$)} & \colhead{($\degr$)} & \colhead{(mag)} &\colhead{(mag)} & \colhead{(mag)} & \colhead{(mag)} &\colhead{(mag)} &\colhead{(mag)} & \colhead{(mag)} & \colhead{} &\colhead{} & \colhead{} &\colhead{mas} & \colhead{kpc}  \\
		\colhead{$(1)$} & \colhead{$(2)$} & \colhead{$(3)$} &\colhead{$(4)$} & \colhead{$(5)$} & \colhead{$(6)$} &\colhead{$(7)$}  & \colhead{$(8)$} & \colhead{$(9)$} &\colhead{$(10)$} & \colhead{$(11)$} & \colhead{$(12)$}  &\colhead{$(13)$}   & \colhead{$(12)$}  &\colhead{$(13)$} 
		}
              \startdata
              J043252.83+511147.9    &     153.9931    &   2.1973    &    12.882    	$\pm$   0.024      &    12.158      $\pm$   0.023     &   10.177    $\pm$  0.068     &    8.338                    &   14.728  $\pm$  0.044  &   13.846   $\pm$  0.030   &   13.249    $\pm$  0.039  &   svmgaia          &    271001136322616576      &    -1.142    &     0.316     &    3.2     \\
J043356.97+503841.3    &     154.5128    &   1.9461    &    13.162    	$\pm$   0.025      &    12.331      $\pm$   0.024     &   9.397     $\pm$  0.040     &    8.155    $\pm$   0.272   &   15.612  $\pm$  0.056  &   14.706   $\pm$  0.064   &   14.032    $\pm$  0.050  &   svmgaia          &    258916369662466944      &    -0.198    &     0.180     &    5.6     \\
J043526.26+520059.8    &     153.6582    &   3.0447    &    10.065    	$\pm$   0.022      &    9.016       $\pm$   0.019     &   6.899     $\pm$  0.019     &    4.258    $\pm$   0.026   &   14.892  $\pm$  0.050  &   13.307   $\pm$  0.042   &   12.331    $\pm$  0.027  &   svmgaia          &    272561339025907328      &    -0.024    &     0.188     &    5.3     \\
J043722.95+504519.0    &     154.7977    &   2.4222    &    12.415    	$\pm$   0.025      &    12.048      $\pm$   0.024     &   10.619    $\pm$  0.089     &    8.206    $\pm$   0.275   &   14.405  $\pm$  0.033  &   13.523   $\pm$  0.033   &   13.072    $\pm$  0.028  &   svmgaia          &    258884891843742464      &    -1.467    &     0.276     &    3.6     \\
J043734.76+505030.2    &     154.7543    &   2.5032    &    12.040    	$\pm$   0.024      &    11.077      $\pm$   0.022     &   8.594     $\pm$  0.033     &    6.238    $\pm$   0.066   &   14.821  $\pm$  0.034  &   13.706   $\pm$  0.037   &   12.979    $\pm$  0.028  &   svmgaia          &    258886541111382784      &    -0.312    &     0.176     &    5.7     \\
\enddata
              \tablenotetext{a}{The marks of  Class$\ast$, svmgaia, and SVMClass$\ast$ indicate the candidate YSOs identified through the scheme of \cite{2014ApJ...791..131K} and obtained from  \cite{2016MNRAS.458.3479M}, and \cite{2019MNRAS.487.2522M}, respectively.
              } 
\end{deluxetable*}
\end{longrotatetable}        

\end{document}